\documentclass[showpacs,amsfonts,aps,prc,floatfix,eqsecnum,groupedaddress,]%
{revtex4}

\usepackage{amsmath,amssymb,amsthm}
\usepackage{graphicx}
\usepackage{bm}
\usepackage{color}

\DeclareMathAlphabet{\mathpzc}{OT1}{pzc}{m}{it}

\begin{document}

\renewcommand{\textfraction}{0.00}


\newcommand{\vAi}{{\cal A}_{i_1\cdots i_n}} 
\newcommand{\vAim}{{\cal A}_{i_1\cdots i_{n-1}}} 
\newcommand{\vAbi}{\bar{\cal A}^{i_1\cdots i_n}}
\newcommand{\vAbim}{\bar{\cal A}^{i_1\cdots i_{n-1}}}
\newcommand{\htS}{\hat{S}} 
\newcommand{\htR}{\hat{R}}
\newcommand{\htB}{\hat{B}} 
\newcommand{\htD}{\hat{D}}
\newcommand{\htV}{\hat{V}} 
\newcommand{\cT}{{\cal T}} 
\newcommand{\cM}{{\cal M}} 
\newcommand{\cMs}{{\cal M}^*}
\newcommand{\vk}{\vec{\mathbf{k}}}
\newcommand{\bk}{\bm{k}}
\newcommand{\kt}{\bm{k}_\perp}
\newcommand{\kp}{k_\perp}
\newcommand{\km}{k_\mathrm{max}}
\newcommand{\vl}{\vec{\mathbf{l}}}
\newcommand{\bl}{\bm{l}}
\newcommand{\bK}{\bm{K}} 
\newcommand{\bb}{\bm{b}} 
\newcommand{\qm}{q_\mathrm{max}}
\newcommand{\vp}{\vec{\mathbf{p}}}
\newcommand{\bp}{\bm{p}} 
\newcommand{\vq}{\vec{\mathbf{q}}}
\newcommand{\bq}{\bm{q}} 
\newcommand{\qt}{\bm{q}_\perp}
\newcommand{\qp}{q_\perp}
\newcommand{\bQ}{\bm{Q}}
\newcommand{\vx}{\vec{\mathbf{x}}}
\newcommand{\bx}{\bm{x}}
\newcommand{\tr}{{{\rm Tr\,}}} 
\newcommand{\bc}{\textcolor{blue}}
\newcommand{\rc}{\textcolor{red}}

\newcommand{\beq}{\begin{equation}}
\newcommand{\eeq}[1]{\label{#1} \end{equation}} 
\newcommand{\ee}{\end{equation}}
\newcommand{\bea}{\begin{eqnarray}} 
\newcommand{\eea}{\end{eqnarray}}
\newcommand{\beqar}{\begin{eqnarray}} 
\newcommand{\eeqar}[1]{\label{#1}\end{eqnarray}}
 
\newcommand{\half}{{\textstyle\frac{1}{2}}} 
\newcommand{\ben}{\begin{enumerate}} 
\newcommand{\een}{\end{enumerate}}
\newcommand{\bit}{\begin{itemize}} 
\newcommand{\eit}{\end{itemize}}
\newcommand{\ec}{\end{center}}
\newcommand{\bra}[1]{\langle {#1}|}
\newcommand{\ket}[1]{|{#1}\rangle}
\newcommand{\norm}[2]{\langle{#1}|{#2}\rangle}
\newcommand{\brac}[3]{\langle{#1}|{#2}|{#3}\rangle} 
\newcommand{\hilb}{{\cal H}} 
\newcommand{\pleft}{\stackrel{\leftarrow}{\partial}}
\newcommand{\pright}{\stackrel{\rightarrow}{\partial}}

\title{Theoretical formalism of radiative jet energy loss in a finite size \\
dynamical QCD medium}
\author{Magdalena Djordjevic}
\email[Correspond to\ ]{mdjordjevic@astate.edu}
\affiliation{Department of Chemistry and Physics, Arkansas State University,
State University, AR 72467}
\affiliation{Physics Department, The Ohio State University,
Columbus, OH 43210, USA}

\begin{abstract}
The computation of radiative energy loss in a finite size QCD medium with dynamical constituents is a key ingredient for obtaining reliable predictions for jet quenching in ultra-relativistic heavy ion collisions. We here present a theoretical formalism for the calculation of the first order in opacity radiative energy loss of a quark jet traveling through a finite size dynamical QCD medium. We show that, while each individual contribution to the energy loss is infrared divergent, the divergence is naturally regulated once all diagrams are taken into account.  Finite size effects are shown to induce a non-linear path length dependence of the energy loss, recovering both the incoherent Gunion-Bertsch limit, as well as destructive Landau-Pomeanchuk-Migdal limit. Finally, our results suggest a remarkably simple general mapping between energy loss expressions for static and dynamical QCD media.
\end{abstract}

\date{\today} 
 

\maketitle

\section{Introduction}
\label{sec1}

Suppression pattern of high transverse momentum hadrons 
is a powerful tool to map out the density of a QCD plasma created in 
ultra-relativistic heavy ion collisions 
\cite{Gyulassy_2002,Gyulassy:1990bh,Gyulassy:1991xb}. Since suppression 
(called jet quenching) results from the energy loss of high energy 
partons moving through the plasma \cite{MVWZ:2004,BDMS,BSZ,KW:2004}, 
reliable theoretical predictions for suppression require
reliable energy loss calculations.

The medium-induced radiative energy loss is, in most studies, computed by 
assuming that the QCD medium consists of randomly 
distributed  {\em static} scattering centers (``static QCD medium''). However, 
in reality, constituents of the medium are dynamical, and we recently showed 
that inclusion of dynamical QCD medium effects are important in the radiative 
energy loss calculations~\cite{DH_Inf,DH_PRL}. Furthermore, calculation of the 
energy loss has to be performed in a finite size QCD medium, since the size of 
the QCD medium created in both RHIC and LHC is finite. While methods for energy loss calculation have been developed for infinite optically thick dynamical QCD medium, no such approach exists for finite optically thin medium. However, it is of crucial importance to develop the energy loss formalism for the case of finite size optically thin medium, in order to make reliable predictions applicable for the range of parameters relevant for RHIC and LHC experiments. 

This paper develops theoretical formalism for the calculation of the 
radiative energy loss in a {\em finite size dynamical} QCD medium, while main 
numerical results of the model are presented elsewhere~\cite{DH_PRL}. Our work 
is based a novel approach, where two hard thermal loops are 
implemented within a finite size QCD model initially introduced by 
Zakharov~\cite{Zakharov}. The computations are presented in the appendices, 
and include analytical calculations of 24 
Feynman diagrams, each of which is individually infrared divergent. However, 
this divergence is naturally regulated when all
the diagrams are taken into account. Furthermore, we obtain an explicit 
analytical expression for the energy loss in dynamical medium, which can be 
directly compared with the equivalent expression obtained in static QCD 
medium. Finally, we discuss the obtained analytical result in the context of 
several qualitative features: {\em i)} Recovery of the static approximation 
for asymptotically high values of energy, {\em ii)} Transition of the 
thickness dependence from Gunion-Bertsch (GB)~\cite{GB} to 
Landau-Pomeanchuk-Migdal (LPM)~\cite{LPM} limit, and  
{\em iii)} Possible extension of the results to higher orders in opacity.

\section{Radiative energy loss in a dynamical QCD medium} 
\label{sec2}

In this Section we outline the computation of the medium induced radiative 
energy loss for a heavy quark to first order in opacity. We 
consider finite QCD medium of size $L$ and assume that the heavy 
quark is produced inside the medium at time $x_0=0$, at the left edge of the 
medium, traveling right.

Medium induced radiative energy loss is caused by the radiation of one 
or more gluons induced by collisional interactions between the quark of 
interest and partons in the medium. The energy loss rate can be expanded
in the number of scattering events suffered by the heavy quark, which is
equivalent to an expansion in powers of the opacity. For a finite medium,
the opacity is given by the product of the density of the medium with the 
transport cross section, integrated along the path of the heavy quark.
The lowest (first) order contribution corresponds to one collisional
interaction with the medium, accompanied by emission of a single gluon. 

We compute the medium induced radiative energy loss for a quark jet to the 
first (lowest) order in number of scattering centers. The finite size medium 
is introduced similarly as in~\cite{MD_Coll}, i.e. 
by starting from the approach described in~\cite{Zakharov}. That is, we assume 
that the medium has a size $L$, and that the collisional interaction has to 
occur inside the medium.     

\begin{figure}[ht]
\vspace*{4.7cm} 
\includegraphics{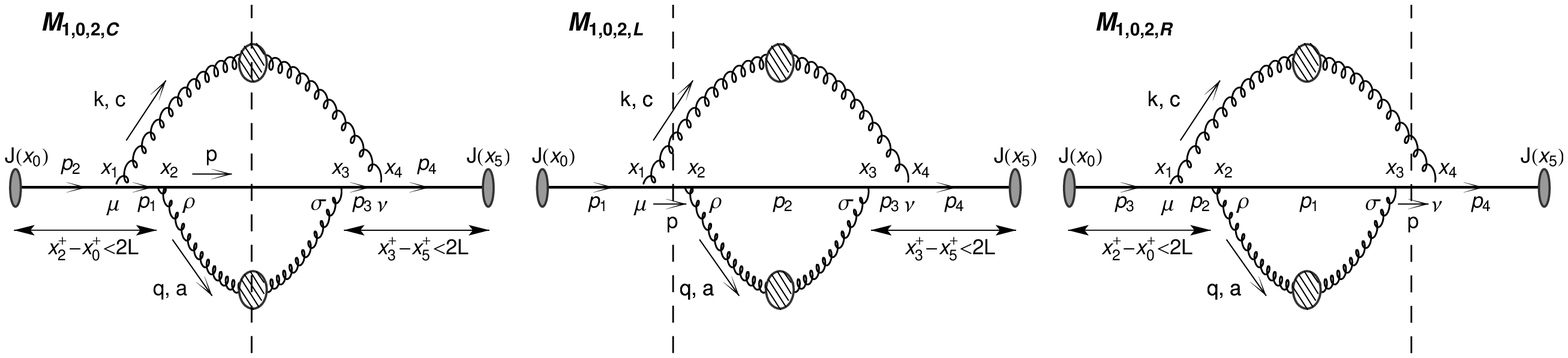}
\caption{Feynman diagrams $M_{1,0,2,C}$, $M_{1,0,2,L}$ and $M_{1,0,2,R}$ 
contribute to the radiative energy loss to the first order in the opacity. On 
each panel, left (right) gray ellipse represent the source $J$, which at time 
$x_0$ ($x_5$) produces an on-shell jet with momentum $p_2$ ($p_5$).
The large dashed circles (``blob'') represent effective HTL
gluon propagators \cite{DG_TM}. A cut gluon propagator with momentum 
$k$ and color $c$ corresponds to the radiated gluon. 
A cut gluon propagator with momentum $q$ and color $a$ corresponds to 
a collisional  interaction with a parton in the medium. Specific time points 
are represented by $x_i$. The diagrams are calculated in 
light cone coordinate system, and $x_2^+-x_0^+ < 2L$ on the first and the 
third panel, as well as $x_3^+-x_5^+ < 2L$ on the first and the second panel,
represent the condition that the distance between collisional interaction and 
jet production has to be smaller than the size $L$ of the medium. Left, middle
 and right panels present three possible cuts (central, left and right, 
respectively) of the same 2-HTL Feynman diagram, all of which contribute to 
the $1^{st}$ order in opacity radiative energy loss. 
}
\label{Diag_Rad}
\end{figure}

As in~\cite{DH_Inf}, we describe the medium by a thermalized
quark-gluon plasma at temperature $T$ and zero baryon density, with $n_f$ 
effective massless quark flavors in equilibrium with the gluons. The formalism 
for computing the energy loss in finite size dynamical QCD medium is 
presented in Appendices~\ref{appM101C}-\ref{appTadpole}, and the diagrams
are evaluated in finite temperature field theory~\cite{Kapusta, Le_Bellac}, 
using HTL resumed propagators~\cite{Le_Bellac} for all gluons.  To outline 
the calculations, in Fig.~\ref{Diag_Rad} we show three typical diagrams that 
have to be computed. The Feynman diagram in left panel of Fig.~\ref{Diag_Rad} 
represents 
the source $J$, which at time $x_0$ produces an on-shell jet with momentum 
$p_2$, and subsequently  radiates a gluon with momentum $k=(\omega,k_z,\bk)$ 
and exchanges a virtual gluon of momentum $q=(q_0,q_z,\bq)$ with a parton in 
the medium. The quark jet emerges with (measured) momentum $p=(E,p_z,\bm{p})$. 
We assume, as in~\cite{GLV}, that $J$ changes slowly with jet momentum, i.e. 
that $J(p + k +q)\approx J(p+k) \approx J(p)$. To incorporate the effect of finite 
size QCD medium, we assume that the distance between the jet production and 
collisional interaction has to be smaller than the size of the medium. 

Since the produced jet can be off-shell, the Feynman diagrams shown in the  
center and right panels of Fig.~\ref{Diag_Rad} also contribute to the $1^{st}$ 
order in opacity radiative energy loss. They are complex conjugates to each 
other and they present the terms which will interfere with the diagram shown 
in the left panel of Fig.~\ref{Diag_Rad}, and consequently lead to the 
appearance of LPM effect 
(in the case of high energy jets), after all relevant contributions are taken 
into account. In addition to the three diagrams shown in Fig.~\ref{Diag_Rad}, 
there are 21 more diagrams that contribute to the $1^{st}$ order in opacity 
radiative energy loss, and their calculation is presented in the 
Appendices~\ref{appM101C}-\ref{appTadpole}. Note that the calculation 
presented in this paper differs from Ref.~\cite{DG_Ind} by the use of HTL 
gluon propagators to describe the interaction of the quark with the medium. 
The difference from Ref.~\cite{DH_Inf} is that in this work we allow the jet 
to be on- or off-shell and the vertices that 
correspond to gluon exchange are restricted to be located inside the medium. 

Since the exchanged gluon momentum is space-like
\cite{MD_Coll,BT,BT_fermions}), only the Landau damping contribution 
($q_0 \le |\vq|$) to the cut HTL effective gluon propagator 
$D(q)$ needs to be taken into account \cite{MD_Coll,TG,BT}. The radiated gluon 
has timelike momentum $k=(\omega,\vk)$, so only the 
quasi-particle contribution at 
$\omega> |\vk|$ from the cut gluon propagator $D(k)$ contributes
\cite{DG_TM,Kapusta,Le_Bellac}.  Since our focus is on heavy quarks with mass $M\gg gT$, we neglect 
the thermal shift of the heavy quark mass. 

The effective gluon propagator has both transverse and longitudinal 
contributions
\cite{Kalash,Klimov,Weldon,Heinz_AP,Pisarski:cs,Rebhan,Gyulassy_Selikhov}. 
The 1-HTL gluon propagator has the form
\beq
 i D^{\mu\nu}(l)=
 \frac{P^{\mu \nu }(l)}{l^2{-}\Pi_T (l)} + 
 \frac{Q^{\mu \nu }(l)}{l^2{-}\Pi_L (l)}\,,
\eeq{dmnMed}
where $l=(l_0, \vl)$ is the 4-momentum of the gluon and 
$P_{\mu \nu}(l)$ and $Q_{\mu \nu}(l)$ are the transverse and
longitudinal projectors, respectively. The transverse and longitudinal 
HTL gluon self energies $\Pi_T$ and $\Pi_L$ are given 
by
\bea
\label{PiT}
\Pi_T (l) &=& \mu^2 \left[ \frac{y^2}{2} + \frac{y (1{-}y^2)}{4} 
\ln\left(\frac{y{+}1}{y{-}1}\right)\right],
\qquad
\Pi_L (l) = \mu^2 \left[ 1 - y^2 - \frac{y (1{-}y^2)}{2} 
\ln\left(\frac{y{+}1}{y{-}1}\right)\right],
\eea
where $y \equiv l_0/|\vl|$ and $\mu=gT\sqrt{N_c/3+N_f/6}$ is the
Debye screening mass.

While the results obtained in this paper are gauge invariant 
\cite{BT_fermions}, the calculation is for simplicity presented in 
Coulomb gauge. In this gauge the only nonzero terms in the transverse 
and longitudinal projectors are
\bea
P^{i j} (l) &=& -\delta^{ij} + \frac{l^i  l^j}{\vl^2},
\label{PQmunu}
\qquad\qquad
Q^{00}(l) = -\frac{l^2}{\vl^2} = 1 -\frac{l_0^2}{\vl^2} = 1-y^2.
\eea

As in~\cite{GLV,Gyulassy_Wang,Wiedemann,WW,DG_Ind,ASW,MD_TR,AMY}, we use the 
same kinematic approximations, i.e. we assume 
validity of the soft gluon ($\omega \ll E$) and soft rescattering 
($\omega \gg |\bk| \sim |\bq| \sim q_0,q_z$) approximations (see 
Appendix~\ref{appa} for details). In Appendices 
\ref{appM101C}-\ref{appTadpole} we compute all the diagrams that 
contribute to the first order in opacity radiative energy loss. Once the 
diagrams are calculated, the interaction rate is given by:
\beqar
\Gamma (E) &=& \frac{1}{N_J} M_{tot}  = \frac{1}{N_J} 
(M_{1,0}  + M_{1,1}  + M_{1,2} ),
\eeqar{Gamma}
here $M_{1,0} $, $M_{1,1} $ and $M_{1,2} $ present the sum of all 
contributions in which zero, one or two (respectively) ends of the exchanged 
gluon $q$ are attached to the radiated gluon $k$. Furthermore, $N_J$ is an 
integrated invariant distribution of jets, created by the effective jet 
source current, and given by~\cite{GLV} (note that $D_R=3$ accounts for the 
jet colors)
\beqar 
N_J =  D_R \int  \frac{d^3p}{(2\pi)^3 2 E} |J(p)|^2.
\eeqar{NJ}

Equations~(\ref{M101C_f}),~(\ref{M102C_f}),~(\ref{M102RL_f}),~(\ref{M1034C_f}),~(\ref{M1034RL_f}) and~(\ref{M1056RL_f}) give final results for the Feynman diagrams contributing to $M_{1,0}$. After adding these expressions $M_{1,0}$ becomes
\beqar
M_{1,0}  &=& 4 L T \, g^4 \, [t_a, t_c]  [t_c, t_a] 
\int  \frac{d^3p}{(2\pi)^3 2 E} \,|J(p)|^2 \, \int
 \frac{d^3k}{(2\pi)^3 2 \omega} 
\, \frac{d^2q}{(2\pi)^2} \, \frac{\mu^2}{\bq^2 (\bq^2{+}\mu^2)} 
 \times \,\frac{\bk^2}{(\bk^2{+}\chi)^2}\,
\left(1- \frac{\sin (\xi L)}{\xi L} \right) \, \nonumber \\
&=& D_R \int  \frac{d^3p}{(2\pi)^3 2 E} |J(p)|^2 \ 
\frac{C_R \alpha_s}{\pi}\, \frac{L}{\lambda_{dyn}}
\int \frac{dx}{x}\, \frac{d^2k}{\pi}\, \frac{d^2q}{\pi} \, 
\frac{\mu^2}{\bq^2 (\bq^2+\mu^2)}   \times \,\frac{\bk^2}{(\bk^2{+}\chi)^2}\,
\left(1- \frac{\sin (\xi L)}{\xi L} \right) \,,
\eeqar{M10_f}  
where $\chi \equiv M^2 x^2 + m_g^2$ and $\xi \equiv \frac{\bk^2+\chi}{x E^{+}}$ are defined by Eq.~(\ref{xi-zeta}) in appendix A.

Here and below we used $[t_a,t_c] \, [t_c, t_a] = C_2 (G) C_R D_R$ 
(with $C_2(G)=3$, $C_R=\frac{4}{3}$, and $D_R=3$) and defined 
a ``dynamical mean free path'' (see~\cite{DH_Inf}) $\lambda_\mathrm{dyn}$ 
through 
\beq
\lambda_\mathrm{dyn}^{-1} \equiv C_2(G) \alpha_s T \; = 3 \alpha_s T\,,
\eeq{lambda_dyn}
where $\alpha_s = \frac{g^2}{4 \pi}$ is coupling constant, and we assumed 
constant coupling $g$. 

Equations~(\ref{M1112C_f}),~(\ref{M1112RL_f}),~(\ref{M1134C_f}) and~(\ref{M1134RL_f}) give final results for the Feynman diagrams contributing to $M_{1,1}$. After adding these expressions $M_{1,1}$ becomes
\beqar
M_{1,1}  &=&  D_R \int  \frac{d^3p}{(2\pi)^3 2 E} |J(p)|^2 \ 
\frac{C_R \alpha_s}{\pi}\, \frac{L}{\lambda_{\mathrm{dyn}}}
\int \frac{dx}{x}\, \frac{d^2k}{\pi}\, \frac{d^2q}{\pi} \, 
\frac{\mu^2}{\bq^2 (\bq^2+\mu^2)} \nonumber \\
&& \; \; \times \,\frac{-\, 2 \,\bk \cdot (\bk{+}\bq)}
{(\bk^2{+}\chi) ((\bk{+}\bq)^2{+}\chi)}\,
\left(1- \frac{\sin (\zeta L)}{\zeta L} \right) \,,
\eeqar{M11_f}  
where $\zeta \equiv \frac{(\bk{+}\bq)^{2}+\chi}{x E^{+}}$ is defined by 
Eq.~(\ref{xi-zeta}) in appendix A.

Equations~(\ref{M12C_f}) and~(\ref{M12RL_f}) give final results for the 
Feynman diagrams contributing to $M_{1,2}$. After adding these expressions 
$M_{1,2}$ becomes
\beqar M_{1,2} 
&=& D_R \int  \frac{d^3p}{(2\pi)^3 2 E} |J(p)|^2 \ 
\frac{C_R \alpha_s}{\pi}\, \frac{L}{\lambda_{\mathrm{dyn}}}
\int \frac{dx}{x}\, \frac{d^2k}{\pi}\, \frac{d^2q}{\pi} \, 
\frac{\mu^2}{\bq^2 (\bq^2+\mu^2)} 
\nonumber \\  && \;\;
\times\
\left[ \frac{2\,(\bk{+}\bq)^2}{((\bk{+}\bq)^2+\chi)^2}
\left(1-\frac{\sin (\zeta L)}{\zeta L}\right)\, - \,
\frac{\bk^2}{(\bk^2{+}\chi)^2}\,
\left(1- \frac{\sin (\xi L)}{\xi L} \right) \right]\, .
\eeqar{M12_f}

By using Eqs.~(\ref{NJ})-(\ref{M12_f}) the interaction rate 
(Eq.~(\ref{Gamma})) reduces to 
\beqar
\Gamma (E) =  \frac{C_R \alpha_s}{\pi}\,\frac{L}{\lambda_{\mathrm{dyn}}}
\int \frac{dx}{x}\, \frac{d^2k}{\pi}\, \frac{d^2q}{\pi} \, 
\frac{\mu^2}{\bq^2 (\bq^2+\mu^2)}  \,\frac{2 \,(\bk{+}\bq)}
{ (\bk{+}\bq)^2{+}\chi}\, \left (\frac{ \,\bk{+}\bq}
{ (\bk{+}\bq)^2{+}\chi}\, -\frac{ \bk}{\bk^2{+}\chi}
\right)
\left(1- \frac{\sin (\zeta L)}{\zeta L} \right) \,.
\eeqar{Gamma_f}  

The heavy quark radiative energy loss per unit length is obtained 
from the above expression for the interaction rate by weighting it with
the lost energy $\omega + q_0$. In the soft 
rescattering approximation $\omega + q_0 \approx \omega $, leading to:
\beqar
\frac{d E_{\mathrm{dyn}}}{dL}
&=& \frac{1}{D_R}  \int d\omega\, \omega \frac{d\Gamma(E)}{d\omega} 
\approx \frac{E}{D_R} \int dx\, x\frac{d\Gamma(E)}{dx} .\quad
\eeqar{dEdl}
This finally leads to 
\beqar
\frac{\Delta E_{\mathrm{dyn}}}{E} 
&=& \frac{C_R \alpha_s}{\pi}\,\frac{L}{\lambda_\mathrm{dyn}}  
    \int dx \,\frac{d^2k}{\pi} \,\frac{d^2q}{\pi} \, 
    \frac{\mu^2}{\bq^2 (\bq^2+\mu^2)} \nonumber \\
&& \; \; \times
\, 2 \,\frac{\bk{+}\bq}
{ (\bk{+}\bq)^2{+}\chi}\,\cdot \left (\frac{ \,\bk{+}\bq}
{ (\bk{+}\bq)^2{+}\chi}\, -\frac{ \bk}{\bk^2{+}\chi}\right)
\left(1- \frac{\sin (\frac{ (\bk{+}\bq)^2{+}\chi}{xE^+}L)}
{\frac{ (\bk{+}\bq)^2{+}\chi}{xE^+} L} \right) \, .
\eeqar{DeltaEDyn}

\medskip
It is important to note that, similarly to~\cite{DH_Inf}, each individual 
diagram that contributes to the energy loss in a finite size dynamical QCD 
medium diverges logarithmically in the limit of zero transverse 
momentum of the exchanged gluon, $\bq{\,\to\,}0$. The reason 
for this divergence is that in a dynamical QCD medium both transverse and 
longitudinal gluon exchange contribute to the radiative energy 
loss~\cite{Wang_Dyn}. While Debye screening 
makes the longitudinal gluon exchange infrared finite, transverse gluon 
exchange causes a well-known logarithmic singularity \cite{Le_Bellac} 
due to the absence of a magnetic screening \cite{fn1}. Remarkably, 
we see from Eq.~(\ref{DeltaEDyn}) that, when the contributions from all 
diagrams are taken into account, the infrared divergences cancel, naturally 
regulating the energy loss rate.

The analytical expression for the energy loss in dynamical medium that we 
obtained, can now be directly compared with the equivalent expression for 
static QCD medium. We below make this comparison in order to study the 
importance of the dynamical QCD medium 
effects. To do that, we 
here rewrite the expression for the DGLV first order in opacity radiative 
energy loss in static QCD medium, which is obtained in~\cite{DG_Ind}:
%
\beqar
\frac{\Delta E_{\mathrm{stat}}}{E} 
&=& \frac{C_R \alpha_s}{\pi}\,\frac{L}{\lambda_\mathrm{stat}}  
    \int dx \,\frac{d^2k}{\pi} \,\frac{d^2q}{\pi} \, 
    \frac{\mu^2}{(\bq^2+\mu^2)^2} \,
     \nonumber \\
    && \; \; \; \times \, 
    2 \, \frac{\bk{+}\bq}{(\bk{+}\bq)^2+\chi}
    \cdot \left(\frac{\bk{+}\bq}{(\bk{+}\bq)^2+\chi}
    - \frac{\bk}{\bk^2+\chi}
    \right) \left(1-\frac{\sin{\frac{(\bk{+}\bq)^2+\chi}{x E^+} \, L}} 
    {\frac{(\bk{+}\bq)^2+\chi}{x E^+} \, L} \right),
\eeqar{DeltaEStat}
%
with~\cite{WHDG,DH_Inf}
\beqar
\frac{1}{\lambda_\mathrm{stat}}=\frac{1}{\lambda_{g}}+\frac{1}{\lambda_{q}}=
6 \frac{1.202}{\pi^2} \frac{1+\frac{n_f}{4}}{1+\frac{n_f}{6}} 3 \alpha_s T
= c(n_f) \; \frac{1}{\lambda_\mathrm{dyn}} \;,
\eeqar{lambda_stat}
where $c(n_f) \equiv 6 \frac{1.202}{\pi^2} \frac{1+n_f/4}{1+n_f/6}$
is a slowly increasing function of $n_f$ that varies between 
$c(0)\approx 0.73$ and $c(\infty)\approx1.09$. For a typical value 
$n_f=2.5$ (which we use in this paper), $c(2.5) \approx 0.84$.

We see that, similarly to the case of infinite QCD medium~\cite{DH_Inf}, 
Eqs.~(\ref{DeltaEDyn}) and (\ref{DeltaEStat}) are remarkably similar, up to 
two important differences: The first is an $\mathcal{O}(15\%)$ 
decrease in the effective mean free path
\beq
\lambda_\mathrm{dyn} \Longleftrightarrow 
\lambda_\mathrm{stat}=\frac{\lambda_\mathrm{dyn}}{c(n_f)}
\eeq{diff_lam}
which increases the energy loss rate in the dynamical medium by 
$\mathcal{O}(20\%)$. The second difference is a change in the effective 
crossection, which in the energy loss rate is reflected by the replacement
\beq
\left[ \frac{\mu^2}{\bq^2 (\bq^2{+}\mu^2)} \right]_\mathrm{dyn} 
\Longleftrightarrow
\left[\frac{\mu^2}{(\bq^2{+}\mu^2)^2}\right]_\mathrm{stat}.
\eeq{diff_cross}
As discussed in~\cite{DH_PRL}, these differences lead to a significant 
increase of the heavy quark energy loss rate in dynamical compared to 
static QCD medium. 

By using the above results, we can now discuss the following two issues: 
1) Comparison between dynamical and static energy loss results, 2) Comparison 
between energy loss results in Bethe-Heitler limit and finite size QCD medium. 
Regarding the first comparison, one should note a remarkably simple mapping 
between dynamical and static QCD medium. That is, the expression for energy 
loss in dynamical QCD medium can be obtained from the expression for the 
energy loss in static QCD medium by simply replacing the effective mean free 
path and the effective crossection by the appropriate expressions given above. 
The simplicity of this substitution rule is surprising, given the complexity
of the calculations and their different structure for static~\cite{DG_Ind}
and dynamical media. In particular, one should note the infrared divergences 
in the dynamical case which cancel only after summing all the diagrams from 
Appendices~\ref{appM101C}-\ref{appTadpole}, but don't arise at all in the 
static case. Regarding the second comparison, the 
expressions for both dynamical and static energy loss in finite size QCD 
medium are significantly different from the corresponding expressions in 
Bethe-Heitler limit~\cite{DH_Inf}. However, despite this difference, the same 
simple substitution rule is found to apply, suggesting a possibly general
mapping between static and dynamic QCD media.

We also note that the study presented here considers a finite, optically
thin dynamical QCD medium (QGP), extending the DGLV approach~\cite{DG_Ind,GLV} 
to include parton recoil. In this sense it is complementary to the work by 
Arnold, Moore and Yaffe~\cite{AMY} who study energy loss in an infinite, 
optically thick QGP. We note that the AMY approach~\cite{AMY} yields
the same form~(\ref{diff_cross}) for the effective cross section in a dynamical
QCD medium as found here (see also~\cite{Aurenche}), further supporting
our conjecture above.

\section{Incoherent Limit}
\label{Incoh}

First of the two relevant limits of the Eq.~(\ref{DeltaEDyn}), which we 
consider in this paper is the incoherent, short formation time limit. This 
limit is relevant for heavy quarks at lower jet energy regions, and it can
be extracted from the Eq.~(\ref{DeltaEDyn}) when 
$\frac{\sin (\frac{ (\bk{+}\bq)^2{+}\chi}{xE^+}L)}
{\frac{ (\bk{+}\bq)^2{+}\chi}{xE^+} L} \rightarrow 0$:
\beqar
\frac{\Delta E_{\mathrm{dyn}}}{E} 
&=& \frac{C_R \alpha_s}{\pi}\,\frac{L}{\lambda_\mathrm{dyn}}  
    \int dx \,\frac{d^2k}{\pi} \,\frac{d^2q}{\pi} \, 
    \frac{\mu^2}{\bq^2 (\bq^2+\mu^2)}
\,\frac{2 \,(\bk{+}\bq)}
{ (\bk{+}\bq)^2{+}\chi}\, \cdot \left (\frac{ \,\bk{+}\bq}
{ (\bk{+}\bq)^2{+}\chi}\, -\frac{ \bk}{\bk^2{+}\chi}\right) \,\nonumber \\
&=& \frac{C_R \alpha_s}{\pi}\,\frac{L}{\lambda_\mathrm{dyn}}  
    \int dx\, d \bk^2 \, d \bq^2 \, 
\frac{\mu^2}{\bq^2 \, (\bq^2+\mu^2)} \; \frac{1}{\bk^2+\chi} \;
\nonumber \\
&\times& 
\left[ 
       \frac{\bk^2+3\bq^2+\chi}{\sqrt{\chi^2 + 2\chi(\bk^2{+}\bq^2)
                                              + (\bk^2{-}\bq^2)^2}}
           - 1
     - \frac{2\, \bq^2 \, ((\bk^2{-}\bq^2)^2 +\chi (5\bk^2 + \bq^2))}
            {\left[\chi^2 + 2\chi(\bk^2{+}\bq^2) 
                          + (\bk^2{-}\bq^2)^2\right]^{\frac{3}{2}}}
\right] ,
\eeqar{DeltaIncohEfi}
where the second step is obtained after angular integration. Furthermore, 
under the assumption that $\alpha_s$ is not running, 
Eq.~(\ref{DeltaIncohEfi}) can 
be further analytically integrated over $0\leq |\bk| \leq \km$ 
where $\km= 2 E \sqrt{x (1-x)}$~\cite{DG_Ind}. We obtain 
\beqar
\frac{\Delta E_{\mathrm{dyn}}}{E} = 
\frac{C_R \alpha_s}{\pi^2}\,\frac{L}{\lambda_\mathrm{dyn}}
\int dx \, d^2q\, \mathcal{J}_{\mathrm{dyn}}(\bq,x)\,,
\eeqar{DeltaEq}
where
%
%
\bea
\mathcal{J}_{\mathrm{dyn}}(\bq,x) &=& \frac{\mu^2}{\bq^2 \, (\bq^2+\mu^2)} 
\Biggl[ -1 +\frac{\bq^2-\km^2+\chi}
                 {\sqrt{\bq^4+2\bq^2(\chi{-}\km^2) +(\km^2{+}\chi)^2}}
\nonumber \\ 
&+&
\ln \left[\frac{\chi+\km^2-\bq^2+\sqrt{\bq^4+2\bq^2(\chi{-}\km^2) +(\km^2{+}\chi)^2}}{2 \left(\chi+\km^2\right)}\right]
\nonumber \\ 
&+&
\frac{(\bq^2+2\chi)}{\bq^2\sqrt{1{+}\frac{4\chi}{\bq^2}}}
\ln\Biggl(\frac{\km^2{+}\chi}{\chi}\,
         \frac{(\bq^2{+}3\chi) + \sqrt{1{+}\frac{4\chi}{\bq^2}}\,
              (\bq^2{+}\chi)}
              {(\bq^2{-}\km^2{+}3\chi) + \sqrt{1{+}\frac{4\chi}{\bq^2}}
               \sqrt{\bq^4+2\bq^2(\chi{-}\km^2)+(\km^2{+}\chi)^2}}
   \Biggr)\Biggr] .
\nonumber
\eeqar{DeltaEDynIncoh}
By comparing Eqs.~(\ref{DeltaIncohEfi})-(\ref{DeltaEDynIncoh}) with the 
equivalent expressions from~\cite{DH_Inf} (see Eqs.~(2.9)-(2.12)), we see 
that, 
though similar, the corresponding expressions are not the same. The reason is 
that, in~\cite{DH_Inf}, we considered the Bethe-Heitler limit, which considers
on-shell quark jets produced at remote past, while in our study, we allow that 
quark jets can be both on- and off-shell.

\section{LPM Limit}
\label{LPM}

The second relevant limit that we consider is the Non-Abelian 
analog of the Landau-Pomeanchuk-Migdal limit. This limit is relevant  
for highly energetic jets, where destructive interference effects 
reduce the energy loss relative to the incoherent limit. The limit is obtained 
from the Eq.~(\ref{DeltaEDyn}) when $E^+ \approx 2 E \rightarrow \infty$. In 
such a limit finite mass effects are negligible. Additionally, 
$\km= 2 E \sqrt{x (1-x)} \rightarrow \infty$ as well, which enables us to 
introduce a substitution $\bk^\prime \equiv \bk + \bq$ in Eq.~(\ref{DeltaEDyn}). With these simplifications, Eq.~(\ref{DeltaEDyn}) becomes
\beqar
\frac{\Delta E_{\mathrm{dyn}}}{E} 
&=& \frac{C_R \alpha_s}{\pi}\,\frac{L}{\lambda_\mathrm{dyn}}  
    \int dx \,\frac{d^2k^\prime}{\pi} \,\frac{d^2q}{\pi} \, 
    \frac{\mu^2}{\bq^2 (\bq^2+\mu^2)} 
\, \frac{ 2 \, \bq \cdot (\bq-\bk^\prime)}
{ \bk^{\prime 2} \, (\bk^\prime-\bq)^2}\, 
\left(1- \frac{\sin (\frac{\bk^{\prime 2} \, L}{2xE})}
{\frac{\bk^{\prime 2} \, L}{xE^+}} \right) \, \nonumber \\
&=& 2 \, \frac{C_R \alpha_s}{\pi}\,\frac{L}{\lambda_\mathrm{dyn}}  
    \int_0^1 dx \, \int_0^\infty d \bk^{\prime 2} \,\int_0^{q_{max}^2} d \bq^2 \, 
    \frac{\mu^2}{\bq^2 (\bq^2+\mu^2)} \, \theta (|\bq| -|\bk^\prime|)
\left(1- \frac{\sin (\frac{\bk^{\prime 2} \, L}{2 xE })}
{\frac{\bk^{\prime 2} \, L}{2xE}} \right) \, ,
\eeqar{DeltaEDyn0}
where in the second step we performed angular integration. To proceed 
further, we observe that the derivative over distance $L$ of the above 
expression (i.e. $\frac{dE}{dL}$) is equal to
\beqar
\frac{1}{E} \frac{dE_{\mathrm{dyn}}}{dL} &=& 2 \, \frac{C_R \alpha_s}{\pi}\,
\frac{1}{\lambda_\mathrm{dyn}}  
    \int_0^1 dx \, \int_0^\infty d \bk^{\prime 2} \, \int_0^{q_{max}^2} 
d \bq^2 \, 
    \frac{\mu^2}{\bq^2 (\bq^2+\mu^2)} \, \theta (|\bq| -|\bk^\prime|)
\left(1- \cos (\frac{\bk^{\prime 2} \, L}{2xE}) \right) \nonumber \\
&=& 2 \, \frac{C_R \alpha_s}{\pi}\,
\frac{1}{\lambda_\mathrm{dyn}}  
    \int_0^1 dx \, \int_0^{q_{max}^2} d \bq^2 \, 
    \frac{\mu^2}{\bq^2 (\bq^2+\mu^2)} \,
\left(\gamma - {\rm Ci}\left(\frac{L \bq^2}{2 E x}\right) + 
\ln \left( \frac{L \bq^2}{2 E x} \right) \right)
\, ,
\eeqar{dEdl_xq} 
where in the second step we performed integral over $\bk^{\prime 2}$ and 
$\gamma \approx 0.577216$ is Euler's constant and ${\rm Ci} (y)$ gives 
the cosine integral function. 

Finally after performing the integration over $x$, we obtain
\beqar
\frac{1}{E} \frac{dE_{\mathrm{dyn}}}{dL} &=&
4 \, \frac{C_R \alpha_s}{\pi}\,
\frac{1}{\lambda_\mathrm{dyn}}  
    \int_0^{q_{max}^2} d \bq^2 \, 
    \frac{\mu^2}{\bq^2 (\bq^2+\mu^2)} \,
\left( Z \left (\frac{L \bq^2}{2 E} \right ) + \pi \frac{L \bq^2}{8 E}
\right)
\, ,
\eeqar{dEdl_LPM} 
where 
\beqar
Z(y) \equiv \gamma - \frac{1}{2} \cos(y)- Ci(y) + \ln(y) + 
\frac{\sin(y)}{2 y} -  \frac{y \, Si(y)}{2} \underset{y \to 0}{\longrightarrow} 0 .
\eeqar{Zlim}

Therefore, for the asymptotically large jet energies, the Eq.~(\ref{dEdl_LPM}) reduces to
\beqar
\frac{1}{E} \frac{dE_{\mathrm{dyn}}}{dL} &=&
\frac{C_R \alpha_s}{2\, E}\,
\frac{L}{\lambda_\mathrm{dyn}}  
    \int_0^{q_{max}^2}  d \bq^2 \, 
    \frac{\mu^2}{\bq^2+\mu^2} \, = 
\frac{C_R \alpha_s}{2\, E}\, \frac{L}{\lambda_\mathrm{dyn}} \, \mu^2 \,  
    \ln \left( \frac{4 E T}{\mu^2} \right) 
\, ,
\eeqar{dEdl_LPM_f} 
where we used $\qm= \sqrt{4 E T}$~\cite{Adil}. Finally, $\frac{\Delta E}{E}$ then becomes
\beqar
\frac{\Delta E_{\mathrm{dyn}}}{E} = 
\frac{C_R \alpha_s}{4\, E}\, \frac{L^2 \mu^2}{\lambda_\mathrm{dyn}} \,   
    \ln \left( \frac{4 E T}{\mu^2} \right)\,.
\eeqar{ELossLPM_dyn}
From the above expressions, we see that at asymptotically large jet energies 
we obtain quadratic thickness dependence for the energy loss, that is we 
recover LPM limit. Therefore, for highly energetic jet, finite size 
corrections 
implemented in the calculation presented in this paper simulate the destructive
effects of LPM interference in an infinite QCD medium~\cite{BDMS}.
This behavior is expected~\cite{GLV} since the nuclear medium
has finite dimensions that may be small compared to the
jet radiation coherence length, especially in the case of
light partons or high jet energies. Due to this, in finite
size media the basic formation time physics developed by
LPM~\cite{LPM} leads to destructive interference effects
on the quark quenching.

Furthermore, by applying the same procedure, it is straightforward to obtain 
that, for asymptotically large jet energies, the radiative energy loss in 
static QCD medium becomes
\beqar
\frac{\Delta E_{\mathrm{stat}}}{E} = 
\frac{C_R \alpha_s}{4\, E}\, \frac{L^2 \mu^2}{\lambda_\mathrm{stat}} \,   
    \left(\ln \left( \frac{4 E T}{\mu^2} \right)-1 \right) \, ,
\eeqar{ELossLPM_stat}
so that ratio between energy losses in dynamical and static QCD medium 
approaches
\beqar
\label{ratio_assymp}
  \lim_{E \rightarrow \infty} \frac{\Delta E_{\mathrm{dyn}}}
  {\Delta E_{\mathrm{stat}}}= \lim_{E \rightarrow \infty}
  \frac{\lambda_\mathrm{stat}} {\lambda_\mathrm{dyn}} 
  \frac{\ln\frac{4 E T}{\mu^2}}{\ln\frac{4 E T}{\mu^2}{-}1} 
  = \frac{\lambda_\mathrm{stat}} {\lambda_\mathrm{dyn}} \,.
\eeqar
Therefore, we conclude that, at asymptotically large jet energies, 
approximation of the medium by a random distribution
of static scatterers becomes valid, up to a multiplicative constant 
$\frac{\lambda_\mathrm{stat}} {\lambda_\mathrm{dyn}}$ which can be 
renormalized. This is consistent with what would be 
expected from established BDMPS results~\cite{BDMS}.

\section{Conclusion}

In this paper, we developed a theoretical formalism for the calculation of the
first order in opacity radiative energy loss of a fast quark traveling through 
a finite dynamical QCD medium. We obtain a closed 
analytical expression for the energy loss in dynamical medium. The obtained 
result is convergent, despite the fact that each individual contribution to 
the energy loss is infrared divergent. Furthermore, the energy loss has a 
nontrivial dependence on the size of the medium, which depends on both mass 
and energy of the quark jet. The finite size effects are found to be most 
important in the ultra-relativistic limit and they effectively reproduce the 
effects of destructive LPM interference. Another interesting limit is 
an incoherent (GB) limit, which is reproduced for heavy quarks 
with moderately small jet energies.

The study presented here considered the radiative energy loss in finite size 
dynamical QCD medium up to the first order in  opacity. On the other hand, 
in static QCD medium, radiative energy loss up to all orders in opacity 
is obtained~\cite{GLV,DG_Ind}. Simplicity of the mapping between the 
$1^{st}$ order in opacity static and dynamical energy loss, implies that the 
same mapping might be generalized to higher order in opacity as well. That is, 
we make a conjecture that, to obtain the energy loss expressions in dynamical 
QCD medium from the existing static QCD medium expressions, one (only) needs to
replace effective mean free path and effective crossection from static QCD 
medium, with the corresponding expressions from dynamical QCD medium.
However, to prove this conjecture is very non-trivial, which will be a subject 
of further research.

The measurement of the heavy flavor suppression is in the current focus of 
intensive experimental efforts, and these measurements are expected to became 
available soon at the upcoming high-luminosity RHIC and LHC experiments. As 
already mentioned, particle suppression is a consequence of the energy loss. 
Our study, which incorporates dynamical effects in realistic finite size QCD 
medium, enables us to provide the most reliable computations of the energy 
loss in QGP. Our future goal is to use these energy loss calculations to make 
accurate theoretical predictions for the heavy flavor suppression. 
These predictions can then be directly compared with the upcoming experimental 
data, in order to test our understanding of QGP, and to further study the 
properties of this novel form of matter.

\begin{acknowledgments}
Part of this work was done while M. D. was supported by the U.S.
Department of Energy, grant DE-FG02-01ER41190.
\end{acknowledgments}

\appendix

\section{Notation and assumptions}
\label{appa}

In the following appendices, the calculation will mostly (with the exception 
of $G_{-+}$ propagators), be done using the light cone coordinate 
system~\cite{Kogut}. This coordinate system is appropriate for systems moving 
with almost the speed 
of light. It is obtained by choosing new spacetime coordinates 
$[x^+, x^-, \bx]$, related to the coordinates in the laboratory frame 
$(t, z, \bx)$ by ($\bx$ is the transverse coordinate)
\beqar
x^+ = (t + z) , \, \,  x^- = (t - z). 
\eeqar{x+-}

In the same way the light cone momentum $[p^+, p^-, \bp]$ is related to the 
momentum in the laboratory frame $[E, p_z, p]$ by ($p$ is the transverse 
momentum)
\beqar
p^+ = (E + p_z) , \, \, p^- = (E - p_z).
\eeqar{p+-}

In this paper we consider a quark of a finite mass $M$, produced inside the 
finite size QGP, at some initial point $x_0$. The jet is assumed to have a 
large spatial momentum $p'\gg M$, and it can be both on-shell and off-shell. 
This is in contrast to the Bethe-Heitler limit, which we considered 
in~\cite{DH_Inf}, where an on-shell jet was created at $-\infty$. 
Further, we choose 
coordinates such that the momentum of the initial quark is along the $z$ axis:
\beqar
p'=\left[E'^+, p'^-,\, \bm{0}\right] \; ,
\eeqar{pprime0}
We are interested in the radiative energy loss to first order in the 
opacity, so we study the case in which the quark exchanges (in arbitrary 
sequence) one virtual gluon with space-like momentum 
\beqar
  q =  [q^+,q^-,\bq]=(q_0,\vq) =(q_0,\,q_z,\,\bq), \qquad q_0 \le |\vq|
\eeqar{q} 
with a parton in the medium and radiates one (medium-modified) real gluon 
with time-like momentum 
\beq
  k = [k^+,k^-,\bk]= ( k_0,\vk) = (\omega,k_z,\bk), \qquad k_0 \geq|\vk|
\eeq{k} 
into the medium. The on-shell quark jet emerges with 4-momentum $p^\mu$. 

\medskip

Similarly to all energy loss formalisms developed so far (see 
e.g.~\cite{BDMS,Gyulassy_Wang,GLV,Wiedemann,WW,DG_Ind,ASW,MD_TR,AMY}), 
the calculations presented in Appendices~\ref{appM101C}-\ref{appTadpole} are 
developed 
under the assumption of a perturbative high-temperature QGP. We note that, 
strictly speaking, this assumption may not be directly applicable to the 
case of a strongly coupled QGP or in cold nuclear matter. The validity of
this assumption for the QGP created in ultra-relativistic heavy-ion 
collisions can be tested by comparison with upcoming experimental data
from the RHIC and LHC heavy-ion programs.

\medskip

For the computation of the Feynman diagrams given in Appendices B-D we 
will need $G_{++}^+(x)$, $G_{--}^-(x)$ and $G_{-+}(x)$ propagators for 
the quark jet $p$, the radiated gluon $k$, and the exchanged gluon $q$. These 
functions are derived in Appendix~\ref{appPropagators}.

As in \cite{Gyulassy_Wang,GLV,Wiedemann,WW,DG_Ind,ASW,MD_TR}, we assume 
validity of the soft gluon ($\omega{\,\ll\,}E$) and soft rescattering 
($|\bq|{\,\sim\,}|\bk|{\,\ll\,}k_z$) approximations. These approximations 
become more reliable as the temperature increases and are expected to hold 
well at the LHC, while at RHIC their application should be further validated.
Together with conservation of energy and momentum ($p'=p+k+q$) they yield
\beqar
k = \Bigl[k^+,\, k^-=\frac{\bk^2{+}m_g^2}{k^+},\,\bk\Bigr] ,
\qquad
p = \Bigl[E^+,\, p^-=\frac{\bp^2+M^2}{E^+},\,\bp\Bigr]
\; ,
\eeqar{kp}

In~\cite{DH_Inf} we showed that it is reasonable to assume that 
$q_z$ has the same order of magnitude as $|\bq|$. Since $|\bk| \ll k_z$ 
and $q_z{\,\sim\,}|\bq|{\,\sim\,}|\bk|$, we then also have
$q_z{\,\ll\,}k_z$. Thus $k_z{+}q_z \approx k_z$ and
$p_z{+}k_z{+}q_z \approx p_z{+}k_z \approx p_z{+}q_z \approx p_z$.
Defining $x \equiv \frac{k^+}{E^+}$, we can therefore also assume that
\beqar
  x \equiv \frac{k^+}{E^+} \approx \frac{(k+q)^+}{(E+q)^+} \; .
\eeqar{x}
By using $x$, we further define $\chi$, $\xi$ and $\zeta$ in the following way
\beqar 
\chi &\equiv& M^2 x^2 + m_g^2 \nonumber \\
\xi &\equiv& \frac{\bk^2+\chi}{x E^{+}}  \nonumber \\
\zeta &\equiv& \frac{(\bk{+}\bq)^{2}+\chi}{x E^{+}} \nonumber \\
\zeta - \xi &=& \frac{(\bk{+}\bq)^{2} - \bk^2}{x E^{+}} \, .
\eeqar{xi-zeta}

Finally, by using that $\xi \, , \zeta \ll |\bk| {\,\sim\,}|\bq|$, and that
$q_z{\,\sim\,}|\bq|{\,\sim\,}|\bk|$, we obtain $\xi \, , \zeta \ll  q_z $ 
leading to
\beqar
q^- + \xi &=& q^0 - (q_z-\xi) \approx  q^0 - q_z \nonumber \\ 
q^- + \zeta &=& q^0 - (q_z-\zeta) \approx  q^0 - q_z \, \nonumber \\
q^- \pm (\zeta - \xi) &=& q^0 - (q_z \pm (\xi - \zeta)) \approx  q^0 - q_z \,.
\eeqar{q-xi-zeta}

\section{Derivation of the propagators in light cone coordinate system}
\label{appPropagators}

In this appendix we present in some detail the derivation of the propagators $G_{\pm \pm}$ in light cone coordinate system, which are needed for the 
calculations presented in the following Appendices. To derive these functions we start from Kallen-Lehmann theorem~\cite{Das}:
\beqar
G_{ab}(x)= \frac{i}{2 \pi} \; 
\int_0^\infty d s \int \frac{d^4 l}{(2 \pi)^4} \left( 
\frac{\rho_{ab} (s, l)}{l^2 - s +i \epsilon}+
\frac{\tilde{\rho}_{ab} (s, l)}{l^2 - s -i \epsilon} \right) e^{-i l \cdot x} \; , 
\eeqar{Gab}
where $a$, $b$ can be $+$ or $-$, and $\rho_{ab} (s, l)$ and 
$\tilde{\rho}_{ab} (s, l)$ are
spectral functions. We can now decompose the 
propagator as
\beqar
G_{ab}(x)= \theta(x^+)G_{ab}^{+}(x)+\theta(-x^+)G_{ab}^{-}(x)
\eeqar{Gab_decomposed}

Note that
\beqar
l \cdot x &=& \frac{1}{2} l^+ x^- + \frac{1}{2} l^- x^+ - \bl \cdot \bx \\
l^2 &=& l^+ l^- - \bl^2 \; ,
\eeqar{lx_l2}

We first concentrate on $G_{ab}^{+}(x)$. To obtain $G_{ab}^{+}(x)$ we take poles from Eq.~(\ref{Gab}). Pole $l^2-s+i\epsilon = 0$ contribute to 
$G_{ab}^{+}(x)$ only if $l^+>0$, while pole $l^2-s-i\epsilon =0$ contribute to $G_{ab}^{+}(x)$ only if $l^+<0$, leading to 
\beqar
G_{ab}^{+}(x) &=& 
\int \frac{d^4 l}{(2 \pi)^4} e^{-i l \cdot x}  \int_0^\infty d s  \; 
\delta (l^2-s) \left( \theta(l^+) \,\rho_{ab} (s, l) - \, \theta(-l^+) \, \tilde{\rho}_{ab} (s, l) \right) = \int \frac{d^4 l}{(2 \pi)^4} e^{-i l \cdot x} \; 
G_{ab}^{+}(l) \; ,
\eeqar{Gab+}
where
\beqar
G_{ab}^{+}(l) = \theta(l^+) \rho_{ab} (l^2,l) - \theta(-l^+) \, 
\tilde{\rho}_{ab} (l^2,l) \; ,
\eeqar{Gab+(l)}
where $\theta(l^+)$ and $\theta(-l^+)$ are unit step functions.

\medskip
To proceed further, we use the following expressions~
\beqar
\rho_{++} (l^2,l) &=& - \tilde{\rho}_{--} (l^2,l) = 
(\theta(l^0)+f(l^0)) \; \rho(l) 
\nonumber \\
\rho_{--} (l^2,l) &=& - \tilde{\rho}_{++} (l^2,l) = 
(\theta(-l^0)+f(l^0)) \; \rho(l) 
\nonumber \\
\rho_{+-} (l^2,l) &=& - \tilde{\rho}_{+-} (l^2,l) = 
f(l^0) \; \rho(l)
\nonumber \\
\rho_{-+} (l^2,l) &=& - \tilde{\rho}_{-+} (l^2,l) = 
(1+f(l^0)) \; \rho(l) \, ,
\eeqar{rhoab}
which can be straightforwardly derived from~\cite{Das}. Here
$\theta (l^0)$ is unit step function, $f(l^0)=(e^{l_0/T}{-}1)^{-1}$, where 
$T$ is the temperature of the medium, and $\rho(l)$ is spectral function. 

Then, by using Eqs.~(\ref{Gab+(l)}) and~(\ref{rhoab}), we obtain
\beqar
G_{++}^+ (l) &=& - \theta(l^+) \; \rho_{++}(l) + \theta(-l^+) \;\rho_{--} (l)=
\left(\theta(l^+) \; \theta(l^0) + \theta(-l^+) \; \theta(-l^0) +f(l^0)
\right) \rho(l) \, 
\nonumber \\
G_{--}^+ (l) &=& - \theta(l^+) \; \rho_{--} (l) + \theta(-l^+) \; 
\rho_{++} (l) =
\left(\theta(l^+) \; \theta(-l^0) + \theta(-l^+) \; \theta(l^0) +f(l^0)
\right) \rho(l) \, 
\nonumber \\
G_{+-}^+ (l) &=& - \theta(l^+) \; \rho_{+-} (l) + \theta(-l^+) \; 
\rho_{+-} (l) =
\rho_{+-} (l) =f(l^0) \; \rho(l) \, 
\nonumber \\
G_{-+}^+ (l) &=& - \theta(l^+) \; \rho_{-+} (l) + \theta(-l^+) \; 
\rho_{-+} (l) =
\rho_{-+} (l) =(1+f(l^0)) \; \rho(l) \, .
\eeqar{Propagators_ab}

Furthermore, it is straightforward to show that
\beqar
G_{++}^+ (l)&=&G_{--}^-(l) \nonumber \\
G_{-+}^- (l)&=&G_{+-}^+(l)
\eeqar{G-}

In this paper we will need only  $G_{++}^+(x)$, $G_{--}^-(x)$ and $G_{-+}(x)$ 
propagators for the quark jet $p$, the radiated gluon $k$, and the 
exchanged gluon $q$.  We will first derive these functions for 
the exchanged gluon $q$. To do this, lets first note that, as noted in the 
previous appendix, in this paper we assume high temperature 
QGP and soft-gluon, soft rescattering approximation. In such a limit, for the 
exchanged gluon $q$, $f(q^0) \approx \frac{T}{q_0} \gg 1$, reducing the 
Eqs.~(\ref{Propagators_ab}) and~(\ref{G-}) to 
\beqar
G_{++}^+ (q)  \approx G_{--}^+ (q) \approx G_{+-}^+ (q) 
\approx G_{-+}^+ (q) \approx G_{--}^-(q) \approx G_{-+}^- (q)
\approx f(q^0) \; \rho(q) \; .
\eeqar{Propagator_virtual}
By using Eq.~(\ref{Propagator_virtual}), it is then straightforward to obtain 
that corresponding propagators for the exchanged gluon are given by
\beqar
D_{++}^{+ \mu \nu} (x_i-x_j) = D_{--}^{- \mu \nu} (x_i-x_j) \approx 
D_{+-}^{\mu \nu} (x_i-x_j) \approx \int \frac{d^4 q} {(2 \pi)^4} D^>(q)
e^{ -i q (x_i-x_j)} \, ,
\eeqar{delta_virtual_gluon}
where $D^>(q)$ is the effective 1-HTL cut gluon propagator for the exchanged 
gluon~\cite{DH_Inf}
\beqar
D^>_{\mu\nu} (q)=\, \theta(1-\frac{q_0^2}{\vq^2})\, (1+f(q_0))\;
2 \, {\rm Im} \left( 
\frac{P_{\mu \nu} (q)}{q^2{-}\Pi_{T}(q)} + 
\frac{Q_{\mu \nu} (q)}{q^2{-}\Pi_{L}(q)}  \right)\; .
\eeqar{exchanged_cut}
Here $\Pi_T(q)$ and $\Pi_L(q)$ (see Eq.~(\ref{PiT}))
are transverse and longitudinal self energies. Note that exchanged gluons 
are spacelike (see Eq.(\ref{q})). Therefore, in the cut 1-HTL exchanged gluon 
propagator, only Landau damping contribution from the gluon spectral 
function contribute to the above expression (for more details, 
see~\cite{DH_Inf}). 

\bigskip
We will now concentrate on the propagators for the radiated gluon and 
quarks jet. For radiated gluon and quark jet 
$f(l_0)=(e^{l_0/T}{-}1)^{-1} \ll 1$, reducing the Eq.~(\ref{Propagators_ab}) to
\beqar
G_{++}^+ (l) &=& \left(\theta(l^+) \; \theta(l^0) + \theta(-l^+) \; 
\theta(-l^0) \right) \rho(l)
\nonumber \\
G_{--}^+ (l) &=& \left(\theta(l^+) \; \theta(-l^0) + \theta(-l^+) \; 
\theta(l^0) \right) \rho(l)
\nonumber \\
G_{+-}^+ (l) &=& 0
\nonumber \\
G_{-+}^+ (l) &=& \rho(l)
\eeqar{Propagators_ab_jet}

To proceed further with propagators for the radiated gluon propagator, 
we note that, for the radiated gluon with momentum $k$, the longitudinal 
contribution can be neglected relative to the transverse one. Also,  for 
the transverse gluon the self energy $\Pi_T(k)$ can be approximated by 
$m_g^2$, where $m_g \approx \mu/\sqrt{2}$ is the 
asymptotic mass (see~\cite{DG_TM}). 
These approximations are true in the soft rescattering limit 
$\omega \gg |\bq|{\,\sim\,}|\bk|{\,\sim\,}g T$ which we use in this 
paper. With these approximations the HTL gluon propagator for the emitted 
gluon can be simplified to~\cite{DG_TM}
\beqar
D_{\mu \nu}(k) \approx -i\, \frac{P_{\mu \nu}(k)}{k^2-m_g^2+i\epsilon}\; ,
\eeqar{rad_contrib}
where $P_{\mu \nu}$ is the transverse projector. 

Then, by using Eqs.~(\ref{Propagators_ab_jet}) and~(\ref{rad_contrib}), 
relevant radiative gluon propagators become (see also~\cite{Kogut}):

\beqar
D_{++}^{+\mu \nu} (x_i-x_j)=D_{--}^{-\mu \nu} (x_i-x_j)=
\int \frac{dk^+ d^2 k} {(2 \pi)^3 2 k^+} 
\theta(k^+) P^{\mu \nu} (k) e^{-i k(x_i-x_j)}
\eeqar{delta+++gluon}

\beqar
D_{-+}^{\mu \nu} (x_i-x_j)=\int \frac{d^3 k} {(2 \pi)^3 2 \omega} 
P^{\mu \nu} (k) e^{-i k(x_i-x_j)} \, 
\eeqar{delta-+gluon}
where $\omega \approx \sqrt{\vk^2 +m_g^2}$ and $k^+ = \omega+k_z$.

Note that we here assume that gluon mass is given by the  
expression $m_g \approx \mu/\sqrt{2}$ regardless whether the gluon is 
radiated inside or outside of the medium. Strictly speaking, 
including the finite size effects on the radiated gluon would assume 
that the above gluon mass is valid only for gluons radiated inside the medium,
while for the gluons radiated outside the medium the gluon mass should 
be equal to zero (see~\cite{MD_TR}). However, based on~\cite{MD_TR}, 
neglecting the finite size effects on the radiated gluon is expected to be a 
reasonable approximation, and we adopt it here to simplify the 
calculations. 

Finally, we now concentrate on the relevant propagators for the quark jet. 
By repeating similar procedure as for the radiated gluon, we obtain  
\beqar
\Delta_{++}^{+} (x_i-x_j)=\Delta_{--}^{-} (x_i-x_j)=
\int \frac{dp^+ d^2 p} {(2 \pi)^3 2 p^+} \theta(p^+) e^{-i p(x_i-x_j)}
\eeqar{delta+++jet}

\beqar
\Delta_{-+} (x_i-x_j)=\int \frac{d^3 p} {(2 \pi)^3 2 E} e^{-i p(x_i-x_j)}
\eeqar{delta-+jet}

\section{Computation of diagram $\bm{M_{1,0,1,C}}$}
\label{appM101C} 

\begin{figure}[ht]
\vspace*{5.8cm} 
\includegraphics{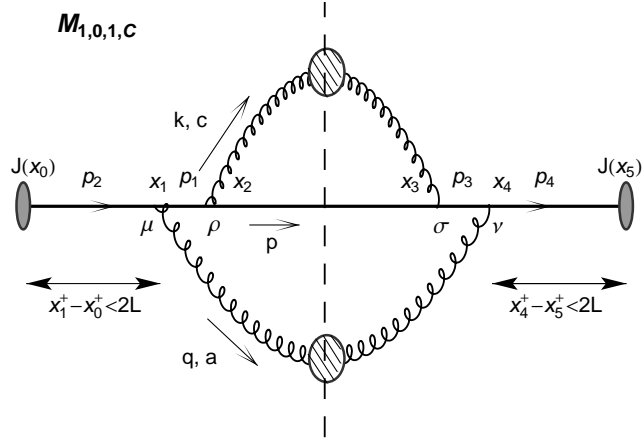}
\caption{Feynman diagram $M_{1,0,1,C}$, contributing to the radiative energy 
loss to first order in opacity, labeled in the same way as Fig.~1.}
\label{DiagM101C}
\end{figure}

In appendices~(\ref{appM101C}) - (\ref{appM1056}) we present in some detail 
the calculation of the diagrams where both ends of the exchanged gluon $q$ are 
attached to the heavy quark, i.e. none is attached to the radiated gluon $k$ 
and no 3-gluon vertex is involved. In this appendix, we start with the 
calculation of the diagram  shown in Fig.~\ref{DiagM101C}. 

Here and later the diagrams are labeled as follows: In $M_{1,i,j,C}$, 
$1$ denotes that these diagrams contribute to the energy loss to first 
order in opacity; $i$ denotes how many ends of the virtual gluon $q$ 
are attached to the radiated gluon $k$; and $j$ labels the specific
diagram in that class. Letter $C$ denotes that we consider central cut of the 
diagram. In the next chapters, letter $R$ ($L$) will denote that we 
consider right (left) cut of the Feynman diagram.

{\bf 1.} We will first calculate the cut diagram 
$M_{1,0,1,C} $:
\beqar
M_{1,0,1,C}  &=& \int \prod_{i=0}^{5} \, d^4 x_i \, J(x_0) \, \Delta_{++}^{+} (x_1-x_0) \,
v_\mu^{+} (x_1) \, D_{-+}^{\mu \nu} (x_4-x_1) \, \Delta_{++}^{+} (x_2-x_1) 
\, v_\rho^{+} (x_2) \, D_{-+}^{\rho \sigma} (x_3-x_2) \, 
\nonumber \\ && \; \; \times\ 
\Delta_{-+} (x_3-x_2) \, v_\sigma^{-} (x_3) \, \Delta_{--}^{-} (x_4-x_3) \,
v_\nu^{-} (x_4) \, \Delta_{--}^{-} (x_5-x_4) \, J(x_5) \,
\nonumber \\
&& \; \; \times\ 
\theta(x_{1}^{+}-x_{0}^{+}) \, \theta(x_{2}^{+}-x_{1}^{+}) \, 
\theta(2 L - (x_{1}^{+}-x_{0}^{+})) \, \theta(x_{4}^{+}-x_{5}^{+}) \, 
\theta(x_{3}^{+}-x_{4}^{+}) \, \theta(2 L - (x_{4}^{+}-x_{5}^{+})) \; , 
\eeqar{M101C_1} 
where $J$ is the source of a jet, $\Delta$ correspond to the jet propagator, 
$D$ to gluon propagators and $v$ to vertices. By using expressions for the 
propagators from the previous section, we obtain:

\beqar
M_{1,0,1,C}  &=& \int \prod_{i=0}^{5} d^4 x_i \, J(x_0) \, 
\int_{-\infty}^{\infty} \int_{0}^{\infty} 
\frac{ dp_{ 2}^{+} d^2 p_{2}} 
{(2 \pi)^3 \, 2 p_{2}^{+}} 
e^{-i p_{2} \cdot (x_{1} -x_{0})}   \,
\int \frac{d^4 q}{(2 \pi)^4} D_{\mu \nu}^{>} (q) e^{- i q (x_{4} -x_{1})}
\nonumber \\
&& \; \; \times\ \int_{-\infty}^{\infty} \int_{0}^{\infty} 
\frac{ dp_{ 1}^{+} d^2 p_{1}} 
{(2 \pi)^3 \, 2 p_{ 1}^{+}} e^{-i p_{ 1} \cdot (x_{2} -x_{1})} 
\left(- i g (p_{2}+p_{1})^\mu t_a \right)
\nonumber \\
&& \; \; \times\ (-1)  \int \frac{d^3k}{(2 \pi)^3 2 \omega} P_{\rho \sigma} (k) 
e^{-i k \cdot (x_{3} -x_{2})} 
\int \frac{d^3p}{(2 \pi)^3 2 E} e^{-i p \cdot (x_{3} -x_{2})} \,
\left(- i g (p_{ 1}+p)^\rho  t_c \right) \, 
\nonumber \\
&& \; \; \times\ \int_{-\infty}^{\infty} \int_{0}^{\infty} 
\frac{ dp_{3}^{+} d^2 p_{3}} {(2 \pi)^3 \, 2 p_{3}^{+}} 
e^{-i p_{3} \cdot (x_{4} -x_{3})} 
\left(i g (p+p_{3})^\sigma t_c \right ) \, 
\int_{-\infty}^{\infty} \int_{0}^{\infty} 
\frac{ dp_{4}^{+} d^2 p_{4}} 
{(2 \pi)^3 \, 2 p_{4}^{+} } e^{-i p_{4} \cdot (x_{5} -x_{4})} \, 
\left(i g (p_{3}+p_{4})^\nu t_a \right) 
\nonumber \\
&& \; \; \times\  J(x_5) \;
\theta(x_{1}^{+}-x_{0}^{+}) \, \theta(x_{2}^{+}-x_{1}^{+}) \, 
\theta(2 L - (x_{1}^{+}-x_{0}^{+})) \, \theta(x_{4}^{+}-x_{5}^{+}) \, \theta(x_{3}^{+}-x_{4}^{+}) \, \theta(2 L - (x_{4}^{+}-x_{5}^{+})) \;  
\nonumber \\
&=& \int_{-\infty}^{\infty} \int_{0}^{\infty} 
\prod_{i=1}^{4} \frac{ dp_{i}^{+} d^2 p_{i}} {(2 \pi)^3 \, 2 p_{i}^{+}} 
\int \frac{d^3k}{(2 \pi)^3 2 \omega} 
\int \frac{d^3p}{(2 \pi)^3 2 E} \, \int \frac{d^4 q}{(2 \pi)^4} 
\, g^4 \, t_a t_c t_c t_a \, 
\nonumber \\
&& \; \; \times\ (p_1+p_2)^{\mu} D_{\mu \nu}^{>} (q) (p_3+p_4)^{\nu} \; 
(p+p_1)^{\rho} P_{\rho \sigma} (k) (p+p_3)^{\sigma} \, I \; , 
\eeqar{M101C_2} 
where
\beqar
I&=& \int \prod_{i=0}^{5} d^4 x_i 
\theta(x_{1}^{+}-x_{0}^{+}) \, \theta(x_{2}^{+}-x_{1}^{+}) \, 
\theta(2 L - (x_{1}^{+}-x_{0}^{+})) \, \theta(x_{4}^{+}-x_{5}^{+}) \, \theta(x_{3}^{+}-x_{4}^{+}) \, \theta(2 L - (x_{4}^{+}-x_{5}^{+})) \; 
\nonumber \\
&& \; \; \times\  e^{-i p_{2} \cdot (x_{1} -x_{0})} \, 
e^{-i q \cdot (x_{4} -x_{1})} \, e^{-i p_{1} \cdot (x_{2} -x_{1})} \,
e^{-i (p + k) \cdot (x_{3} -x_{2})} \, e^{-i p_{3} \cdot (x_{4} -x_{3})} \,
e^{-i p_{4} \cdot (x_{5} -x_{4})} \, J(x_0) \,  J(x_5) \nonumber \\
&=& |J(p)|^2 \, 
(2 \pi)^3 \, \delta ((p_2 - p-k-q)^+) \delta^2 (\bp_2 - \bp-\bk-\bq) \, 
(2 \pi)^3 \, \delta ((p_1 - p-k)^+) \delta^2 (\bp_1 - \bp-\bk) \,
\nonumber \\ && \; \; \times\
(2 \pi)^3 \, \delta ((p_3 - p-k)^+) \delta^2 (\bp_3 - \bp-\bk) \,
(2 \pi)^3 \, \delta ((p_4 - p-k-q)^+) \delta^2 (\bp_4 - \bp-\bk-\bq) \, 
I_1 \; ,
\eeqar{I_101C}
where 
\beqar
I_1 = \int_0^\infty d x_2^{\prime +} 
e^{-\frac{i}{2} (p_1 - p-k)^- x_2^{\prime +}}  
\int_0^{2 L} d x_1^{\prime +} 
e^{-\frac{i}{2} (p_2 - p-k-q)^- x_1^{\prime +}}  
\int_0^\infty d x_3^{\prime +} 
e^{\frac{i}{2} (p_3 - p-k)^- x_3^{\prime +}} 
\int_0^{2 L} d x_4^{\prime +} 
e^{\frac{i}{2} (p_4 - p-k-q)^- x_4^{\prime +}}.
\eeqar{I1_101C}
Here we defined $x_1^{\prime} = x_1-x_0$, $x_2^{\prime} = x_2-x_1$,  
$x_3^{\prime} = x_3-x_4$, $x_4^{\prime} = x_4-x_5$.

By applying $\delta$ functions from Eq.~(\ref{I_101C}), and by using 
%
\beqar
p_{i}^{-}&=&\frac{\bp_{i}^{2}+M^2}{p_{i}^{+}} \, \nonumber \\
k^{-}&=&\frac{\bk^{2}+m_g^2}{k^{+}} \, 
\eeqar{pk-}
we obtain (note $\bp+\bk{+}\bq =0 \rightarrow \bp+\bk=-\bq$):
\beqar
p_{1}^{-}&=&p_{3}^{-}= \frac{\bq^{2}+M^2}{(p+k)^{+}} \,
\eeqar{p1p3}
Then in the soft gloun, soft rescattering limit we obtain 
(note $x\equiv \frac{k^+}{E^+}$, and that $\xi$ and $\zeta$ are defined in 
Eq.~(\ref{xi-zeta})): 
\beqar
(p_{1}-p-k)^{-}&=&(p_{3}-p-k)^{-} = 
\frac{\bk^2+\chi}{x E^{+}} = - \xi\, \nonumber \\
(p_{2}-p-k)^{-}&=&\frac{(\bk{+}\bq)^{2}+\chi}{x E^{+}} = - \zeta 
\eeqar{p-soft}
leading to
\beqar
I_1 &=& \frac{16}{((p_1-p-k)^{-})^2 }\, 
4 \frac{\sin^2 \left((p_2-p-k-q)^{-} \frac{L}{2} \right)}
{\left((p_2-p-k-q)^{-} \right)^2} 
\eeqar{I1_101C_2}
In the paper~\cite{MD_Coll}, it was shown that finite size effects on collisions are negligible, i.e. for collisional parts 
\beqar
&& 4 \frac{\sin^2 \left((p_2-p-k-q)^{-} \frac{L}{2} \right)}
{\left((p_2-p-k-q)^{-} \right)^2} \approx 2\pi L \, \delta ((p_2-p-k-q)^{-})
\approx 
2\pi L \, \delta (q^- + \zeta)
\approx 2\pi L \,  \delta (q^0-q_z)
\eeqar{delta_col}
where in the last step we used soft gloun, soft rescattering approximation, 
i.e. Eq.~(\ref{q-xi-zeta}).

By using Eq.~(\ref{delta_col}), Eq.~(\ref{I1_101C_2}) reduces to
\beqar
I_1 \approx 
\frac{16 x^2 E^{+ 2}}{(\bk^2+\chi)^2} \, 2\pi L \delta (q^0-q_z) \; .
\eeqar{I1_101C_final} 

Similarly as in~\cite{DH_Inf}, for highly energetic jets
\beqar
(p{+}p_1)^\rho P_{\rho\sigma}(k) (p{+}p_3)^\sigma
&\approx& - 4\, \frac{\bk^2}{x^2} \nonumber \\ 
(p_1{+}p_2)^\mu P_{\mu\nu}(q) (p_3{+}p_4)^\nu
&\approx& - (p_1{+}p_2)^\mu Q_{\mu\nu}(q) (p_3{+}p_4)^\nu
\approx - \, E^{+2} \frac{\bq^2}{\vq^2} \;, 
\eeqar{PQ}

By using Eq.~(\ref{PQ}) and Eqs.~(\ref{exchanged_cut}), 
$(p_1{+}p_2)^\mu D_{\mu\nu}^>(q) (p_3{+}p_4)^\nu$ becomes
\beqar
(p_1{+}p_2)^\mu D_{\mu\nu}^>(q) (p_3{+}p_4)^\nu &\approx& (p_1{+}p_2)^\mu \, 2 \, {\rm Im} \left( 
\frac{P_{\mu \nu} (q)}{q^2{-}\Pi_{T}(q)} + 
\frac{Q_{\mu \nu} (q)}{q^2{-}\Pi_{L}(q)}  \right) (p_3{+}p_4)^\nu 
\nonumber \\
&\approx&  \theta(1-\frac{q_0^2}{\vq^2}) f(q_0) E^{+2} \, \frac{\bq^2}{\vq^2}
\;  2 \, {\rm Im} 
\left( \frac{1}{q^2{-}\Pi_{L}(q)} - \frac{1}{q^2{-}\Pi_{T}(q)} \right)
\eeqar{coll_contrib1}

Finally, by using Eqs.~(\ref{I_101C}), (\ref{I1_101C}), (\ref{I1_101C_final}), 
(\ref{PQ}) and~(\ref{coll_contrib1}), and after performing integrations over $p_1$, $p_2$, $p_3$ and $p_4$, Eq.~(\ref{M101C_2}) reduces to
\beqar
M_{1,0,1,C}  &=& 4 L \, g^4 \, t_a t_c t_c t_a \,  \int
\frac{d^3 p}{(2 \pi)^3 2 E}  |J(p)|^2 \,  \frac{d^3 k}{(2 \pi)^3 2 \omega}
\frac{\bk^2}{(\bk^2+\chi)^2} \, I_q \, ,
\eeqar{M101C_3}
where 
\beqar
I_q=\int \frac{d^4q}{(2\pi)^4} 2\pi\delta(q_0-q_z)\, f(q_0) \, 
\frac{\bq^2}{\vq^2} \, 2 \, {\rm Im} 
\left( \frac{1}{q^2{-}\Pi_{L}(q)} - \frac{1}{q^2{-}\Pi_{T}(q)}   \right) \, .
\eeqar{Iq}
In high temperature limit $I_q$ reduces to (see Appendix C in~\cite{DH_Inf}):
\beqar
I_q &=& T \int \frac{d^2q}{(2\pi)^2}\, \frac{\mu^2}{\bq^2 (\bq^2{+}\mu^2)} \; .
\eeqar{Iq_f}

Finally, Eq.~(\ref{M101C_3}) becomes 
\beqar
M_{1,0,1,C}   = 4 L T \, g^4 \, t_a t_c t_c t_a 
\int \frac{d^3p}{(2\pi)^3 2 E} \,|J(p)|^2 \int
\frac{d^3k}{(2\pi)^3 2 \omega} \, \frac{d^2q}{(2\pi)^2} \,
\frac{\mu^2}{\bq^2 (\bq^2{+}\mu^2)} \,
\frac{\bk^2}{(\bk^2{+}\chi)^2}\, .
\eeqar{M101C_f}

\bigskip

\section{Computation of diagrams $\bm{M_{1,0,2,C}}$, $\bm{M_{1,0,2,R}}$ and 
$\bm{M_{1,0,2,L}}$}
\label{appM102} 

\begin{figure}[ht]
\vspace*{4.7cm} 
\includegraphics{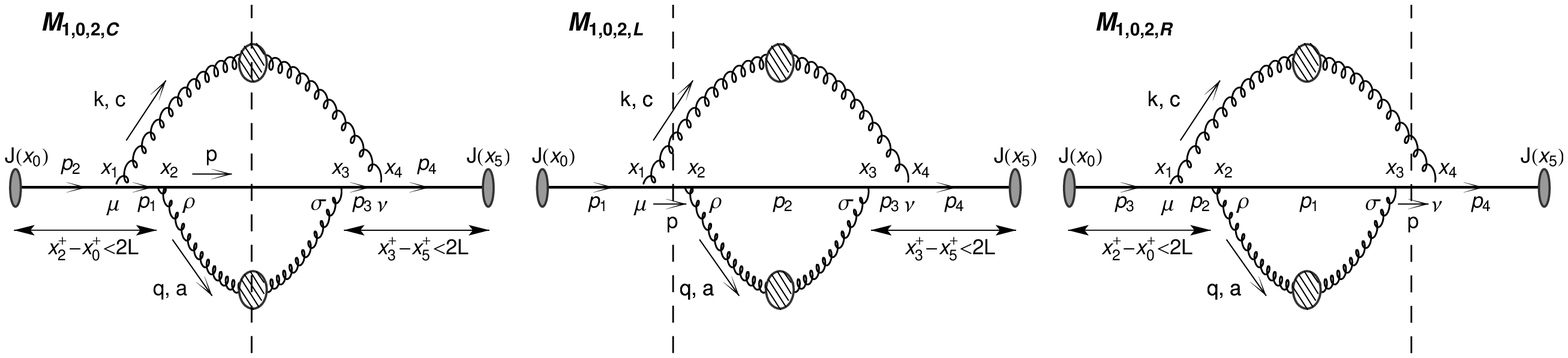}
\caption{Feynman diagrams $M_{1,0,2,C}$, $M_{1,0,2,L}$ and $M_{1,0,2,R}$ 
contributing to the radiative energy loss to first order in opacity, labeled 
in the same way as Fig.~1. The same figure is also presented as Fig.~1 in 
the main text, and is repeated here for completeness.}
\label{DiagM102}
\end{figure}

In this appendix we present in some detail the calculation of the diagrams 
shown in Fig.~\ref{DiagM102}. 

{\bf 1.} We will first calculate the cut diagram $M_{1,0,2,C} $, shown in the 
left panel of Fig.~\ref{DiagM102}:

\beqar
M_{1,0,2,C}  &=& \int \prod_{i=0}^{5} \, d^4 x_i \, J(x_0) \, \Delta_{++}^{+} (x_1-x_0) \,
v_\mu^{+} (x_1) \, D_{-+}^{\mu \nu} (x_4-x_1) \, \Delta_{++}^{+} (x_2-x_1) 
\, v_\rho^{+} (x_2) \, D_{-+}^{\rho \sigma} (x_3-x_2) \, 
\nonumber \\ && \; \; \times\ 
\Delta_{-+} (x_3-x_2) \, v_\sigma^{-} (x_3) \, \Delta_{--}^{-} (x_4-x_3) \,
v_\nu^{-} (x_4) \, \Delta_{--}^{-} (x_5-x_4) \, J(x_5) \,
\nonumber \\
&& \; \; \times\ 
\theta(x_{1}^{+}-x_{0}^{+}) \, \theta(x_{2}^{+}-x_{1}^{+}) \, 
\theta(2 L - (x_{2}^{+}-x_{0}^{+})) \, \theta(x_{4}^{+}-x_{5}^{+}) \, \theta(x_{3}^{+}-x_{4}^{+}) \, \theta(2 L - (x_{3}^{+}-x_{5}^{+}))  
\nonumber \\
&=& - \int_{-\infty}^{\infty} \int_{0}^{\infty} 
\prod_{i=1}^{4} \frac{ dp_{i}^{+} d^2 p_{i}} {(2 \pi)^3 \, 2 p_{i}^{+}} 
\int \frac{d^3k}{(2 \pi)^3 2 \omega} 
\int \frac{d^3p}{(2 \pi)^3 2 E} \, \int \frac{d^4 q}{(2 \pi)^4} 
\, g^4 \, t_c t_a t_a t_c \, 
\nonumber \\
&& \; \; \times\ (p_1+p_2)^{\mu} P_{\mu \nu}(k) (p_3+p_4)^{\nu} \; 
(p+p_1)^{\rho} D_{\rho \sigma}^{>}  (q) (p+p_3)^{\sigma} \, I \; , 
\eeqar{M102C_1} 
where
\beqar
I&=& \int \prod_{i=0}^{5} d^4 x_i 
\theta(x_{1}^{+}-x_{0}^{+}) \, \theta(x_{2}^{+}-x_{1}^{+}) \, 
\theta(2 L - (x_{2}^{+}-x_{0}^{+})) \, \theta(x_{4}^{+}-x_{5}^{+}) \, \theta(x_{3}^{+}-x_{4}^{+}) \, \theta(2 L - (x_{3}^{+}-x_{5}^{+})) \; 
\nonumber \\
&& \; \; \times\  e^{-i p_{2} \cdot (x_{1} -x_{0})} \, 
e^{-i k \cdot (x_{4} -x_{1})} \, e^{-i p_{1} \cdot (x_{2} -x_{1})} \,
e^{-i (p + q) \cdot (x_{3} -x_{2})} \, e^{-i p_{3} \cdot (x_{4} -x_{3})} \,
e^{-i p_{4} \cdot (x_{5} -x_{4})} \, J(x_0) \,  J(x_5) \nonumber \\
&=& |J(p)|^2 \, 
(2 \pi)^3 \, \delta ((p_2 - p_1-k)^+) \delta^2 (\bp_2 - \bp_1-\bk) \, 
(2 \pi)^3 \, \delta ((p_1 - p-q)^+) \delta^2 (\bp_1 - \bp-\bq) \,
\nonumber \\ && \; \; \times\
(2 \pi)^3 \, \delta ((p_3 - p-q)^+) \delta^2 (\bp_3 - \bp-\bq) \,
(2 \pi)^3 \, \delta ((p_4 - p_3-k)^+) \delta^2 (\bp_4 - \bp_3-\bk) \, 
I_{1} \; ,
\eeqar{I_102C}
and where 
\beqar
I_{1} = \int_0^{2 L} d x_2^{\prime +} 
e^{-\frac{i}{2} (p_1 - p-q)^- x_2^{\prime +}} 
\int_0^{x_2^{\prime +}} d x_1^{\prime +}
e^{-\frac{i}{2} (p_2 - p_1-k)^- x_1^{\prime +}} 
\int_0^{2 L} d x_3^{\prime +} 
e^{\frac{i}{2} (p_3 - p-q)^- x_3^{\prime +}} 
\int_0^{x_3^{\prime +}} d x_4^{\prime +}
e^{\frac{i}{2} (p_4 - p_3-k)^- x_4^{\prime +}} 
\eeqar{I1_102C}
Here we defined $x_1^{\prime} = x_1-x_0$, $x_2^{\prime} = x_2-x_0$,  
$x_3^{\prime} = x_3-x_5$, $x_4^{\prime} = x_4-x_5$.

\medskip 
By applying $\delta$ functions from Eq.~(\ref{I_102C}), and by using 
Eq.~(\ref{pk-}), we obtain $p_1^-=p_3^-$ and $p_2^-=p_4^-$.

$I_{1}$ then becomes
\beqar
I_{1} &=& \frac{16}{(p_2-p_1-k)^{-})^2 }\, 
\{ \frac{e^{-i (p_2-p-k-q)^{-} L} -1}{(p_2-p-k-q)^{-}} 
\left( \frac{e^{i (p_2-p-k-q)^{-} L} -1}{(p_2-p-k-q)^{-}}
- \frac{e^{i (p_1-p-q)^{-} L} -1}{(p_1-p-q)^{-}} \right)
\nonumber \\
&& \; \; +
\frac{e^{-i (p_1-p-q)^{-} L} -1}{(p_1-p-q)^{-}}
\left( \frac{e^{i (p_2-p-k-q)^{-} L} -1}{(p_2-p-k-q)^{-}}
- \frac{e^{i (p_1-p-q)^{-} L} -1}{(p_1-p-q)^{-}} \right) \}
\nonumber \\ &=& \frac{16}{(p_2-p_1-k)^{-})^2 }\, 
\{ 4 \frac{\sin^2 \left((p_2-p-k-q)^{-} \frac{L}{2} \right)}
{\left((p_2-p-k-q)^{-} \right)^2} 
\left(1-  \frac{(p_2-p-k-q)^{-}}{e^{i (p_2-p-k-q)^{-} L} -1}\frac{e^{i (p_1-p-q)^{-} L} -1}{(p_1-p-q)^{-}} \right)
\nonumber \\
&& \; \;  + 4 \frac{\sin^2 \left((p_1-p-q)^{-} \frac{L}{2} \right)}
{\left((p_1-p-q)^{-} \right)^2} 
\left( 1 - \frac{e^{i (p_1-p-q)^{-} L} -1}{(p_1-p-q)^{-}} 
\frac{(p_2-p-k-q)^{-}}{e^{i (p_2-p-k-q)^{-} L} -1}
\right) \}
\eeqar{I1_102C_2}

In the paper~\cite{MD_Coll}, it was shown that finite size effect are small for collisional energy loss, i.e. for collisional parts the following approximation can be used
\beqar
4 \frac{\sin^2 \left((p_2-p-k-q)^{-} \frac{L}{2} \right)}
{\left((p_2-p-k-q)^{-} \right)^2} &\approx& 2\pi L \delta ((p_2-p-k-q)^{-})
\nonumber \\
\frac{\sin^2 \left((p_1-p-q)^{-} \frac{L}{2} \right)}
{\left((p_1-p-q)^{-} \right)^2} &\approx& 2\pi L \delta ((p_1-p-q)^{-})
\eeqar{delta_col_M102C}

By using Eq.~(\ref{delta_col_M102C}), Eq.~(\ref{I1_102C_2}) reduces to
\beqar
I_1 &=&  \frac{32\pi L}{(p_1+k-p_2)^{-})^2 }\, 
\left \{ \delta ((p_2-p-k-q)^{-})
\left(1-  \frac{e^{i (p_1+k-p_2)^{-} L} -1}{i (p_1+k-p_2)^{-} L} \right)
+ \delta ((p_1-p-q)^{-})
\left( 1 - \frac{e^{- i (p_1+k-p_2)^{-} L} -1}{-i (p_1+k-p_2)^{-} L} \right) \right \}
\nonumber \\
&=&  \frac{32\pi L}{(p_1+k-p_2)^{-})^2 }\, 
\Big \{ \left(\delta ((p_2-p-k-q)^{-})+ \delta ((p_1-p-q)^{-}) \right)
\left(1-  \frac{\sin ((p_1+k-p_2)^{-} L)}{(p_1+k-p_2)^{-} L} \right)
\nonumber \\
&& \hspace*{3cm} + 
i \left(\delta ((p_2-p-k-q)^{-}) - \delta ((p_1-p-q)^{-}) \right)
\frac{\cos ((p_1+k-p_2)^{-} L)-1}{(p_1+k-p_2)^{-} L} \Big \} .
\eeqar{I1_102C_3}

In soft gluon, soft rescattering approximation 
\beqar
&& (p_1-p)^- \approx \frac{\bk^2{-}(\bk{+}\bq)^2}{E^+} \nonumber \\
&& \hspace*{0.4cm} (p_1+k-p_2)^- \approx \xi\, .
\eeqar{p1+k-p2}
Note that $(p_1-p)^- \ll \xi \ll |\bk|, |\bq|, q_z$, leading to 
\beqar
\left(\delta ((p_2-p-k-q)^{-}) + \delta ((p_1-p-q)^{-}) \right) &\approx&
2 \delta ((p_2-p-k-q)^{-}) \approx 2 \delta (q_0 - q_z) \, , \nonumber \\
\left(\delta ((p_2-p-k-q)^{-}) - \delta ((p_1-p-q)^{-}) \right) &\approx& 
\delta (q_0 - q_z + \xi) - \delta (q_0 - q_z) \, .
\eeqar{deltas_M102C}

By using Eq.~(\ref{deltas_M102C}), Eq.(\ref{I1_102C_3}) finally reduces to  
\beqar
I_1 \approx 
2\pi L \, \frac{16 x^2 E^{+ 2}}{(\bk^2+\chi)^2} 
\left \{ 2 \delta (q^0-q_z) \, 
\left(1- \frac{\sin (\xi L)}{{ \xi L}} \right) + 
i \Big(\delta (q_0 - q_z + \xi) - \delta (q_0 - q_z) \Big) 
\frac{\cos (\xi L)-1}{ \xi L} \right \} \, .
\eeqar{I1_102C_final} 

Similarly as in previous section, for highly energetic jets and in high temperature limit
\beqar
&& (p_1{+}p_2)^\mu P_{\mu\nu}(k) (p_3{+}p_4)^\nu
\approx - 4\, \frac{\bk^2}{x^2}\,, \nonumber \\ 
(p{+}p_1)^\rho D_{\rho\sigma}^>(q) (p{+}p_3)^\sigma &\approx&  
\theta(1-\frac{q_0^2}{\vq^2}) \frac{T}{q_0} E^{+2} \, \frac{\bq^2}{\vq^2}
\;  2 \, {\rm Im} 
\left( \frac{1}{q^2{-}\Pi_{L}(q)} - \frac{1}{q^2{-}\Pi_{T}(q)} \right) 
\equiv F(q_0,q_z,\bq) 
\eeqar{coll_contrib1_102C}
By using Eqs.~(\ref{I_102C}), (\ref{I1_102C_final}) and~(\ref {coll_contrib1_102C}), and after performing integrations over $p_1$, $p_2$, $p_3$ and $p_4$, Eq.~(\ref{M102C_1}) becomes 
\beqar
M_{1,0,2,C}  &=& 4 L T \, g^4 \, t_c t_a t_a t_c 
\int  \frac{d^3p}{(2\pi)^3 2 E} \, |J(p)|^2 \,
\frac{d^3k}{(2\pi)^3 2 \omega} \, \frac{\bk^2}{(\bk^2{+}\chi)^2}\, 
\nonumber \\
&& \; \; \times \,
\left \{ 2 \left(1- \frac{\sin (\xi L)}{\xi L} \right) \, I_q + 
i \, \frac{\cos (\xi L)-1}{ \xi L} \, J_q \right \} \, ,
\eeqar{M102C_2}
where $I_q$ is given by Eq.~(\ref{Iq_f}), and 
\beqar
J_q=\int \frac{d^4 q}{(2\pi)^4} \Big(\delta (q_0 - q_z + \xi) - \delta (q_0 - q_z) \Big) F(q_0, q_z, \bq) = \xi \int \frac{d^3 q}{(2\pi)^3} 
\frac{dF(q_0, q_z, \bq)}{dq_0}|_{q_0=q_z} = 0 \, . 
\eeqar{J_q_102C}
In the last step we used that 
$\frac{dF(q_0, q_z, \bq)}{dq_0}|_{q_0=q_z}$ is an odd function of $q_z$.

Finally, by using Eqs.~(\ref{Iq_f}) and (\ref{J_q_102C}), Eq.~(\ref{M102C_2}) reduces to
\beqar
M_{1,0,2,C}  &=& 8 L T \, g^4 \, t_c t_a t_a t_c 
\int \frac{d^3p}{(2\pi)^3 2 E} \, |J(p)|^2 \int 
\frac{d^3k}{(2\pi)^3 2 \omega} \,\frac{d^2q}{(2\pi)^2} \, 
\frac{\mu^2}{\bq^2 (\bq^2{+}\mu^2)} \,
 \frac{\bk^2}{(\bk^2{+}\chi)^2}\, 
\left(1- \frac{\sin (\xi L)}{\xi L} \right)\, .
\eeqar{M102C_f} 

{\bf 2.} We will now calculate the cut diagrams 
$M_{1,0,2,L} $ and $M_{1,0,2,R} $, shown in the central and right panel of 
Fig.~(\ref{DiagM102}), respectively. We start with $M_{1,0,2,R} $:
\beqar
M_{1,0,2,R}  &=& \int \prod_{i=0}^{5} \, d^4 x_i \, J(x_0) \, \Delta_{++}^{+} (x_1-x_0) \,
v_\mu^{+} (x_1) \, D_{-+}^{\mu \nu} (x_4-x_1) \, \Delta_{++}^{+} (x_2-x_1) 
\, v_\rho^{+} (x_2) \, D_{++}^{+ \rho \sigma} (x_3-x_2) \, 
\nonumber \\ && \; \; \times\ 
\Delta_{++}^{+} (x_3-x_2) \, v_\sigma^{+} (x_3) \, 
\Delta_{-+} (x_4-x_3) \,
v_\nu^{-} (x_4) \, \Delta_{--}^{-} (x_5-x_4) \, J(x_5) \,
\nonumber \\
&& \; \; \times\ 
\theta(x_{1}^{+}-x_{0}^{+}) \, \theta(x_{2}^{+}-x_{1}^{+}) \, 
\theta(x_{3}^{+}-x_{2}^{+}) \,
\theta(2 L - (x_{2}^{+}-x_{0}^{+})) \, \theta(x_{4}^{+}-x_{5}^{+})
\nonumber \\
&=& \int_{-\infty}^{\infty} \int_{0}^{\infty} 
\prod_{i=1}^{4} \frac{ dp_{i}^{+} d^2 p_{i}} {(2 \pi)^3 \, 2 p_{i}^{+}} 
\int \frac{d^3k}{(2 \pi)^3 2 \omega} 
\int \frac{d^3p}{(2 \pi)^3 2 E} \, \int \frac{d^4 q}{(2 \pi)^4} 
\, g^4 \, t_c t_a t_a t_c \, 
\nonumber \\
&& \; \; \times\ (p_2+p_3)^{\mu} P_{\mu \nu} (k) (p_+p_4)^{\nu} \; 
(p_1+p_2)^{\rho} D_{\rho \sigma}^> (q) (p+p_1)^{\sigma} \, I \; , 
\eeqar{M102R_1}
where  
\beqar
I&=& \int \prod_{i=0}^{5} d^4 x_i 
\theta(x_{1}^{+}-x_{0}^{+}) \, \theta(x_{2}^{+}-x_{1}^{+}) \, 
\theta(x_{3}^{+}-x_{2}^{+}) \,
\theta(2 L - (x_{2}^{+}-x_{0}^{+})) \, \theta(x_{4}^{+}-x_{5}^{+}) \; 
\nonumber \\
&& \; \; \times\  e^{-i p_{3} \cdot (x_{1} -x_{0})} \, 
e^{-i k \cdot (x_{4} -x_{1})} \, e^{-i p_{2} \cdot (x_{2} -x_{1})} \,
e^{-i (p_1 + q) \cdot (x_{3} -x_{2})} \, e^{-i p \cdot (x_{4} -x_{3})} \,
e^{-i p_{4} \cdot (x_{5} -x_{4})} \, J(x_0) \,  J(x_5) \nonumber \\
&=& |J(p)|^2 \, 
(2 \pi)^3 \, \delta ((p_3 - p_2-k)^+) \delta^2 (\bp_3 - \bp_2-\bk) \, 
(2 \pi)^3 \, \delta ((p_2 - p)^+) \delta^2 (\bp_2 - \bp) \,
\nonumber \\ && \; \; \times\
(2 \pi)^3 \, \delta ((p_1 - p+q)^+) \delta^2 (\bp_1 - \bp+\bq) \,
(2 \pi)^3 \, \delta ((p_4 - p-k)^+) \delta^2 (\bp_4 - \bp-\bk) \, 
I_{1} \; ,
\eeqar{I_102R}
and where 
\beqar
I_{1} = \int_0^{2 L} d x_2^{\prime +} 
e^{-\frac{i}{2} (p_2 - p)^- x_2^{\prime +}} 
\int_0^{x_2^{\prime +}} d x_1^{\prime +}
e^{-\frac{i}{2} (p_3 - p_2-k)^- x_1^{\prime +}} 
\int_0^{\infty} d x_3^{\prime +} 
e^{-\frac{i}{2} (p_1 - p+q)^- x_3^{\prime +}} 
\int_0^{\infty} d x_4^{\prime +}
e^{\frac{i}{2} (p_4 - p-k)^- x_4^{\prime +}} 
\eeqar{I1_102R_1}
Here we defined $x_1^{\prime} = x_1-x_0$, $x_2^{\prime} = x_2-x_0$,  
$x_3^{\prime} = x_3-x_2$, $x_4^{\prime} = x_4-x_5$.

\medskip 
By applying $\delta$ functions from Eq.~(\ref{I_102R}), and by using 
Eq.~(\ref{pk-}), we obtain
\beqar
p_2^- =p^- \rightarrow (p_2-p)^-&=&0, \nonumber \\
p_3^-=p_4^- = \frac{(\bp +\bk)^2+M^2}{(p+k)^+} &\approx& \frac{M^2}{E^+} \, 
\nonumber \\
&& \hspace*{-3.5cm}(p+k-p_3)^- =  \xi \nonumber \\
&& \hspace*{-3.5cm}(p_1 - p+q)^- \approx q^-.
\eeqar{p_102R}
Note that in the second equation we used $\bp=-\bk$ and $(p+k)^+ \approx E^+$.

$I_{1}$ then becomes
\beqar
I_{1} &=& \int_0^{2 L} d x_2^{\prime +}  
\int_0^{x_2^{\prime +}} d x_1^{\prime +}
e^{\frac{i}{2} \xi x_1^{\prime +}} 
\int_0^{\infty} d x_3^{\prime +} 
e^{-\frac{i}{2} (p_1 - p+q)^- x_2^{\prime +}} 
\int_0^{\infty} d x_4^{\prime +}
e^{-\frac{i}{2} \xi x_4^{\prime +}} 
\nonumber \\ 
&=& \frac{8L}{\xi^2 }\, 
\left ( 1- \frac{\sin \xi L}{ \xi L} - i \frac{1- \cos \xi L}{ \xi L}\right) \int_0^{\infty} d x_3^{\prime +} 
e^{-\frac{i}{2} q^- x_3^{\prime +}} 
\eeqar{I1R_102_2}

By using Eqs.~(\ref{I1R_102_2}), (\ref{I_102R}) and (\ref{M102R_1}) 
$M_{1,0,2,R} $ becomes ($y \equiv \frac{x_3^{\prime +}}{2}$)
\beqar
M_{1,0,2,R}  &=& -4 L \, g^4 \, t_c t_a t_a t_c \, 
\int \frac{d^3p}{(2 \pi)^3 2 E} \, |J(p)|^2 \,
\int \frac{d^3k}{(2 \pi)^3 2 \omega}  \frac{\bk^2}{(\bk^2 + \chi)^2}  
\nonumber \\ && \; \times\ \left ( 1- \frac{\sin \xi L}{ \xi L} - i \frac{1- \cos \xi L}{ \xi L}\right)\, \int \frac{d^4 q}{(2 \pi)^4} 
(p+p_1)^{\rho} D_{\rho \sigma}^{>} (q) (p+p_1)^{\sigma} \,
\int_0^{\infty} d y 
e^{- i q^- y} ,
\eeqar{M102R_2}
where, as in the previous sections, we used 
$(p{+}p_3)^\mu P_{\mu\nu}(k) (p{+}p_3)^\nu \approx - 4\, \frac{\bk^2}{x^2}$.

In the same way, it can be obtained that $M_{1,0,2,L}$ is equal to
\beqar
M_{1,0,2,L}  &=& -4  L \, g^4 \, t_c t_a t_a t_c \, 
\int \frac{d^3p}{(2 \pi)^3 2 E} \, |J(p)|^2 \,
\int \frac{d^3k}{(2 \pi)^3 2 \omega}  \frac{\bk^2}{(\bk^2 + \chi)^2} 
\nonumber \\ && \;\times\ \left ( 1- \frac{\sin \xi L}{ \xi L} - i \frac{1- \cos \xi L}{ \xi L}\right)\,  \int \frac{d^4 q}{(2 \pi)^4} 
(p+p_1)^{\rho} D_{\rho \sigma}^{>} (q) (p+p_1)^{\sigma} \,
\int_0^{\infty} d y 
e^{ i q^- y} .
\eeqar{M102L_2}

$M_{1,0,2,R}  + M_{1,0,2,L} $ then becomes
\beqar
M_{1,0,2,R}  + M_{1,0,2,L}  &=& -4 L \, g^4 \, t_c t_a t_a t_c \, 
\int \frac{d^3p}{(2 \pi)^3 2 E} \, |J(p)|^2 \,
\int \frac{d^3k}{(2 \pi)^3 2 \omega}  \frac{\bk^2}{(\bk^2 + \chi)^2} 
\; \nonumber \\ && \; \times\ \int \frac{d^4 q}{(2 \pi)^4} 
(p+p_1)^{\rho} D_{\rho \sigma}^{>} (q) (p+p_1)^{\sigma} \, I_2 \, ,
\eeqar{M102R+L}

where
\beqar
I_2 &=&
\left ( 1- \frac{\sin \xi L}{ \xi L} - i \frac{1- \cos \xi L}{ \xi L}\right)\, 
\int_0^{\infty} \, d y \,
e^{-i y q^-} 
+ \left ( 1- \frac{\sin \xi L}{ \xi L} + i \frac{1- \cos \xi L}{ \xi L}\right)\, 
\int_0^{\infty} \, d y \, e^{i y q^-} \nonumber \\ 
&=&
\left ( 1- \frac{\sin \xi L}{ \xi L} \right)\, 
\int_{-\infty}^{\infty} \, d y \,
e^{-i y q^- } 
- 2 \, \frac{1- \cos \xi L}{ \xi L}\, 
\int_0^{\infty} \, d y \, 
\sin\left( q^- y \right).
\nonumber \\ 
& \approx &
\left ( 1- \frac{\sin \xi L}{ \xi L} \right)\, 
2 \pi \delta (q^-) 
- 2 \, \frac{1- \cos \xi L}{ \xi L}\, 
\int_0^{\infty} \, d y \, 
\sin\left( q^- y \right) \, .
\eeqar{I2_1} 

Similarly as in previous section, for highly energetic jets
\beqar
(p{+}p_1)^\rho D_{\rho\sigma}^>(q) (p{+}p_1)^\sigma &\approx&  
\theta(1-\frac{q_0^2}{\vq^2}) f(q_0) E^{+2} \, \frac{\bq^2}{\vq^2}
\;  2 \, {\rm Im} 
\left( \frac{1}{q^2{-}\Pi_{L}(q)} - \frac{1}{q^2{-}\Pi_{T}(q)} \right)
\eeqar{coll_contrib1_102RL} 

Finally, by using Eqs.(\ref{I2_1}) and (\ref{coll_contrib1_102RL}), Eq.~(\ref{M102R+L}) becomes
\beqar
M_{1,0,2,R}  + M_{1,0,2,L}  &=& - 4 L \, g^4 \, t_c t_a t_a t_c \, 
\int \frac{d^3p}{(2 \pi)^3 2 E} \, |J(p)|^2 \, 
\int \frac{d^3k}{(2 \pi)^3 2 \omega} 
\frac{\bk^2}{(\bk^2 + \chi)^2}  
\left ( 1- \frac{\sin \xi L}{ \xi L} \right)  I_{q} \nonumber \\ 
&& + 8 L \, g^4 \, t_c t_a t_a t_c \, 
\int \frac{d^3p}{(2 \pi)^3 2 E} \, |J(p)|^2 \, 
\int \frac{d^3k}{(2 \pi)^3 2 \omega} \frac{\bk^2}{(\bk^2 + \chi)^2}  
\frac{1 - \cos \xi L}{ \xi L} \, J_{q} \, ,
\eeqar{M102R+L_2}
where $I_q$ is given by Eq.~(\ref{Iq_f}), and $J_q$ is given by
\beqar
J_q= \int_0^{\infty} d z \int \frac{d q_0 \, d q_z \, d^2 q}{ (2 \pi)^4}
\left(\sin (q_0 z) \cos (q_z z) - \cos (q_0 z) \sin (q_z z) \right) 
F( q_0, q_z, \bq),
\eeqar{J_q1}
In the above equation, we defined $F( q_0, q_z, \bq)$ as 
\beqar
F( q_0, q_z, \bq) \equiv \frac{\bq^2}{\vq^2} \frac{T}{q_0} 
\left ( \frac{2 {\rm Im} \Pi_L (q)}{(q^2 -{\rm Re} \Pi_L (q))^2 + {\rm Im} \Pi_L (q)^2} - 
\frac{2 {\rm Im} \Pi_T (q)}{(q^2 -{\rm Re} \Pi_T (q))^2 + {\rm Im} \Pi_T (q)^2} \right)
\eeqar{F}
$F( q_0, q_z, \bq)$ is even function of both $q_0$ and $q_z$, leading to
\beqar
\int_{-\infty}^{\infty} dq_0 \sin (q_0 z) F( q_0, q_z, \bq) &=& 0 \, \nonumber \\ 
\int_{-\infty}^{\infty} dq_z \sin (q_z z) F( q_0, q_z, \bq) &=& 0 \, ,
\eeqar{even_odd_int}
which consequently leads to
\beqar
J_q=0.
\eeqar{J_q2}

Finally, by using Eqs.~(\ref{Iq_f}) and (\ref{J_q2}), Eq.~(\ref{M102R+L_2}) becomes
\beqar
M_{1,0,2,R}  + M_{1,0,2,L}  &=& - 4 L T\, g^4 \, t_c t_a t_a t_c \, 
\int \frac{d^3p}{(2 \pi)^3 2 E} \, |J(p)|^2 \, 
\int \frac{d^3k}{(2 \pi)^3 2 \omega} \, \frac{d^2q}{(2\pi)^2} \,
\frac{\mu^2}{\bq^2 (\bq^2{+}\mu^2)}
 \, \nonumber \\
&& \; \; \times\ \frac{\bk^2}{(\bk^2 + \chi)^2} 
\left ( 1- \frac{\sin (\xi L)}{ \xi L} \right) \,.
\eeqar{M102RL_f}

\section{Computation of diagrams $\bm{M_{1,0,3,C}}$ and $\bm{M_{1,0,4,C}}$ } 
\label{appM103CM104C} 
 
We will now calculate cut diagrams $M_{1,0,3,C} $ and $M_{1,0,4,C} $, 
shown in the Fig.~(\ref{DiagM1034C}). We start with $M_{1,0,3,C} $:

\begin{figure}[ht]
\vspace*{5.2cm} 
\includegraphics{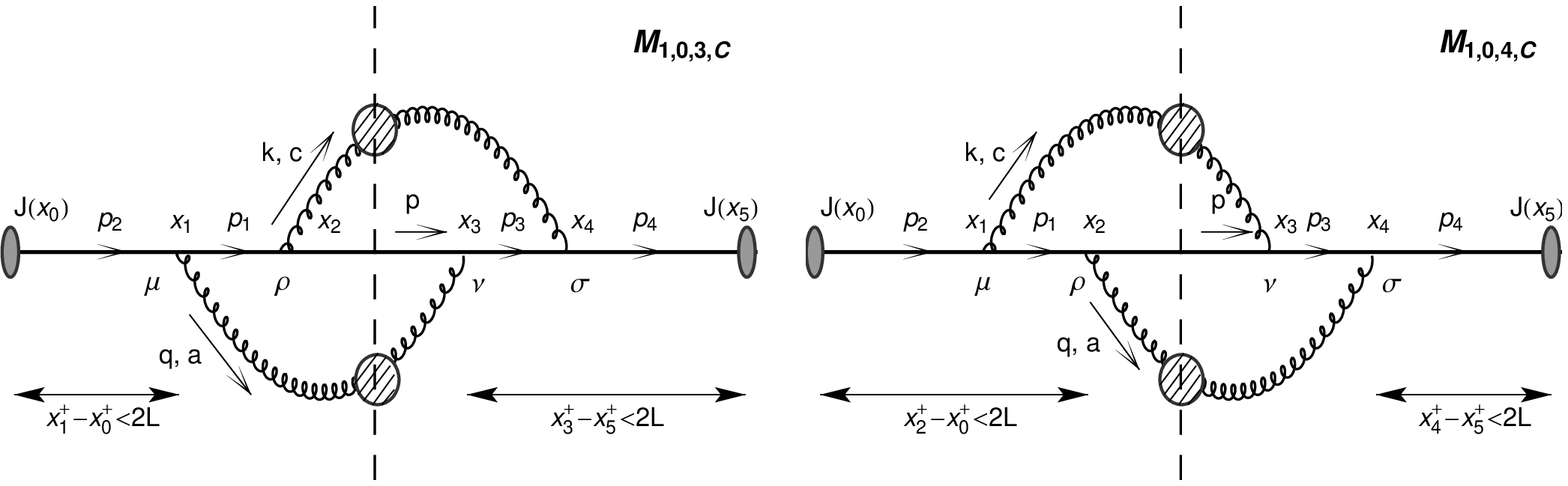}
\caption{Feynman diagrams $M_{1,0,3,C}$ and $M_{1,0,4,C}$ 
contributing to the radiative energy loss to first order in opacity, labeled 
in the same way as Fig.~1.}
\label{DiagM1034C}
\end{figure}
%
\beqar M_{1,0,3,C}  &=& \int 
\prod_{i=0}^{5} \, d^4 x_i \, J(x_0) \, \Delta_{++}^{+} (x_1-x_0) 
\, v_\mu^{+} (x_1) \, D_{-+}^{\mu \nu} (x_3-x_1) \, 
\Delta_{++}^{+} (x_2-x_1) \, v_\rho^{+} (x_2) \, D_{-+}^{\rho 
\sigma} (x_4-x_2) \, \nonumber \\ && \; \; \times\ \Delta_{-+} 
(x_3-x_2) \, v_\nu^{-} (x_3) \, \Delta_{--}^{-} (x_4-x_3) \, 
v_\sigma^{-} (x_4) \, \Delta_{--}^{-} (x_5-x_4) \, J(x_5) \, 
\nonumber \\ 
&& \; \; \times\ \theta(x_{1}^{+}-x_{0}^{+}) \, 
\theta(x_{2}^{+}-x_{1}^{+}) \, \theta(2 L - (x_{1}-x_{0})^{+}) \, 
\theta(x_{4}^{+}-x_{5}^{+}) \, \theta(x_{3}^{+}-x_{4}^{+}) \, 
\theta(2 L - (x_{3}-x_{5})^{+}) \nonumber \\ 
&=& - \int_{-\infty}^{\infty} \int_{0}^{\infty} \prod_{i=1}^{4} 
\frac{ dp_{i}^{+} d^2 p_{i}} {(2 \pi)^3 \, 2 p_{i}^{+}} \int 
\frac{d^3k}{(2 \pi)^3 2 \omega} \int \frac{d^3p}{(2 \pi)^3 2 
E} \, \int \frac{d^4 q}{(2 \pi)^4} \, g^4 \, t_a t_c t_a t_c \, 
\nonumber \\ 
&& \; \; \times\ (p_1+p_2)^{\mu} D_{\mu \nu}^{>} (q) 
(p+p_3)^{\nu} \; (p+p_1)^{\rho} P_{\rho \sigma} (k) \, 
(p_3+p_4)^{\sigma} \, I \; , 
\eeqar{M103C_1} 
where 
\beqar 
I&=& \int \prod_{i=0}^{5} d^4 x_i 
\theta(x_{1}^{+}-x_{0}^{+}) \, \theta(x_{2}^{+}-x_{1}^{+}) \, 
\theta(2 L - (x_{1}-x_{0})^{+}) \, \theta(x_{4}^{+}-x_{5}^{+}) \, 
\theta(x_{3}^{+}-x_{4}^{+}) \, \theta(2 L - (x_{3}-x_{5})^{+}) \; 
\nonumber \\ 
&& \; \; \times\  e^{-i p_{2} \cdot (x_{1} -x_{0})} \, e^{-i q 
\cdot (x_{3} -x_{1})} \, e^{-i p_{1} \cdot (x_{2} -x_{1})} \, 
e^{-i k \cdot (x_{4} -x_{2})} \, e^{-i p \cdot (x_{3} -x_{2})} \,
e^{-i p_{3} \cdot (x_{4}-x_{3})} \, 
e^{-i p_{4} \cdot (x_{5} -x_{4})} \, J(x_0) \,  J(x_5) \nonumber \\ 
&=& |J(p)|^2 \, (2 \pi)^3 \, \delta ((p_2 - p-k-q)^+) \delta^2 
(\bp_2 - \bp-\bk-\bq) \, (2 \pi)^3 \, \delta ((p_1 - p-k)^+) 
\delta^2 (\bp_1 - \bp-\bk) \, \nonumber \\ && \; \; \times\ (2 
\pi)^3 \, \delta ((p_3 - p-q)^+) \delta^2 (\bp_3 - \bp-\bq) \, (2 
\pi)^3 \, \delta ((p_4 - p_3-k)^+) \delta^2 (\bp_4 - \bp_3-\bk) \, 
I_1 \; , 
\eeqar{I_103C} 
and where 
\beqar 
I_1 &=& \int_0^\infty d x_2^{\prime +} e^{-\frac{i}{2} (p_1 - 
p-k)^- x_2^{\prime +}} \int_0^{2 L} d x_1^{\prime +} 
e^{-\frac{i}{2} (p_2 - p-k-q)^- x_1^{\prime +}} \int_0^{2L} d 
x_3^{\prime +} e^{\frac{i}{2} (p_3 - p-q)^- x_3^{\prime +}} 
\int_0^{x_3} d x_4^{\prime +} e^{\frac{i}{2} (p_4 - p_3-k)^- 
x_4^{\prime +}} \nonumber \\ 
&=& \; \; \frac{-16}{(p_1-p-k)^-(p_4-p_3-k)^-}\, 
\frac{e^{-i(p_2-p-k-q)L} - 1}{(p_2-p-k-q)^-} \, 
\left \{\frac{e^{i(p_4 - p - k - q)^{-}L} - 1}{(p_4 - p - k - q)^{-}} - 
\frac{e^{i(p_3-p-q)^{-}L} - 1}{(p_3-p-q)^{-}}\right \} 
\eeqar{I1_103C_1} 
 
Similarly as in previous sections, by using $\delta$ functions from 
Eqs.~(\ref{I_103C}) and~(\ref{pk-}), we obtain $p_4^- = p_2^-$ and 
\beqar 
(p_1-p-k)^- \approx (p_4-p_3-k)^- =- \xi \, .
\eeqar{103C_moments} 

$I_1$ then becomes 
\beqar 
I_1 &-& -\frac{16}{\xi^2} \, 
\frac{4 \sin^2((p_2-p-k-q)^-\frac{L}{2})}{((p_2-p-k-q)^-)^2} \, 
\left\{1-\frac{(p_2-p-k-q)^-}{e^{-i(p_2-p-k-q)^-L} - 1}
\frac{e^{i (p_3-p-q)^- L}-1}{(p_3-p-q)^-}\right\}\, 
\nonumber \\ 
& \approx & -\frac{16}{\xi^2} \, 2\pi L \, \delta ((p_2-p-k-q)^-) \, 
\Big\{1-\lim\limits_{(p_2-p-k-q)^-\rightarrow 0} 
\frac{(p_2-p-k-q)^-}{e^{-i(p_2-p-k-q)^-L} - 1}\, 
\frac{e^{i\{(p_2-p-k-q)^-+ \xi\}L}-1}{(p_2-p-k-q)^-+ \xi}\Big\}\; \nonumber \\
& \approx &  -\frac{16}{\xi^2} \, 2\pi L \delta (q_0-q_z) \, \left \{1-\frac{\sin(\xi L)}{\xi L}+
i\frac{\cos(\xi L)-1}{\xi L} \right \} \, . 
\eeqar{I1_103C_f} 
where similarly to previous sections, in the second step we used 
Eq.~(\ref{q-xi-zeta}) and the fact that finite size effects are negligible for 
collisional interactions. That is, we used 
$\frac{4 \sin^2((p_2-p-k-q)^-\frac{L}{2})}{((p_2-p-k-q)^-)^2} 
\approx 2\pi L\, \delta (q^- - \xi) \approx 2\pi L \, \delta (q_0 -q_z)$.

\medskip

Finally, by using Eqs.~(\ref{I_103C}) and (\ref{I1_103C_f}), Eq.~(\ref{M103C_1}) becomes 
\beqar 
M_{1,0,3,C}  & = & -4 L T \, g^4 \, t_a t_c t_a t_c \int 
 \frac{d^3p}{(2\pi)^3 2 E} \,  |J(p)|^2 \, 
\int \frac{d^3k}{(2\pi)^3 2 \omega} \,
\frac{\bk^2}{(\bk^2{+}\chi)^2}\,  \nonumber \\ 
&& \; \; \times \, \int 
\frac{d^2q}{(2\pi)^2} \, \frac{\mu^2}{\bq^2 (\bq^2{+}\mu^2)} 
\left \{1-\frac{\sin(\xi L)}{\xi L}-i 
\frac{1-\cos(\xi L)}{\xi L} \right \} \, . 
\eeqar{M103C_f} 
 
Since $M_{1,0,4,C}  =\left( M_{1,0,4,C}  \right)^*$, we obtain
\beqar 
M_{1,0,3,C}  + M_{1,0,4,C}  & = & -8 L T \, g^4 \, t_a t_c t_a t_c \int 
 \frac{d^3p}{(2\pi)^3 2 E} \, 
|J(p)|^2 \, \int \frac{d^3k}{(2\pi)^3 2 \omega} \, \frac{d^2q}{(2\pi)^2} \,
 \frac{\mu^2}{\bq^2 (\bq^2{+}\mu^2)} 
\nonumber \\ 
&& \; \; \times \, 
\frac{\bk^2}{(\bk^2{+}\chi)^2}\,
\left (1-\frac{\sin(\xi L)}{\xi L} \right ) \, . 
\eeqar{M1034C_f}

\section{Computation of diagrams $\bm{M_{1,0,3,R}}$ and $\bm{M_{1,0,4,L}}$} 
\label{appM103RM104L} 

We will now calculate the cut diagrams $M_{1,0,3,R} $ and $M_{1,0,4,L} $, 
shown in the Fig.~(\ref{DiagM103R4L}). We start with $M_{1,0,3,R} $:

\begin{figure}[ht]
\vspace*{5.2cm} 
\includegraphics{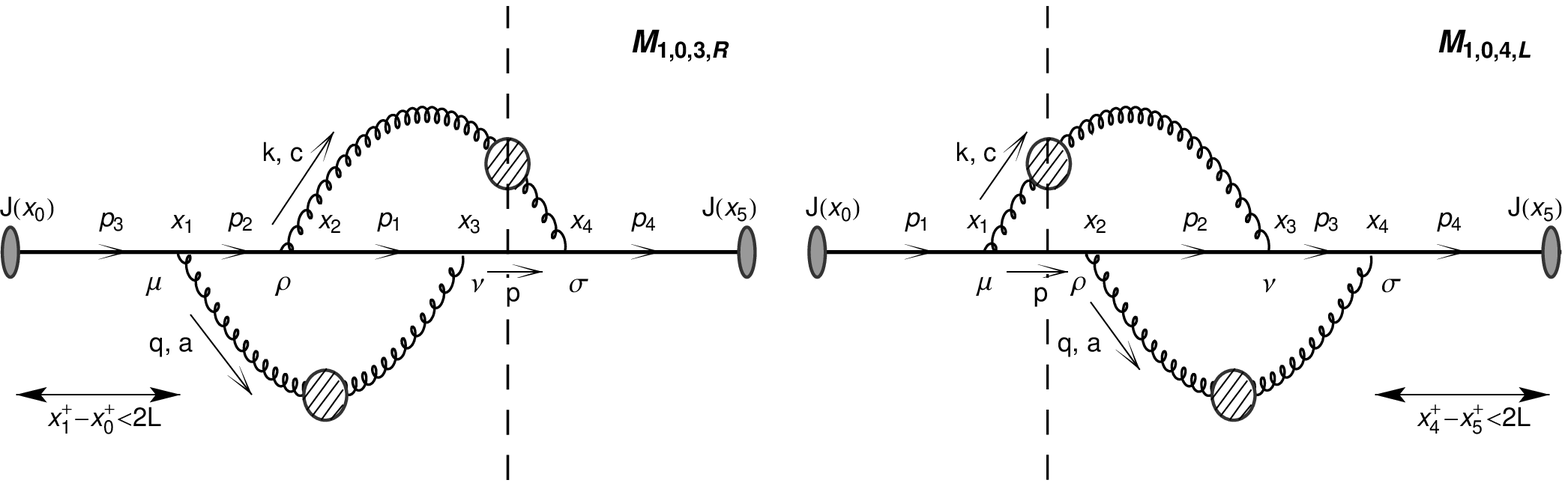}
\caption{Feynman diagrams $M_{1,0,3,R}$ and $M_{1,0,4,L}$ 
contributing to the radiative energy loss to first order in opacity, labeled 
in the same way as Fig.~1.}
\label{DiagM103R4L}
\end{figure}
%
 \beqar M_{1,0,3,R}  &=& \int 
\prod_{i=0}^{5} \, d^4 x_i \, J(x_0) \, \Delta_{++}^{+} (x_1-x_0) 
\, v_\mu^{+} (x_1) \, D_{++}^{+\mu \nu} (x_3-x_1) \, 
\Delta_{++}^{+} (x_2-x_1) \, v_\rho^{+} (x_2) \, D_{-+}^{\rho 
\sigma} (x_4-x_2) \, \nonumber \\ && \; \; \times\ \Delta_{++}^{+} 
(x_3-x_2) \, v_\lambda^{+} (x_3) \, \Delta_{-+} (x_4-x_3) \, 
v_\sigma^{-} (x_4) \, \Delta_{--}^{-} (x_5-x_4) \, J(x_5) \, 
\nonumber \\ 
&& \; \; \times\ \prod_{i=0}^{3} \,\theta(x_{i+1}^{+}-x_{i}^{+}) 
\, \theta(x_{4}^{+}-x_{5}^{+}) \, \theta(2 L - (x_{1}-x_{0})^{+}) 
\nonumber \\
&=& \int_{-\infty}^{\infty} \int_{0}^{\infty}\frac{ dp_{i}^{+} 
d^2 p_{i}} {(2 \pi)^3 \, 2 p_{i}^{+}} \int \frac{d^3 k}{(2 
\pi)^3 2 \omega} \, \frac{d^3 p}{(2 \pi)^3 2 E} \,  \frac{d^4 
q}{(2 \pi)^4} \, g^4 \, t_a t_c t_a t_c \, 
\nonumber \\ 
&& \; \; \times\ (p_2+p_3)^{\mu} D_{\mu \lambda}^{>} (q) 
(p+p_1)^{\lambda} \; (p_1+p_2)^{\rho} P_{\rho \sigma} (k) 
(p+p_4)^{\sigma} \, I \; , 
\eeqar{M103R_1} 
where 
\beqar I&=& \int \prod_{i=0}^{5} d^4 x_i\, \prod_{j=0}^{3} 
\theta(x_{j+1}^{+}-x_{j}^{+}) \, \theta(2 L - (x_{1}-x_{0})^{+}) 
\, \theta(x_{4}^{+}-x_{5}^{+}) \, J(x_0)J(x_5) 
\nonumber \\ 
&& \; \; \times\  e^{-i p_{3} \cdot (x_{1} -x_{0})} \, e^{-i q 
\cdot (x_{3} -x_{1})} \, e^{-i p_{2} \cdot (x_{2} -x_{1})} \, 
e^{-ip_1\cdot (x_{3} -x_{2})} \, e^{-ip\cdot (x_{4} -x_{3})} \, 
e^{-i k \cdot (x_{4} -x_{1})} \, e^{-ip_{4}\cdot (x_{5} -x_{4})} \, \nonumber \\ 
&=& |J(p)|^2 \, (2 \pi)^3 \, \delta ((p_3 - p-k)^+) \delta^2 
(\bp_3 - \bp-\bk) \, (2 \pi)^3 \, \delta ((p_2 - p_1-k)^+) 
\delta^2 (\bp_2 - \bp_1-\bk) \,\nonumber \\ 
&& \; \; \times\ (2 \pi)^3 \, \delta ((p_1 + q-p)^+) \delta^2 (\bp_1 
+ \bq-\bp) \, (2 \pi)^3 \, \delta ((p_4 - p-k)^+) \delta^2 (\bp_4 - \bp-\bk) 
\, I_1 \; . 
\eeqar{I_103R} 
Here 
\beqar 
I_1 = \int_0^\infty d x_3^{\prime +} e^{-\frac{i}{2} (p_1 + 
q-p)^- x_3^{+}} \int_0^{x_3^{+}} d x_2^{\prime +} e^{-\frac{i}{2} 
(p_2 - p_1-k)^- x_2^{\prime +}} \int_0^{2L} d x_1^{\prime +} 
e^{\frac{-i}{2} (p_3 - p-k)^- x_1^{\prime +}} \int_0^{\infty} d 
x_4^{\prime +} e^{\frac{i}{2} (p_4 - p-k)^- x_4^{\prime +}}. 
\eeqar{I1_101}
where we defined $x_1^{\prime}=x_1-x_0$, $x_2^{\prime}=x_2-x_1$, 
$x_3^{\prime}=x_3-x_1$ and $x_4^{\prime}=x_4-x_5$. By using $\delta$ 
functions from Eqs.~(\ref{I_103R}) and~(\ref{pk-}), we obtain in soft 
gluon soft rescattering approximation: 

\beqar 
(p_{3}-p-k)^{-}&=& (p_4 - p-k)^- 
\approx (p_{2}-k-p_1)^{-} \approx  - \xi 
\eeqar{p103R-soft} 
\beqar (p_1 - p)^{-} &\approx&  \frac{\bq^2+ 2 \bk \bq}{p^+} \ll \xi  
\ll |\bq|, |\bk|, q_z , \mbox{ leading to } \, (p_1 - p)^{-} + q^- \approx q^- \, . 
\eeqar{p1p_103R}
$I_1$ then becomes 
\beqar I_1 &=& 
\frac{-8}{\xi^3}(\sin  \xi L + i\,(1- \cos  \xi L) )
\int_0^{\infty} dx_3^{\prime 
+}(e^{-\frac{i}{2}(q^- -  \xi )x_3^{\prime +}} - 
e^{-\frac{i}{2}q^-x_3^{\prime +}})\, .
\eeqar{I1_103R_f} 

Finally, by using Eqs.~(\ref{I_103R}) and (\ref{I1_103R_f}) Eq.~(\ref{M103R_1}) becomes 
\beqar M_{1,0,3,R}  &=& 2 L g^4 \, t_a t_c t_a t_c \, \int 
\frac{d^3 p}{(2 \pi)^3 2 E} \,  |J(p)|^2 \, 
\int \frac{d^3 k}{(2 \pi)^3 2 \omega} \, 
\frac{\bk^2}{(\bk^2+\chi)^2} \,\nonumber \\ 
&& \; \; \times\ \int \frac{d^4q}{(2\pi)^4} \,  
f(q_0) \, \frac{\bq^2}{\vq^2} \, 2 \, {\rm Im} 
\left (\frac{1}{q^2{-}\Pi_{L}} - \frac{1}{q^2{-}\Pi_{T}} \right)
\, \left(\frac{\sin  \xi L}{\xi L} + i\frac{1-\cos  \xi L}{\xi L} \right)\, 
\nonumber \\ 
&& \; \; \times\ \int_0^{\infty}dx_3^{\prime 
+}(e^{-\frac{i}{2}(q^- -  \xi )x_3^{\prime +}} - 
e^{-\frac{i}{2}q^-x_3^{\prime +}})\,.
\eeqar{M_103R_f} 
 
Note that $M_{1,0,4,L}  = (M_{1,0,3,R} )^*$, leading to
 
\beqar 
M_{1,0,3,R}  + M_{1,0,4,L}  &=& 4 L g^4 \, t_a t_c t_c t_a 
\, \int \frac{d^3 p}{(2 \pi)^3 2 E}  |J(p)|^2 \int 
\frac{d^3 k}{(2 \pi)^3 2 \omega} 
\frac{\bk^2}{(\bk^2+\chi)^2}\, \nonumber \\ 
&& \; \; \times\ \int \frac{d^4q}{(2\pi)^4}\, f(q_0) \, 
\frac{\bq^2}{\vec { {\bf q}}^2} \, 2 \, {\rm Im} \Big(\frac{1}{q^2{-}\Pi_{L}} 
- \frac{1}{q^2{-}\Pi_{T}}\Big)\, 
\theta\Big(1-\frac{q_0^2}{\vec { {\bf q}}^2}\Big) \, \nonumber \\ 
&& \; \; \times\ \Big\{\frac{\sin  \xi L}{ \xi L}(\delta(q^- - 
 \xi )-\delta(q^-))+2 \,\frac{1-\cos  \xi L}{ \xi L}\, 
\int_0^{\infty}dy (\sin(q^- - 
 \xi )y -\sin(q^-y ))\Big\}\, ,
\eeqar{M103R+M104L}
where $y \equiv \frac{ x_3^{\prime}}{2}$. Lets define 
\beqar F(q_0,q_z^2,\bq^2) = f(q_0)\,\frac{\bq^2}{\vec { {\bf q}}^2}\, 
2 \, {\rm Im} \Big(\frac{1}{q^2{-}\Pi_{L}(q)} - 
\frac{1}{q^2{-}\Pi_{T}(q)}\Big)\, 
\theta\Big(1-\frac{q_0^2}{\vec { {\bf q}}^2}\Big)\, 
\eeqar{F_q} 
\\\\ 
In high temperature limit $F(q_0,q_z^2,\bq^2)$ is even function 
of $q_0$ and $q_z$. Then $\frac{\partial F(q_0,q_z^2,\bq^2)}
{\partial q_0}\Big|_{q_0=q_z}$ is an odd function of $q_z$, leading to 
\beqar 
\int \frac{d^4q}{(2\pi)^4}\,F(q_0,q_z^2,\bq^2)\,
(\delta(q_0-q_z-\xi)-\delta(q_2-q_z))\,\nonumber 
= \xi \int \frac{d^3q}{(2\pi)^3}\frac{\partial 
F(q_0,q_z^2,\bq^2)}{\partial q_0}\Big|_{q_0=q_z}\, = 0 \, .
\eeqar{partialF}

We will now compute the second part of the integral in Eq.(\ref{M103R+M104L}): 
\beqar 
\int \frac{d^4q}{(2\pi)^4}\, F(q_0,q_z^2,\bq^2)\,
\int_0^{\infty}dy \Big(\sin(q_0-q_z - \xi )y - \sin(q_0 - q_z)y \Big) \eeqar{I103R104L_2nd} 
To do this we first concentrate on  
\beqar 
&&\int_0^{\infty} dy\, \int_0^{\infty}\frac{d^4q}{(2\pi)^4}\,
F(q_0,q_z^2,\bq^2)\, \sin((q_0-q_z - \xi )y)\,\nonumber \\ 
\; \; &=& \int_0^{\infty}dy\,\int_0^{\infty}\frac{d^4q}{(2\pi)^4}\,
F(q_0,q_z^2,\bq^2)\,
\Big (\sin(q_0y)\cos((q_z- \xi )y)-\cos(q_0y)\sin((q_z- \xi )y) \Big )\,\nonumber\\ 
 \; \; &=& \int_0^{\infty}dy\,\int_0^{\infty}\frac{d^3q}{(2\pi)^4}\,
\cos((q_z- \xi )y)\,\int dq_0\, F(q_0,q_z^2,\bq^2)\,\sin(q_0y) \,\nonumber \\ 
 \; \; && \; \; -\int_0^{\infty}dy\,
\int\frac{dq_0\,d^2q}{(2\pi)^4}\,\cos(q_0y)\,\int 
dq_z\, g(q_0,q_z^2,\bq^2)\,\sin((q_z- \xi )y)\, .
\eeqar{I103R104L_2nd_1} 
Since $ \xi \ll q_z,|\bq|$ we can assume $q_z- \xi \approx q_z$. Therefore
$\sin((q_z- \xi )y)\approx \sin q_zy $, leading to $\int 
dq_z\,F(q_0,q_z^2,\bq^2)\,\sin((q_z- \xi )y)\,\approx \int 
dq_z\,F(q_0,q_z^2,\bq^2)\,\sin(q_zy) = 0 $. 
Similarly, for the second part of the integral in Eq.~(\ref{I103R104L_2nd})
we also obtain 0, which finally leads to 
\beqar 
M_{1,0,3,R}  + M_{1,0,4,L} \approx 0 \, .
\eeqar{M1034RL_f} 

\section{Computation of diagrams $\bm{M_{1,0,5,R}}$ and $\bm{M_{1,0,6,L}}$} 
\label{appM1056} 

We will now calculate the cut diagrams $M_{1,0,5,R} $ and $M_{1,0,6,L} $, 
shown in the Fig.~(\ref{DiagM1056}). We start with $M_{1,0,5,R} $:

\begin{figure}[ht]
\vspace*{5.2cm} 
\includegraphics{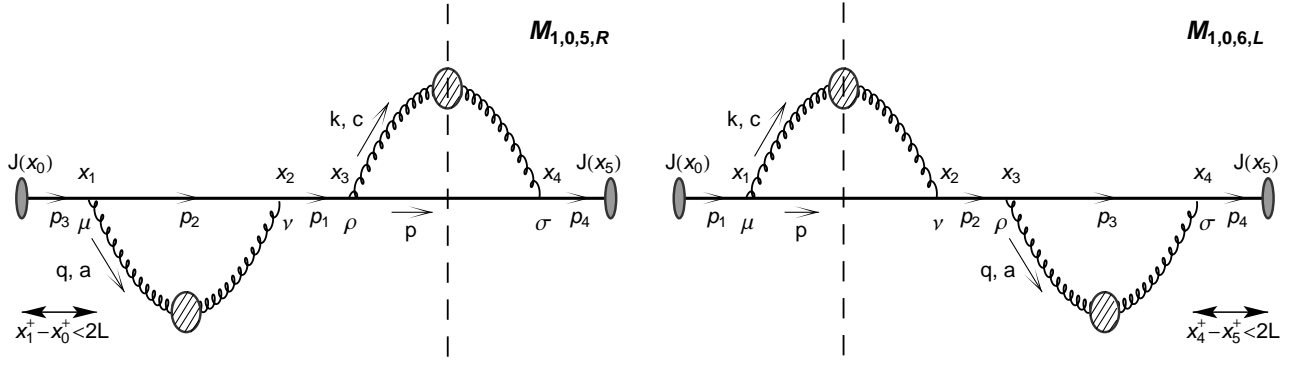}
\caption{Feynman diagrams $M_{1,0,5,R}$ and $M_{1,0,6,L}$ 
contributing to the radiative energy loss to first order in opacity, labeled 
in the same way as Fig.~1.}
\label{DiagM1056}
\end{figure}
%
 \beqar 
M_{1,0,5,R}  &=& \int \prod_{i=0}^{5} \, d^4 x_i \, J(x_0) 
\, \Delta_{++}^{+} (x_1-x_0) \, v_\mu^{+} (x_1) \, D_{++}^{+\mu 
\nu} (x_2-x_1) \, \Delta_{++}^{+} (x_2-x_1) \, 
v_\nu^{+} (x_2) \, \Delta_{++}^{+} (x_3-x_2) \, \nonumber \\ 
&& \; \; \times\  v_\rho^{+} (x_3) \, D_{-+}^{\rho\sigma} 
(x_4-x_3) \, \Delta_{-+} (x_4-x_3) \, v_{\sigma}^{-}(x_4)\, 
\Delta_{--}^{-} (x_5-x_4)\, J(x_5) \nonumber \\
&& \; \; \times\  \; \theta(x_{1}^{+}-x_{0}^{+}) \, 
\theta(x_{2}^{+}-x_{1}^{+}) \, \theta(x_{3}^{+}-x_{2}^{+}) \, 
\theta(x_{5}^{+}-x_{4}^{+})\, \theta(2 L - (x_{1}-x_{0})^{+}) \, 
\nonumber \\ 
&=& g^{4}t_at_at_ct_c \int_{0}^{\infty} \int_{-\infty}^{\infty} 
\prod_{i=1}^{4} \frac{ dp_{i}^{+} d^{2} p_{i}} {(2 \pi)^3 \, 2 
p_{i}^{+}} \int \frac{d^3 k}{(2 \pi)^3 2 \omega}  \frac{d^3 p}{(2 \pi)^3 2 E} \, \int \frac{d^4 q}{(2 \pi)^4} \, 
\nonumber \\ 
&& \; \; \times\ \{(p_2+p_3)^{\mu} D_{\mu \nu}^{>} (q) 
(p_1+p_2)^{\nu}\} \; \{(p+p_1)^{\rho} P_{\rho \sigma} (k) 
(p+p_4)^{\sigma}\} \, I \; , 
\eeqar{M105} 
where 
\beqar 
I&=& \int \prod_{i=0}^{5} d^4 x_i 
\theta(x_{1}^{+}-x_{0}^{+}) \, \theta(x_{2}^{+}-x_{1}^{+}) \,\, 
\theta(x_{3}^{+}-x_{2}^{+}) \, \theta(x_{4}^{+}-x_{5}^{+}) 
\theta(2 L - (x_{1}^{+}-x_{0}^{+})) \, 
\nonumber \\ 
&& \; \; \times\  e^{-i p_{3} \cdot (x_{1} -x_{0})} \, e^{-i q 
\cdot (x_{2} -x_{1})} \, e^{-i p_{2} \cdot (x_{2} -x_{1})} \, 
e^{-i p_1  \cdot (x_{3} -x_{2})} \, e ^{-i k (x_4 - x_3)} \, e^{-i 
p \cdot (x_{4} -x_{3})} \, 
e^{-i p_{4} \cdot (x_{5} -x_{4})} \, J(x_0) \,  J(x_5) \nonumber \\ 
&=& |J(p)|^2 \, (2 \pi)^3 \, \delta ((p_3 - p-k)^+) \delta^2 
(\bp_3 - \bp-\bk) \, (2 \pi)^3 \, \delta ((p_2 + q - p-k)^+) 
\delta^2 (\bp_2 +\bq - \bp-\bk) \, \nonumber \\ && \; \; \times\ 
(2 \pi)^3 \, \delta ((p_1 - p-k)^+) \delta^2 (\bp_1 - \bp-\bk) \, 
(2 \pi)^3 \, \delta ((p_4 - p-k)^+) \delta^2 (\bp_4 - \bp-\bk) \, 
I_1 \; , 
\eeqar{I_105} 
and where 
\beqar 
I_1 = \int_0^{2L} d x_1^{\prime +} e^{-\frac{i}{2} (p_3 - 
p-k)^- x_1^{\prime +}} \int_0^{\infty} d x_4^{\prime +} 
e^{\frac{i}{2} (p_4 - p-k)^{-} x_4^{\prime +}} \int_0^\infty d 
x_3^{\prime +} e^{-\frac{i}{2} (p_1 - p-k)^{-} x_3^{\prime +}} 
\int_0^{\infty} d x_2^{\prime +} e^{-\frac{i}{2} (p_2 + q - p - 
k)^{-} x_2^{\prime +}}. 
\eeqar{I1_105} 
Here we defined $x_1^{\prime} = x_1-x_0$, $x_2^{\prime} = x_2-x_1$, 
$x_3^{\prime} = x_3-x_2$, $x_4^{\prime} = x_4-x_5$.

\medskip

By using $\delta$ functions from Eqs.~(\ref{I_105}) and~(\ref{pk-}) we obtain
$p_1^-=p_3^-=p_4^-$ and: 
\beqar
(p_3 - p - k)^{-} = (p_1 - p - k )^{-} = (p_4 - p - k)^{-} = -  \xi  
\eeqar{pk_105-} 
Eq.~(\ref{I1_105}) then becomes 
\beqar I_1 &=& \frac{16L}{ \xi ^2}\, (\frac{e^{-i \xi L}-1}{-i \xi L}) 
\int_0^{\infty} dy\, e^{-iy(p_2 - p - k + q)^{-}}
\approx \frac{16L}{ \xi ^2}\, (\frac{\sin  \xi L}{ \xi L}
- i\, \frac{\cos  \xi L - 1}{ \xi L}) \int_0^{\infty} dy\, e^{-iy q^{-}}\, , 
\eeqar{I1_105_f} 
where $y=x_2^\prime/2$, and in the last step we used $(p_2-p-k+q)^- = q^- -\xi \approx q^-$ for small $\xi$. 

\bigskip
Finally, by using Eqs.~(\ref{I_105}) and (\ref{I1_105_f}), 
Eq.~(\ref{M105}) becomes
\beqar 
M_{1,0,5,R} & = & -4 L T \, g^4 \, t_a t_a t_c t_c 
\int \frac{d^3p}{(2\pi)^3 2E}\,  |J(p)|^2\, 
\int \frac{d^3 k}{(2\pi)^3 2\omega}\, \frac{\bk^2}{(\bk^2+ \chi)^2}\,
\nonumber \\ 
&& \; \; \times\  \int \frac{d^{4}q}{(2\pi)^4}\, 
\theta(1-\frac{q_0^2}{{\vec q}^2})\frac{1}{q_0}\, \frac{\bq^2}{{\vec q}^2}\, 
2 {\rm Im} \left(\frac{1}{q^2-\Pi_L(q)}-\frac{1}{q^2-\Pi_T(q)}\right)\, 
\nonumber \\ 
&& \; \; \times\ \Big(\frac{\sin  \xi L}{ \xi L}-i\frac{\cos  \xi L - 
1}{ \xi L}\Big)\, \int_0^{\infty} dy\, e^{-iyq^-} 
\eeqar{M105_final} 
 
Since $M_{1,0,6,L} = (M_{1,0,5,R})^*$, it is straightforward to obtain 
(note that, similarly to Appendix~\ref{appM102},  
$\left(\frac{\cos  \xi L - 1}{ \xi L} \right)$ part will vanish under integration over $q$):
\beqar  
M_{1,0,5,R} + M_{1,0,6,L} &= & -4 L T \, g^4 \, t_a t_a t_c t_c 
\, \int \frac{d^3 p}{(2 \pi)^3 2E} \,  |J(p)|^2  \, 
\int \frac{d^3 k}{(2 \pi)^3 2 \omega} \,
\frac{d^2 q}{(2 \pi)^2 } \, \frac{\mu^2}{q^2 (q^2+\mu^2)}
 \nonumber \\ 
&& \hspace*{3cm} \times\ \frac{\bk^2}{(\bk^2+\chi)^2}\, 
\frac{\sin  \xi L}{ \xi L} \, .
\eeqar{M1056RL_f}

\section{Computation of diagrams $\bm{M_{1,1,1,C} }$ and $\bm{M_{1,1,2,C} }$}
\label{appM1112C}

\begin{figure}
\vspace*{5.cm} 
\includegraphics{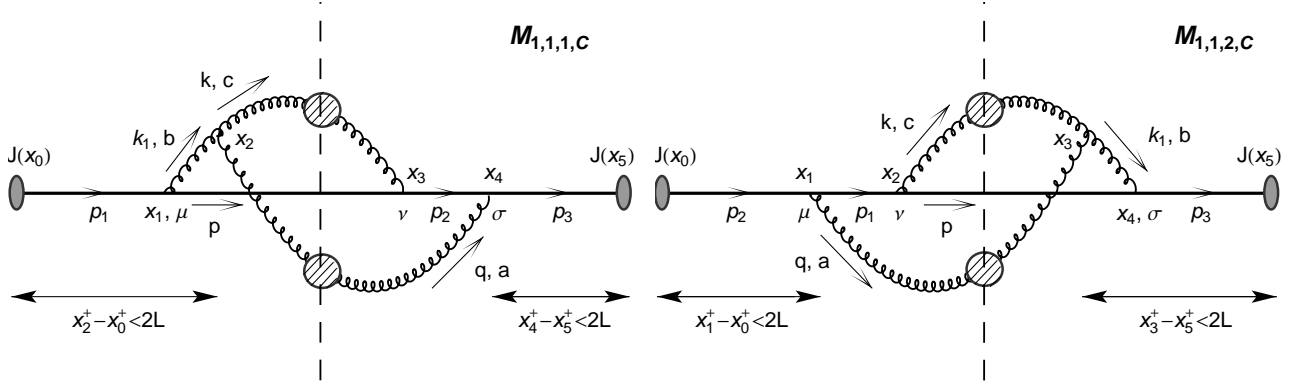}
\caption{Feynman diagrams $M_{1,1,1,C}$ and $M_{1,1,2,C}$ 
contributing to the radiative energy loss to first order in opacity, labeled 
in the same way as Fig.~1.}
\label{DiagM1112C}
\end{figure}

In appendices~(\ref{appM1112C}) - (\ref{appM1134RL}) we present in some detail 
the calculation of the diagrams where only one end of the exchanged gluon $q$ 
is attached to the heavy quark, i.e. one end is attached to the radiated gluon 
$k$ and consequently one 3-gluon vertex is involved in the process. In this appendix, we start with the calculation of the diagrams shown in Fig.~\ref{DiagM1112C}. 

We will first calculate the cut diagram $M_{1,1,1,C} $:
\beqar M_{1,1,1,C}  &=& \int
\prod_{i=0}^{5} \, d x_i \, J(x_0) \, \Delta_{++}^{+} (x_1-x_0) \,
v_\mu^{+} (x_1) \, D_{++}^{+\mu \rho_1} (x_2-x_1) \,
v_{\rho_1\rho_2\rho_3}^{+}(x_2) \Delta_{-+} (x_3-x_1) \,
\nonumber \\
&& \; \; \times\ D_{-+}^{\rho_3 \nu} (x_3-x_2) \,
v_\nu^{-} (x_3) \, D_{-+}^{\rho_2\sigma} (x_4-x_2) \,
v_\sigma^{-} (x_4) \, \Delta_{--}^{-} (x_4-x_3) \, \Delta_{--}^{-}
(x_5-x_4) \, J(x_5) \,
\nonumber \\
&&\; \; \times  \ \theta(x_1^+ -x_0^+) \theta(x_2^ + -x_1^+)
\theta(2L-(x_2-x_0)^+)\theta(x_3^+ -x_4^+)\theta(x_4^+ -x_5^+)
\theta(2L-(x_4-x_5)^+)
\nonumber \\
&=& -\int_{-\infty}^{\infty} \int_{0}^{\infty} \prod_{i=1}^{3} \,
\frac{d p_i^{ +} d^{2} p_i} {(2 \pi)^3 \, 2 p_i^{+}} \frac{ d
k_{1}^{+} d^{2} k_{1}} {(2 \pi)^3 \, 2 k_{1}^{+}} \int \frac{
d^{3} k} {(2 \pi)^3 \, 2 \omega} \frac{ d^{3} p} {(2 \pi)^3 \, 2
E} \frac{ d^{4} q} {(2 \pi)^4 }\, (-i) g^4 \,f^{cba}\ t_c t_b
t_a
\nonumber \\
&&\; \; \times \ (p+p_{1})^{\mu}\, P_{\mu\rho_1}(k_{1})
\,(g^{\rho_1\rho_3}(k_{1}+k)^{\rho_2}+g^{\rho_2\rho_3}(q-k)^{\rho_1}
+g^{\rho_1\rho_2}(-k_{1}-q)^{\rho_3})\,
\nonumber \\
&& \; \; \times\ P_{\rho_3 \nu}(k)\ D_{\rho_2 \sigma}^{>}(q)(p+p_2)^\nu
(p_2+p_3)^\sigma \, I, 
\eeqar{M111C_1}
where 
\beqar 
I &=& \int \prod_{i=0}^{5}\ dx_i \theta(x_1^+ -x_0^+)
\theta(x_2^ + -x_1^+)\theta(2L-(x_2-x_0)^+)\theta(x_3^+
-x_4^+)\theta(x_4^+
-x_5^+)\theta(2L-(x_4-x_5)^+) \,  \nonumber \\
&& \; \; \times\ e^{-ip_{1} (x_{1} -x_{0})} e^{-i k_{1}(x_{2}
-x_{1})} e^{-i p(x_{3} -x_{1})}  e^{-i k(x_{3} -x_{2})} e^{-i
q (x_{4} -x_{2})} e^{-ip_{2}(x_{4} -x_{3})} e^{-i p_3(x_{5}
-x_{4})}\, J(x_0)\ J(x_5) \,
\nonumber \\
&=& \; \; |J(p)|^2(2\pi)^3
\delta((p_{1}-p-k_{1})^+)\delta^2(\bp_{1}-\bp-\bk_1) (2\pi)^3
\delta((k_{1}-k-q)^+)\delta^2(\bk_{1}-\bk-\bq)
\nonumber \\
&& \; \; \times\ (2\pi)^3 \delta((p+k-p_2)^+)\delta^2(\bp+\bk-\bp_2)
(2\pi)^3 \delta((p+k+q-p_2)^+)\delta^2(\bp+\bk{+}\bq-\bp_2)\ I_1 \, ,
\eeqar{I111C_1} 
and where 
\beqar 
I_1 &=& \int_{0}^{2L}
dx_2^{\prime+}\ e^{-\frac{i}{2}(k_{1}-k-q)^- x_2^{\prime+}}
\int_{0}^{x_2^{\prime+}} dx_1^{\prime+}\
e^{-\frac{i}{2}(p_{1}-p-k_{1})^- x_1^{\prime+}}
 \nonumber \\
&& \; \; \times\ \int_{0}^{\infty} dx_3^{\prime+}\
e^{-\frac{i}{2}(p+k-p_2)^- x_3^{\prime+}} \int_{0}^{2L}
dx_4^{\prime+}\ e^{-\frac{i}{2}(p+k+q-p_3)^-x_4^{\prime+}}\,
\nonumber \\
&=& \; \; \frac{1}{-\frac{i}{2}(p_1-k_1-p)^-}\Big\{\frac{e^{-i(p_1
- p-k-q)^{-}L}-1}{-\frac{i}{2}(p_1 - p-k-q)^-} - \frac{e^{-i(k_1 -
k-q)^{-}L}-1}{-\frac{i}{2}(k_1-k-q)^-}\Big\}\, \nonumber \\
&=& \; \; \frac{1}{-\frac{i}{2}(p_1-k_1-p)^-}\,
\frac{e^{-i(p_1 -p-k-q)^{-}L-1}}{-\frac{i}{2}(p_1-p-k-q)^-}
\Big\{ 1- \frac{-\frac{i}{2}(p_1-p-k-q)^{-}}{e^{-i(p_1-p-k-q)^{-}L}-1} 
\frac{e^{-i(k_1 -k-q)^{-}L}-1}{-\frac{i}{2}(k_1-k-q)^-}\Big\} . 
\eeqar{I111C_1_0}

\bigskip
By applying $\delta^3$ functions from Eq.~(\ref{I111C_1}), and
by using $p_i^-=\frac{p_i^2+M^2}{p_i^+}$ we obtain
\beqar p_1^-&=&p_{3}^{-}
=\frac{M^2}{p_{1}^{+}}=\frac{M^2}{(p+k+q)^{+}}
\nonumber \\
p^-&=&\frac{(\bk{+}\bq)^2+M^2}{p^+},\ \ \ k^-=\frac{\bk^2+m_g^2}{k^+},\ \
\ \ k_{1}^{-}=\frac{(\bk{+}\bq)^2+m_g^2}{(k+q)^+}\ \ \ \
p_{2}^{-}=\frac{\bq^2+M^2}{(p+q)^+}\, , 
\eeqar{pk111C} 
which leads to 
\beqar 
(p+k-p_{2})^-&=&\frac{\bk^2+\chi}{2xE} \equiv \xi
\eeqar{p1pk_111C} 
(derived for $x=\frac{k^+}{(p+k^+)}$  and $x \ll 1$) and 
\beq (p_{1}-k_{1}-p)^{-}\approx
-\frac{(\bk{+}\bq)^2+ \chi}{2xE} \equiv - \zeta, 
\eeq{p3p2k_111C} (derived for
$x_1=\frac{(k+q)^+}{(p+k+q)^+}\approx \frac{k}{(p+k)^+}\approx x$, 
and $x \ll 1$).

\bigskip
By using Eqs.~(\ref{p1pk_111C}) and~(\ref{p3p2k_111C}) and assuming (as in 
previous sections) that finite size effects are negligible for
collisional contribution, we obtain
\beqar 
I_1 &=& \frac{-16x^2E^{+2}}{(\bk^{2}+\chi)
((\bk{+}\bq)^{2}+\chi)}
\Big (1- \frac{e^{- i \zeta L}-1}{-i \zeta L}\Big ) \,  
2\pi L \,\delta(q_1-q_z) \, ,
\eeqar{I111C_1_f} 
where we used $\delta((p+k+q-p_1)^-) \approx \delta(q_1-q_z)$. 

\bigskip
Lets now compute
\beqar 
&&(p_1+p_2)^\mu(p+p_2)^\nu(p_2+p_3)^\sigma
P_{\mu\rho_1}(k_1)P_{\rho_3\nu}(k)D_{\rho_2\sigma}^>(q) \,
\Big (g^{\rho_1\rho_3}(k+k_{1})^{\rho_2}+g^{\rho_2\rho_3}(q-k)^{\rho_1}
+g^{\rho_1\rho_2}(-k_{1}-q)^{\rho_3} \Big)\, \nonumber \\
&& \hspace*{2.5cm} \approx  \Big\{(p_{1}+p_{2})^{\mu}\, P_{\mu\rho_1}(k_{1})
\,P_{\nu}^{\rho_1}(k)(p+p_2)^\nu\Big\}\Big\{(k+k_1)^{\rho}
D_{\rho\sigma}^{>}(q)(p_2+p_3)^{\sigma}\Big\} \,
\nonumber \\
&& \hspace*{1.8cm} \approx \Big\{-4\frac{\bk \cdot \bk_1}{x^2}\Big\}\Big\{
\theta \Big(1-\frac{q_0^2}{\vq^2} \Big)\ f(q^0)\ E^+k^+\
\frac{\bq^2}{\vq^2}\
2{\rm Im}\Big(\frac{1}{q^2-\Pi_L(q)}-\frac{1}{q^2-\Pi_T(q)}
\Big)\Big\}.\eeqar{*111C} 
(for more details see Eqs. (C1-C4) in~\cite{DH_Inf}).
Also, note that
\beqar
- i \,f^{abc}t_a t_b t_c = \frac{1}{2} [t_a,t_c][t_c,t_a]\cdot  
\eeqar{tcab}
\bigskip
Finally, by using Eqs.~(\ref{I111C_1}),~(\ref{I111C_1_f}),~(\ref{*111C}) and
(\ref{tcab}), Eq.~(\ref{M111C_1}) reduces to
\beqar M_{1,1,1,C}  &=& -2LT \, g^4 \, [t_a,t_c]\ [t_c,t_a] \,
\int \frac{d^3p}{(2\pi)^3 2E} \, |J(p)|^2 \, \int
\frac{d^3k}{(2\pi)^3 2 \omega } \, 
\frac{d^2q}{(2\pi)^2}\ \frac{\mu^2}{\bq^2(\bq^2+\mu^2)}
 \nonumber \\
&& \; \times  \frac{\bk \cdot
(\bk{+}\bq)}{(\bk^2+ \chi)\big ((\bk{+}\bq)^2+ \chi \big )}
\Big\{1-\frac{\sin \zeta L}{\zeta L} + i \frac{1-\cos \zeta
L}{\zeta L}\Big\} \, .
\eeqar{M111C_f}

\bigskip

Since $M_{1,1,2,C} $ is a complex conjugate of $M_{1,1,1,C} $,
one finally obtains 
\beqar 
M_{1,1,1,C} +M_{1,1,2,C} &=&
 -4LT \, g^4 \, [t_a,t_c] \ [t_c,t_a] \,
 \int \frac{d^3p}{(2\pi)^3 2E} \, |J(p)|^2 \, 
\int \frac{d^3k}{(2\pi)^3 2\omega} 
\frac{d^2q}{(2\pi)^2}\ \frac{\mu^2}{\bq^2(\bq^2+\mu^2)}
\nonumber \\
&& \; \times \frac{\bk \cdot
(\bk{+}\bq)}{(\bk^2+ \chi)\big ((\bk{+}\bq)^2+ \chi \big )} \,
\Big (1-\frac{\sin \zeta L}{\zeta L} \Big ) \, .
\eeqar{M1112C_f}

\section{Computation of diagrams $\bm{M_{1,1,2,R}} + \bm{M_{1,1,1,L}}$}
\label{appM111L2R}

In this appendix we will calculate cut diagrams $M_{1,1,1,L} $ and 
$M_{1,1,2,R} $, shown in the Fig.~(\ref{DiagM111L112R}). We start with 
$M_{1,1,2,R} $:

\begin{figure}[ht]
\vspace*{5.cm} 
\includegraphics{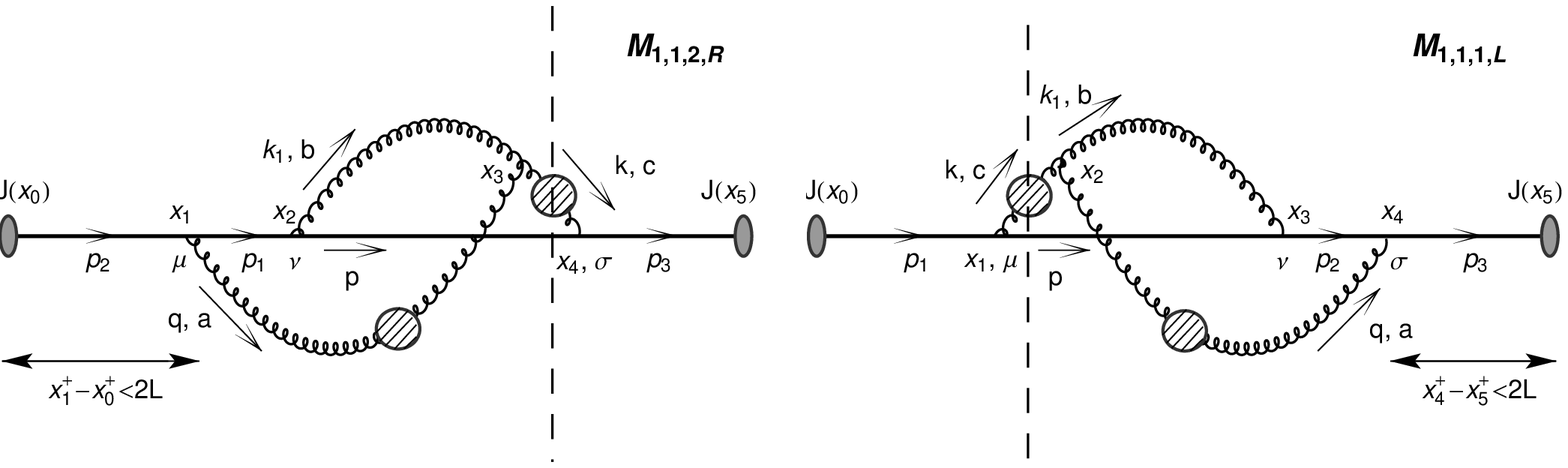}
\caption{Feynman diagrams $M_{1,1,1,L}$ and $M_{1,1,2,R}$ 
contributing to the radiative energy loss to first order in opacity, labeled 
in the same way as Fig.~1.}
\label{DiagM111L112R}
\end{figure}
%
\beqar 
M_{1,1,2,R}  &=& \int
\prod_{i=0}^{5} \, d x_i \, J(x_0) \, \Delta_{++}^{+} (x_1-x_0) \,
v_\mu^{+} (x_1) \, D_{++}^{+\mu \rho_2} (x_3-x_1) \,
v_{\rho_1\rho_2\rho_3}^{+}(x_3) \Delta_{++}^{+} (x_2-x_1) \,
\nonumber \\
&& \; \; \times\ D_{++}^{\rho_1 \nu} (x_3-x_2) \,
v_\nu^{+} (x_2) \, D_{-+}^{\rho_3\sigma} (x_4-x_2) \,
v_\sigma^{-} (x_4) \, \Delta_{-+} (x_4-x_2) \, v_{--}^{-}
(x_5-x_4) \, J(x_5) \,
\nonumber \\
&&\; \; \times  \ \theta(x_1^+ -x_0^+) \, \theta(x_2^+ -x_1^+) \, 
\theta(2L-(x_1-x_0)^+) \, \theta(x_3^+ -x_2^+) \, \theta(x_4^+ -x_5^+) 
\nonumber \\
&=& \int_{-\infty}^{\infty} \int_{0}^{\infty}
\prod_{i=1}^{3} \, \frac{d p_i^{+} d^{2} p_i} {(2 \pi)^3 \, 2
p_i^{+}} \frac{ d k_{1}^{+} d^{2} k_{1}} {(2 \pi)^3 \, 2 k_{1}^{+}}
\int \frac{ d^{3} k} {(2 \pi)^3 \, 2 \omega} \frac{ d^{3} p} {(2
\pi)^3 \, 2 E} \frac{ d^{4} q} {(2 \pi)^4 }\, (-i) g^4
\,f^{abc}\ t_a t_b t_c
\nonumber \\
&&\;\;\times\
(p_{1}+p_{2})^{\mu}\,(p+p_{1})^\nu\,(p+p_{3})^\sigma\,
D_{\mu \rho_2}^{>}(q) P_{\nu\rho_1}(k_{1})\,P_{\rho_3 \sigma }(k)\nonumber \\
&& \; \; \times\
g^{\rho_1\rho_3}(k+k_1)^{\rho_2}+g^{\rho_2\rho_3}(-k-q)^{\rho_1}
+g^{\rho_1\rho_2}(q-k_1)^{\rho_3}\, I, 
\eeqar{M112R_1}

where 
\beqar 
I &=& \int \prod_{i=0}^{5}\ dx_i \theta(x_1^+ -x_0^+)
\theta(x_2^ + -x_1^+)\theta(2L-(x_1-x_0)^+)\theta(x_3^+
-x_2^+)\theta(x_4^+
-x_5^+)\,  \nonumber \\
&& \; \; \times\ e^{-ip_{2} (x_{1} -x_{0})} e^{-i q (x_{3}
-x_{1})} e^{-i p_1(x_{2} -x_{1})}  e^{-i k_1(x_{3} -x_{2})} e^{-i
k (x_{4} -x_{3})} e^{-ip(x_{4} -x_{2})} e^{-i p_3(x_{5} -x_{4})}\,
J(x_0)\ J(x_5) \,
\nonumber \\
&=& \; \; |J(p)|^2 \, (2\pi)^3
\delta((p_{2}-p-k)^+)\delta^2(\bp_{2}-\bp-\bk) (2\pi)^3
\delta((p_{1}-p-k_{1})^+)\delta^2(\bp_{1}-\bp-\bk_{1})
\nonumber \\
&& \; \; \times\ (2\pi)^3 \delta((k_1+q-k)^+)\delta^2(\bk_1+\bq-\bk)
(2\pi)^3 \delta((p_3-p-k)^+)\delta^2(\bp_3-\bp-\bk)\ I_1 \, ,
\eeqar{I_112R_1} 
and where 
\beqar I_1 &=& \int_{0}^{2L}
dx_1^{\prime+}\ e^{-\frac{i}{2}(p_{2}-p-k)^- x_1^{\prime+}}
 \int_{0}^{\infty}
dx_4^{\prime+}\ e^{\frac{i}{2}(p_3-p-k)^-x_4^{\prime+}}\,
 \nonumber \\
&& \; \; \times\ \int_{0}^{\infty} dx_3^{\prime+}\
e^{-\frac{i}{2}(k_1+q-k)^- x_3^{\prime+}}
\int_{0}^{x_3^{\prime+}} dx_2^{\prime+}\
e^{-\frac{i}{2}(p_{1}-p-k_{1})^- x_2^{\prime+}}
\eeqar{I1_112R_1}
In the last equation we defined $x_1^{\prime}=x_1-x_0$, $x_2^{\prime}=x_2-x_1$, $x_3^{\prime}=x_3-x_1$ and $x_4^{\prime}=x_4-x_5$.

\bigskip 
By applying $\delta^3$ functions from Eq.~(\ref{I_112R_1}), and
by using $p_i^-=\frac{p_i^2+M^2}{p_i^+}$ we obtain 
$ p_2^-=p_{3}^{-} =\frac{M^2}{(p+k)^{+}}$, leading to (in soft gluon soft 
rescattering approximation)  
\beqar
(p_2-p-k)^- &=&(p_3-p-k)^-=  - \, \xi
\nonumber \\
&& \hspace*{-1.3cm}(p_{1}-p-k_1)^{-} \approx  
- \, \zeta\, , 
\eeqar{pkq_112R_relations}
where in the second relation we used $x_1= \frac{(k-q)^+}{(p+k-q)^+} \approx
\frac{k^+}{(p+k)^+}=x$.

$I_1$ then becomes
\beqar 
I_1 &= & \frac{8 L}{\xi \, \zeta} \frac{1-e^{i \xi L}}{i \xi L} \,  
\int_{0}^{\infty}dx_3^{\prime +}
\Big\{e^{-\frac{i}{2}(q^- -\xi)x_3^{\prime +}}-
e^{-\frac{i}{2}(q^- + \zeta-\xi) x_3^{\prime +}}\Big\} \, .
\eeqar{I1_112R_2}

Furthermore, similarly as in~\cite{DH_Inf}, for highly energetic jets
\beqar  && \hspace*{0.5cm} \Big\{(p+p_{1})^{\nu}\,
P_{\nu\rho_1}(k_{1})
\,P_{\sigma}^{\rho_1}(k)(p+p_3)^\sigma\Big\}\Big\{(p_1+p_2)^{\mu}
D_{\mu\sigma}(q)(k+k_1)^{\rho}\Big\} \,
\nonumber \\
&\approx& \Big\{-4 \, \frac{\bk \cdot (\bk -\bq)}{x^2}\Big\}\Big\{
E^+ k^+ \, \theta(1-\frac{q_0^2}{\vq^2}) f(q_0) \, \frac{\bq^2}{\vq^2}
\;  2 \, {\rm Im} 
\left(\frac{1}{q^2{-}\Pi_{L}(q)} - \frac{1}{q^2{-}\Pi_{T}(q)} \right)
\Big\}.
\eeqar{PQ_112R}

By using Eqs.~(\ref{I_112R_1}),~(\ref{I1_112R_2}),
(\ref{PQ_112R}) and
(\ref{tcab}), Eq.~(\ref{M112R_1}) finally reduces to
\beqar 
M_{1,1,2,R}  &=& L g^4 \, [t_a,t_c]\ [t_c,t_a] \, \int
\frac{d^3p}{(2\pi)^3 2E} |J(p)|^2 \, \int \frac{d^3k}{(2\pi)^3 2\omega} \,
\frac{d^4q}{(2\pi)^4}\ 
 \nonumber \\
&& \; \times  \frac{\bk \cdot (\bk-\bq)}{(\bk^2+ \chi)((\bk-\bq)^2+ \chi)
}\,\theta(1-\frac{q_0^2}{\vq^2}) f(q_0) \, \frac{\bq^2}{\vq^2}
\;  2 \, {\rm Im} 
\left(\frac{1}{q^2{-}\Pi_{L}(q)} - \frac{1}{q^2{-}\Pi_{T}(q)} \right)
 \nonumber \\
&& \; \times \Big(i \, \frac{1-\cos \xi L}{\xi L} + \frac{\sin
\xi L}{\xi L}\Big) \, \int_{0}^{\infty}dx_3^{\prime +
}\Big\{e^{-\frac{i}{2}(q_0-q_z-\xi)x_3^{\prime
+}}-e^{-\frac{i}{2}(q_0-q_z+ \zeta - \xi )x_3^{\prime
+}}\Big\}
\eeqar{M112R_2}

Since $M_{1,1,1,L} $ is a complex conjugate of $M_{1,1,2,R} $, it is straightforward to obtain ($y \equiv x_3^+/2$)
\beqar 
&& M_{1,1,1,L} +M_{1,1,2,R}  = 2L g^4 \, [t_a,t_c] \ [t_c,t_a] \,
 \int \frac{d^3p}{(2\pi)^3 2E}\, |J(p)|^2 \, \int \frac{d^3k}{(2\pi)^3 2\omega}
\frac{d^4q}{(2\pi)^4}\ 
\nonumber \\
&& \hspace*{2cm} \times  
\frac{\bk \cdot (\bk-\bq)}{(\bk^2+ \chi)((\bk-\bq)^2+ \chi)
}\,\theta(1-\frac{q_0^2}{\vq^2}) f(q_0) \, \frac{\bq^2}{\vq^2}
\;  2 \, {\rm Im} 
\left(\frac{1}{q^2{-}\Pi_{L}(q)} - \frac{1}{q^2{-}\Pi_{T}(q)} \right)
 \nonumber \\
&&  \hspace*{2cm}\times\ \Big\{\frac{\sin \xi L}{\xi L}\Big(\delta(q_0-q_z-\xi)-\delta(q_0-q_z+\zeta - \xi)\Big)\,\nonumber \\
&&  \hspace*{2.5cm} + \ 2 \ \frac{1-\cos \xi L}{\xi L}
\int_0^{\infty}dy \Big\{\sin(q_0-q_z-\xi )y -\sin(q_0-q_z +\zeta-\xi)y\Big\}\, .\eeqar{M1112RL_f}

By applying the same procedure as in Appendix~(\ref{appM103RM104L}), we obtain that 
$ M_{1,1,1,L} +M_{1,1,2,R} = 0$.

\section{Computation of diagrams $\bm{M_{1,1,3,C} }$ and $\bm{M_{1,1,4,C} }$}
\label{appM1134C} 

In this appendix we will calculate the cut diagrams $M_{1,1,3,C} $ and 
$M_{1,1,4,C} $, shown in the Fig.~(\ref{DiagM1134C}). We start with 
$M_{1,1,3,C} $:

\begin{figure}[ht]
\vspace*{4.cm} 
\includegraphics{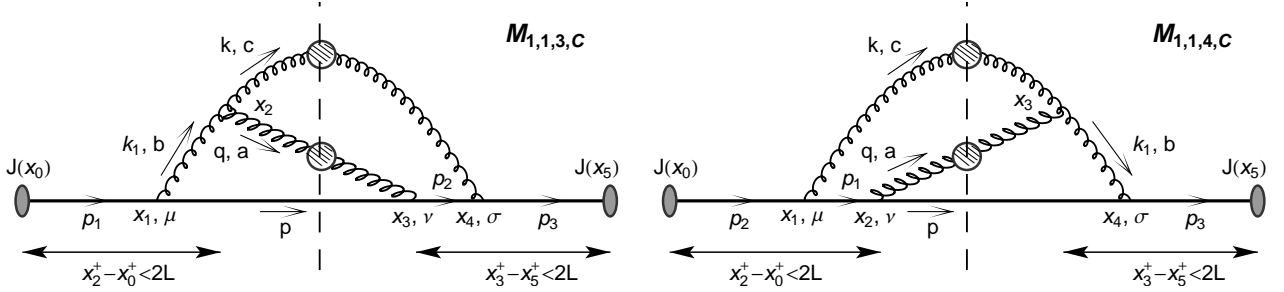}
\caption{Feynman diagrams $M_{1,1,3,C}$ and $M_{1,1,4,C}$ 
contributing to the radiative energy loss to first order in opacity, labeled 
in the same way as Fig.~1.}
\label{DiagM1134C}
\end{figure}

\beqar M_{1,1,3,C}  &=& \int \prod_{i=0}^{5} \, d x_i \, J(x_0) \, 
\Delta_{++}^{+} (x_1-x_0) \, v_\mu^{+} (x_1) \, D_{++}^{+\mu \rho_1} 
(x_2-x_1) \, v_{\rho_1\rho_2\rho_3 }^{+} (x_2) \Delta_{-+} (x_3-x_1) 
\, \nonumber \\ 
 && \; \; \times\ D_{-+}^{\rho_2 \lambda} 
(x_3-x_2) \, v_{\lambda}^{-} (x_3) \Delta_{--}^{-} (x_4-x_3) \, 
D_{-+}^{\rho_3 \sigma} (x_4-x_2) \, v_{\sigma}^{-} (x_4) 
\Delta_{--}^{-} (x_5-x_4) \, J(x_5) \, 
\nonumber \\ 
&& \; \; \times \ \int_{-\infty}^{\infty} \int_{0}^{\infty} 
\frac{ d p_{3}^{+} d^{2} p_{3}} 
{(2 \pi)^3 \, 2 p_{3}^{+}} e^{-i p_{3} (x_{5} -x_{4})}\, 
(ig (p_{2}+p_{3})^{\sigma}\ t_b) \, J(x_5) 
\nonumber \\ 
&&\; \; \times  \ \theta(x_1^+ -x_0^+) \theta(x_2^ + 
-x_1^+)\theta(2L-(x_2-x_0)^+)\theta(x_3^+ -x_4^+)\theta(x_4^+ 
-x_5^+)\theta(2L-(x_3-x_5)^+) 
\nonumber \\ 
&=& \int_{-\infty}^{\infty} \int_{0}^{\infty} \prod_{i=1}^{3} \, 
\frac{d p^{i +} d^{2} p^i} {(2 \pi)^3 \, 2 p^{i+}} \frac{ d k_{1}^{+} 
d^{2} k_{1}} {(2 \pi)^3 \, 2 k_{1}^{+}} \int \frac{ d^{3} k} 
{(2 \pi)^3 \, 2 \omega} \frac{ d^{3} p} {(2 \pi)^3 \, 2 E} 
\frac{ d^{4} q} {(2 \pi)^4 }\ g^4 \ (-i) f^{bac}\ t_b t_a t_c 
\nonumber \\ 
&&\; \; \times \ (p+p_{1})^{\mu}\ (p+p_{2})^{\nu}\ (p_{2}+p_{3})^{\sigma}\ 
P_{\mu\rho_1}(k_{1}) \ D_{\rho_2 \nu}^{>}(q) \ P_{\rho_3 \sigma}(k)\ 
\nonumber \\ 
&& \; \; \times \ 
(g^{\rho_1\rho_3}(k_{1}+k)^{\rho_2}+g^{\rho_2\rho_3}(q-k)^{\rho_1} 
+g^{\rho_1\rho_2}(-k_{1}-q)^{\rho_3})\ I, 
\eeqar{M113C_1}
where 
\beqar I &=& \int \prod_{i=0}^{5}\ dx_i \theta(x_1^+ -x_0^+) 
\theta(x_2^ + -x_1^+)\theta(2L-(x_2-x_0)^+)\theta(x_3^+ 
-x_4^+)\theta(x_4^+ 
-x_5^+)\theta(2L-(x_3-x_5)^+) \ J(x_0)\ J(x_5) \nonumber \\ 
&& \; \; \times e^{-ip_{1} (x_{1} -x_{0})} 
e^{-i k_{1}(x_{2} -x_{1})} 
e^{-i p(x_{3} -x_{1})}  e^{-i q(x_{3} -x_{2})} e^{-i 
p_{2} (x_{4} -x_{3})} e^{-ip_{3}(x_{5} 
-x_{4})} e^{-i k(x_{4} -x_{2})} 
\nonumber \\ 
&=& |J(p)|^2 \, (2\pi)^3 
\delta((p_{1}-p-k_{1})^+)\delta^2(\bp_{1}-\bp-\bk) (2\pi)^3 
\delta((k_{1}-k-q)^+)\delta^2(\bk_{1}-\bk-\bq) 
\nonumber \\ 
&& \; \; \times (2\pi)^3 \delta((p_{2}-p-q)^+)\delta^2(\bp_{2}-\bp-\bq) 
(2\pi)^3 
\delta((p_{3}-p_{2}-k)^+)\delta^2(\bp_{3}-\bp_{2}-\bk)\ I_1 \, ,  
\eeqar{I113C_1} 
and where 
\beqar I_1 &=& \int_{0}^{2L} dx_2^{\prime+}\ 
e^{-\frac{i}{2}(k_{1}-k-q)^- x_2^{\prime+}} 
\int_{0}^{2x_2^{\prime+}} dx_1^{\prime+}\ 
e^{-\frac{i}{2}(p_{1}-p-k_{1})^- x_1^{\prime+}} 
 \nonumber \\ 
&& \; \; \times \int_{0}^{2L} dx_3^{\prime+}\ 
e^{-\frac{i}{2}(p_{2}-p-q)^- x_3^{\prime+}} 
\int_{0}^{2x_3^{\prime+}} dx_4^{\prime+}\ 
e^{-\frac{i}{2}(p_{3}-p_{2}-k)^-x_4^{\prime+}}. 
\eeqar{I113C_1_1} 
Here we defined $x_1^{\prime+}= x_1-x_0$, $x_2^{\prime+}= x_2-x_0$, 
$x_3^{\prime+}= x_3-x_5$ and $x_4^{\prime+}= x_4-x_5$.

By applying $\delta^3$ functions from Eq.~(\ref{I113C_1}), and by using 
$p_i^-=\frac{\bp_i^2+M^2}{p_i^+}$ we obtain 
\beqar 
p^-&=&\frac{(\bk{+}\bq)^2+M^2}{p^+},\ \ \ \
p_{2}^{-}=\frac{\bk^2+M^2}{(p+q)^+}, \ \ \ \
p_{3}^{-} = p_{1}^{-}=\frac{M^2}{p_{1}^{+}}=\frac{M^2}{(p+k+q)^{+}}, 
\nonumber \\ 
&&k^-=\frac{\bk^2+m_g^2}{k^+},\ \ \ \ 
k_{1}^{-}=\frac{(\bk{+}\bq)^2+m_g^2}{(k+q)^+} , \ \ \ \ 
\eeqar{pk113C}
which leads to
\beqar
(p_{1}-p-k_{1})^{-}&=&-\zeta, 
\eeqar{p1pk_113C} 
(derived for $x_1=\frac{(k+q)^+}{(p+k+q)^+}\approx 
\frac{k^+}{(p+k)^+}\approx x \ll 1)$ 
and 
\beq (p_{3 }-p_{2}-k)^{-}=-\xi, 
\eeq{p3p2k_113C} 
(derived for $x_2=\frac{k^+}{(p+k+q)^+}\approx 
\frac{(k+q)^+}{(p+k+q)^+}\approx \frac{k^+}{(p+k)^+} \approx x \ll 1)$. 

\bigskip
By using Eqs.~(\ref{pk113C})-(\ref{p3p2k_113C}), Eq.~(\ref{I113C_1_1}) becomes
\beqar 
I_1 &=& 
 \frac{16}{\zeta \xi}\Big\{ \frac{4 
\sin^2(p_{1}-p-k-q)^-L/2}{((p_{1}-p-k-q)^-)^2}\Big(1-\frac{(p_{1}-p-k-q)^-} 
{e^{-i(p_{1}-p-k-q)^-L}-1}\cdot 
\frac{e^{-i(k_{1}-k-q)^-L}-1}{(k_{1}-k-q)^-} \Big) 
\nonumber \\ 
&& - \frac{4 
\sin^2(p_{2}-p-q)^-L/2}{((p_{2}-p-q)^-)^2}\Big(-\frac{(p_{2}-p-q)^-} 
{e^{-i(p_{2}-p-q)^-L}-1}\cdot 
\frac{e^{-i(k_{1}-k-q)^-L}-1}{(k_{1}-k-q)^-} 
\nonumber \\ 
&&+\frac{(p_{2}-p-q)^-} 
{e^{-i(p_{2}-p-q)^-L}-1}\cdot\frac{e^{-i(p_{1}-p-k-q)^-L}-1}{p_{1}-p-k-q)^-} 
\Big) 
  \Big\} 
  \nonumber \\ 
&\approx&  \frac{16}{\zeta\xi}2\pi L \Big\{ 
\delta(q^- + \zeta)\Big ( 1-\frac 
{e^{-i \zeta L}-1}{-i \zeta L} 
\Big) 
-\delta(p_{1}-p-q)^-)\Big ( \frac 
{e^{ i \xi L}-1}{i \xi L} 
- \frac 
{e^{-i(\zeta - \xi) L}-1}{-i(\zeta - \xi) L} 
\Big) 
 \Big \}  \nonumber \\
 &\approx& \frac{16}{\zeta \xi}2\pi L\ \delta(q^-)\Big 
[1-\frac{e^{-i\zeta L}-1}{-i\zeta L} -\frac{e^{i\xi L}-1}
{i\xi L}+\frac{e^{-i(\zeta-\xi) L}-1}{-i(\zeta-\xi) L}\Big] ,
\eeqar{I113C_1_2} 
where, as in the previous sections, we assumed that finite size effects are 
negligible for collisional contribution, and in the last step we used Eq.~(\ref{q-xi-zeta}). 

\medskip
Analogously to calculations performed in~\cite{DH_Inf} (see Appendix C 
in~\cite{DH_Inf}) one has 
\beqar 
&& (p+p_{1})^{\mu}(p+p_{2})^{\nu}(p_{2}+p_{3})^{\sigma}\ 
P_{\mu \rho_1}(k_{1})\ D^>_{\rho_2 \lambda}(q)\ P_{\rho_3 \sigma}(k) 
\big (g^{\rho_1 \rho_3}(k+k_{1})^{\rho_2}+g^{\rho_2 
\rho_3}(q-k)^{\rho_1}+g^{\rho_1 
\rho_2}(-k_{1}-q)^{\rho_3}\big ) 
\nonumber \\ 
&& \hspace*{3cm} \approx  (p+p_{1})^{\mu}\ P_{\mu \rho_1}(k_{1})\ 
P^{\rho_1}_{\sigma}(k)\ 
(p_{2}+p_{3})^{\sigma}\big [ 
(k+k_{1})^{\rho}\ D^>_{\rho 
\sigma}(q)(p+p_{2})^{\nu}\big] 
\nonumber \\ 
&&\hspace*{2.3cm} \approx -4\frac{\bk \cdot (\bk{+}\bq)}{x^2}\ 
\theta\Big(1-\frac{q_0^2}{\vq^2} 
\Big)\ f(q^0)\ E^+k^+\ \frac{\bq^2}{\vq^2}\ 
2{\rm Im}\Big(\frac{1}{q^2-\Pi_L(q)}-\frac{1}{q^2-\Pi_T(q)} \Big). 
\eeqar{propagators113C} 

By using Eqs.~(\ref{I113C_1}),~(\ref{I113C_1_2}), (\ref{propagators113C}) and
(\ref{tcab}), Eq.~(\ref{M113C_1}) finally reduces to
\beqar  
M_{1,1,3,C}  &=& -2LT \, g^4 \, [t_a,t_c]\ [t_c,t_a] \, \int 
\frac{d^3p}{(2\pi)^3 2E}   |J(p)|^2 \, \int \frac{d^3k}{(2\pi)^3 2\omega} \,
\frac{d^2q}{(2\pi)^2}\ \frac{\mu^2}{\bq^2(\bq^2+\mu^2)}  
 \nonumber \\ 
&& \; \times 
\frac{\bk \cdot (\bk{+}\bq)}{(\bk^2+ \chi)\big ((\bk{+}\bq)^2+ \chi 
\big )} 
\Big [1-\frac{e^{-i\zeta L}-1}{-i\zeta L} -\frac{e^{i\xi L}-1}{i\xi L}
+\frac{e^{-i(\zeta-\xi) L}-1}{-i(\zeta-\xi) L}\Big]. 
\eeqar{M113C_f} 

Since $M_{1,1,4,C} $ is a complex conjugate of $M_{1,1,3,C} $, one finally obtains 
\beqar 
M_{1,1,3,C} +M_{1,1,4,C} &=& 
 -4LT \, g^4 \, [t_a,t_c] \ [t_c,t_a] \, 
 \int \frac{d^3p}{(2\pi)^3 2E} \, |J(p)|^2 \, 
\int \frac{d^3k}{(2\pi)^3 2\omega} \, 
\frac{d^2q}{(2\pi)^2}\, \frac{\mu^2}{\bq^2(\bq^2+\mu^2)} 
\nonumber \\ 
&& \hspace*{-0.3cm}\times 
\frac{\bk \cdot(\bk{+}\bq)}{(\bk^2+ \chi)\big ((\bk{+}\bq)^2+ \chi 
\big )} 
\left(1-\frac{\sin\zeta L}{\zeta L} -\frac{\sin \xi 
L}{\xi L}+\frac{\sin(\zeta-\xi) L}{(\zeta-\xi) L}\right ) . 
 \eeqar{M1134C_f}

\section{Computation of diagrams $\bm{M_{1,1,3,R} }\, , \bm{M_{1,1,3,L} }\, ,
\bm{M_{1,1,4,L} }$ and $\bm{M_{1,1,4,R} }$ 
}
\label{appM1134RL}

In this appendix we will calculate the cut diagrams $M_{1,1,3,R} $, 
$M_{1,1,3,L} $, $M_{1,1,4,R} $ and $M_{1,1,4,L} $ shown in the 
Fig.~(\ref{DiagM1134RL}). We start with 
$M_{1,1,3,R} $:

\begin{figure}[ht]
\vspace*{8.cm} 
\includegraphics{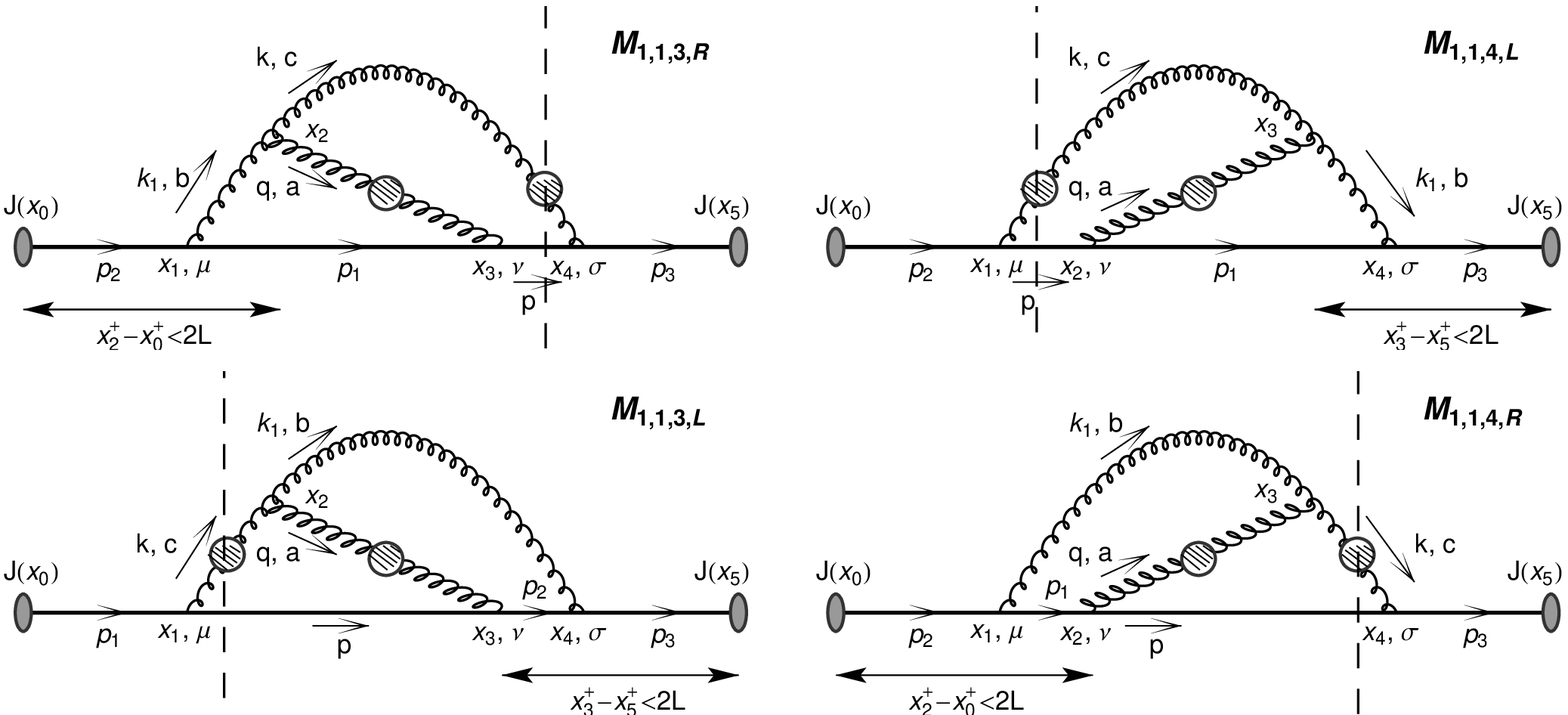}
\caption{Feynman diagrams $M_{1,1,3,R} $, 
$M_{1,1,3,L} $, $M_{1,1,4,R} $ and $M_{1,1,4,L} $ 
contributing to the radiative energy loss to first order in opacity, labeled 
in the same way as Fig.~1.}
\label{DiagM1134RL}
\end{figure}
%
\beqar M_{1,1,3,R}  &=& \int
\prod_{i=0}^{5} \, d x_i \, J(x_0) \, \Delta_{++}^{+} (x_1-x_0) \,
v_\mu^{+} (x_1) \, D_{++}^{+\mu \rho_1} (x_2-x_1) \,
v_{\rho_1\rho_2\rho_3}^{+}(x_2) \Delta_{++}^{+} (x_3-x_1) \,
\nonumber \\
&& \; \; \times\ D_{++}^{+ \rho_2 \nu} (x_3-x_2) \,
v_\nu^{+} (x_3) \, D_{-+}^{\rho_3\sigma} (x_4-x_2) \,
v_\sigma^{-} (x_4) \, \Delta_{-+} (x_4-x_3) \, \Delta_{--}^{-}
(x_5-x_4) \, J(x_5) \,
\nonumber \\
&&\; \; \times  \ \theta(x_1^+ -x_0^+) \theta(x_2^ +
-x_1^+)\theta(2L-(x_2-x_0)^+)\theta(x_3^+ -x_2^+)\theta(x_4^+
-x_5^+)\nonumber \\
&\approx& -\int_{-\infty}^{\infty} \int_{0}^{\infty}
\prod_{i=1}^{3} \, \frac{d p_i^{+} d^{2} p_i} {(2 \pi)^3 \, 2
p_i^{+}} \frac{ d k_{1}^{+} d^{2} k_{1}} {(2 \pi)^3 \, 2 k_{1}^{+}}
\int \frac{ d^{3} k} {(2 \pi)^3 \, 2 \omega} \frac{ d^{3} p} 
{(2 \pi)^3 \, 2 E} \frac{ d^{4} q} {(2 \pi)^4 }\, (-i) g^4
\,f^{cba}\ t_c t_b t_a
\nonumber \\
&&\; \; \times \ \Big\{(p_{1}+p_{2})^{\mu}\, P_{\mu\rho_1}(k_{1})
\,P_\sigma^{\rho_1}(k)(p+p_3)^\sigma\Big\}\Big\{(k+k_1)^{\rho_2}\,D_{\rho_2
\sigma}^{>}(q)(p+p_1)^\nu\Big\} \, I \, ,
\eeqar{M113R_1} 
where we used that in soft gluon soft rescattering approximation (see 
Eqs. (C1-C4) in~\cite{DH_Inf})
\beqar &&
(p_{1}+p_{2})^{\mu}\,(p+p_1)^\nu(p+p_3)^\sigma\,
P_{\mu\rho_1}(k_{1})\, D_{\rho_2 \nu}^{>}(q)\,
P_{\rho_3 \sigma}(k)\,
(g^{\rho_1\rho_3}(k_{1}+k)^{\rho_2}+g^{\rho_2\rho_3}(q-k)^{\rho_1}
+g^{\rho_1\rho_2}(-k_{1}-q)^{\rho_3})
\,\nonumber \\
&& \hspace*{2.5cm}\approx
\Big\{(p_{1}+p_{2})^{\mu}\,P_{\mu\rho_1}(k_{1})\,
P_\sigma^{\rho_1}(k)(p+p_3)^\sigma\,\Big\}
\Big\{(k+k_1)^{\rho_2}\, D_{\rho_2 \nu}^{>}(q)(p+p_1)^\nu\Big\} \;.
\eeqar{c_113R}
In Eq.~(\ref{M113R_1}) $I$ corresponds to  
\beqar 
I &=& \int \prod_{i=0}^{5}\ dx_i \theta(x_1^+ -x_0^+)
\theta(x_2^ + -x_1^+)\theta(2L-(x_2-x_0)^+)\theta(x_3^+ - x_2^+)
\theta(x_4^+ -x_5^+)\,  \nonumber \\
&& \; \; \times\ e^{-ip_{2} (x_{1} -x_{0})} e^{-i k_{1}(x_{2}
-x_{1})} e^{-i p_1(x_{3} -x_{1})}  e^{-i q(x_{3} -x_{2})} e^{-i k
(x_{4} -x_{2})} e^{-ip(x_{4} -x_{3})} e^{-i p_3(x_{5} -x_{4})}\,
J(x_0)\ J(x_5) \,
\nonumber \\
&=& \; \; |J(p)|^2 \, (2\pi)^3
\delta((p_{2}-p_1-k_{1})^+)\delta^2(\bp_{2}-\bp_1-\bk_1) (2\pi)^3
\delta((k_{1}-k+p_1-p)^+)\delta^2(\bk_{1}-\bk+\bp_1-\bp)
\nonumber \\
&& \; \; \times\ (2\pi)^3 \delta((p_1+q-p)^+)\delta^2(\bp_1+\bq-\bp)
(2\pi)^3 \delta((p_3-p-k)^+)\delta^2(\bp_3-\bp-\bk)\ I_1 \, ,
\eeqar{I113R_1} 
where 
\beqar 
I_1 &=& \int_{0}^{2L}
dx_2^{\prime+}\ e^{-\frac{i}{2}(k_{1}-k-p_1-p)^- x_2^{\prime+}}
\int_{0}^{x_2^{\prime+}} dx_1^{\prime+}\
e^{-\frac{i}{2}(p_{2}-p_1-k_{1})^- x_1^{\prime+}}
 \nonumber \\
&& \; \; \times\ \int_{0}^{\infty} dx_3^{\prime+}\
e^{-\frac{i}{2}(p_1+q-p)^- x_3^{\prime+}} \int_{0}^{\infty}
dx_4^{\prime+}\ e^{\frac{i}{2}(p_3-p-k)^-x_4^{\prime+}}\,
\eeqar{I1_113R_1}

 By applying $\delta^3$ functions from Eq.~(\ref{I113R_1}), and
by using $p_i^-=\frac{\bp_i^2+M^2}{p_i^+}$ we obtain in soft gluon soft 
rescattering approximation (see also~(\ref{q-xi-zeta})) 
\beqar 
&&p_2^- = p_{3}^{-} = \frac{M^2}{E^{+}} \, ,  \nonumber \\
&&(p_2-p_1-k_{2})^- = -\zeta  \, , \nonumber \\
&&(p_{3}-p-k)^{-} = (p_2-p-k)^- = -\xi \, , \nonumber \\
&&(k_{1}-k+p_1-p)^{-} =  \zeta - \xi \, , \nonumber \\
&&(q+p_1-p)^- \approx q_0-q_z\, .
\eeqar{zeta_chi_113R} 

By using Eq.~(\ref{zeta_chi_113R}), 
Eq.~(\ref{I1_113R_1}) becomes ($y \equiv \frac{x_3^{\prime +}}{2}$)
\beqar 
I_1 &= & - \frac{16L}{\zeta \xi}
\Big[\frac{e^{i \xi L}-1}{i \xi L}-\frac{e^{i(\xi-\zeta)L}-1}
{i(\xi-\zeta)L}\Big]\,
\int_{0}^{\infty}e^{-iq^-y}dy \, .
\eeqar{I1_113R_f}

Similarly to previous appendices, Eq.~(\ref{c_113R}) is equal to
\beqar  
&& \hspace*{0.3cm} \Big\{(p_{1}+p_{2})^{\mu}\,
P_{\mu\rho_1}(k_{1})
\,P_{\sigma}^{\rho_1}(k)(p+p_2)^\sigma \Big\}\Big\{(k+k_1)^{\rho_2}
D_{\rho_2\nu}^{>}(q)(p_2+p_3)^{\nu}\Big\} \,
\nonumber \\
&\approx& 
-4\frac{\bk \cdot (\bk{+}\bq)}{x^2}\ 
\theta\Big(1-\frac{q_0^2}{\vq^2} 
\Big)\ f(q^0)\ E^+k^+\ \frac{\bq^2}{\vq^2}\ 
2{\rm Im}\Big(\frac{1}{q^2-\Pi_L(q)}-\frac{1}{q^2-\Pi_T(q)} \Big). 
\eeqar{c_113R_f} 

Finally, by using Eqs.~(\ref{I113R_1}),~(\ref{I1_113R_f})-(\ref{tcab}), Eq.~(\ref{M113R_1}) reduces to
\beqar M_{1,1,3,R}  &=& -2L \, g^4 \, [t_a,t_c]\ [t_c,t_a] \,
\int \frac{d^3p}{(2\pi)^3 2E} \, |J(p)|^2 \, \int \frac{d^3k}{(2\pi)^3 2\omega}
\frac{d^4q}{(2\pi)^4}\  \,  
 \nonumber \\
&& \; \; \times  \frac{\bk \cdot (\bk{+}\bq)}{((\bk{+}\bq)^2+ \chi)\big
((\bk^2+ \chi \big )}\, \nonumber \\
&& \; \; \times\ \theta\Big(1-\frac{q_0^2}{\vq^2}
\Big)\, f(q_0) \, \frac{\bq^2}{\vq^2}\
2 \, {\rm Im}\Big(\frac{1}{q^2-\Pi_L(q)}-\frac{1}{q^2-\Pi_T(q)} \Big)
 \nonumber \\
&& \; \; \times \Big[\frac{e^{i \xi L}-1}{i \xi L}-\frac{e^{i(\xi-\zeta)L}-1}
{i(\xi-\zeta)L}\Big] \int_{0}^{\infty}e^{-iq^-y}dy \, .
\eeqar{M113R_f}

Since $M_{1,1,4,L} $ is a complex conjugate of $M_{1,1,3,R} $, one finally 
obtains
\beqar 
M_{1,1,3,R} +M_{1,1,4,L} &=& 
 -2LT \, g^4 \, [t_a,t_c] \ [t_c,t_a] \,
 \int \frac{d^3p}{(2\pi)^3 2E}\,  |J(p)|^2 \, 
\int \frac{d^3k}{(2\pi)^3 2\omega} \,
\frac{d^2q}{(2\pi)^2}\,\frac{\mu^2}{\bq^2(\bq^2+\mu^2)}
\nonumber \\
&& \hspace*{-2cm} \times 
\frac{\bk\cdot(\bk{+}\bq)}{((\bk{+}\bq)^2+ \chi)(\bk^2+ \chi)}
\Big(\frac{\sin \xi L}{ \xi L}-
\frac{\sin (\zeta-\xi) L}{(\zeta-\xi) L}\Big) \, .
\eeqar{M113R+M114L_f} 

Similarly 
$M_{1,1,3,L} +M_{1,1,4,R}  = M_{1,1,3,R} +M_{1,1,4,L}  $ 
leading to 
\beqar
M_{1,1,3,L} + M_{1,1,3,R} +M_{1,1,4,R}  +M_{1,1,4,L}  &=& 
 -4LT \, g^4 \, [t_a,t_c] \ [t_c,t_a] \,
 \int \frac{d^3p}{(2\pi)^3 2E} \, |J(p)|^2 \, 
\int \frac{d^3k}{(2\pi)^3 2\omega} \, 
\frac{d^2q}{(2\pi)^2}\ \,
\nonumber \\
&& \hspace*{-5cm} \times  \frac{\mu^2}{\bq^2(\bq^2+\mu^2)} \,
\frac{\bk\cdot(\bk{+}\bq)}{((\bk{+}\bq)^2+ \chi)(\bk^2+ \chi)}
\Big(\frac{\sin \xi L}{ \xi L}-
\frac{\sin (\zeta-\xi) L}{(\zeta-\xi) L}\Big) \, .
\eeqar{M1134RL_f}

\section{Computation of diagram $\bm{M_{1,2,C}}$}
\label{appM12C}

\begin{figure}[ht]
\vspace{4.3cm} 
\includegraphics{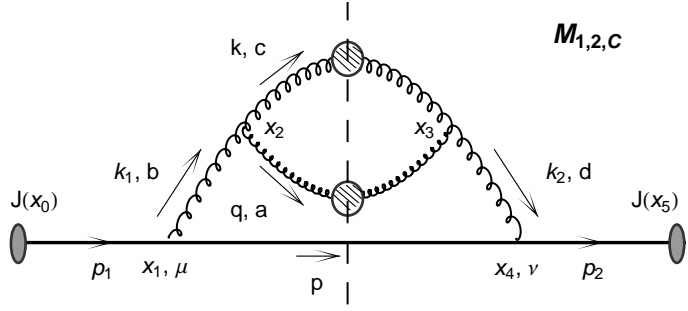}
\caption{Feynman diagram $M_{1,2}$ contributing to the radiative energy loss 
to first order in opacity, labeled in the same way as Fig.~1.}
\label{DiagM12C}
\end{figure}

In the Appendices~(\ref{appM12C}) and~(\ref{appM12RL}) we present in some 
detail the calculation of the diagrams where both ends of the exchanged gluon 
$q$ are attached to the radiated gluon $k$, i.e. there are two 3-gluon 
vertices involved in the process. In this appendix, we start with the 
calculation of the diagram $M_{1,2,C} $ shown in Fig.~\ref{DiagM12C}.  

%
\beqar 
M_{1,2,C}  &=& \int
\prod_{i=0}^{5} \, d x_i \, J(x_0) \, \Delta_{++}^{+} (x_1-x_0) \,
v_\mu^{+} (x_1) \, D_{++}^{+\mu \rho_1} (x_2-x_1) \,
v_{\rho_1\rho_2\rho_3}^{+}(x_2) D_{-+}^{\rho_3\sigma_3} (x_3-x_2)
\,
\nonumber \\
&& \; \; \times\ D_{-+}^{\rho_2 \sigma_2} (x_3-x_2) \,\Delta_{-+}
(x_4-x_1) v_{\sigma_1\sigma_2\sigma_3}^{-} (x_3) \,
D_{--}^{-\sigma_1\nu} (x_4-x_3) \, v_\nu^{-} (x_4) \, \,
\Delta_{--}^{-}(x_5-x_4) \, J(x_5) \,
\nonumber \\
&&\; \; \times\ \theta(x_1^+ -x_0^+) \theta(x_2^+
-x_1^+)\theta(2L-(x_2-x_0)^+)\theta(x_3^+ -x_4^+)\theta(x_4^+
-x_5^+)\theta(2L-(x_3-x_5)^+)\nonumber \\
&=&  - \int_{0}^{\infty}\int_{-\infty}^{\infty}
\prod_{i=1}^{2} \, \frac{d p_i^{+} d^{2} p_i} {(2 \pi)^3 \, 2
p_i^{+}} \frac{ d k_{i}^{+} d^{2} k_{i}} {(2 \pi)^3 \, 2
k_{i}^{+}} \int \frac{ d^{3} k} {(2 \pi)^3 \, 2 \omega} \frac{
d^{3} p} {(2 \pi)^3 \, 2 E} \frac{ d^{4} q} {(2 \pi)^4 }\,  g^4
\,f^{bac}\ t_b \, f^{dac}t_d \nonumber \\
&& \; \; \times \,  
\Big ( (k+k_{1})^{\rho_2} \, D_{\rho_2\sigma_2}^{>}(k+k_{2})^{\sigma_2}\Big )\,
\Big ( (p+p_{1})^\mu \, P_{\mu\rho_1}(k_1) \, P^{\sigma_1\rho_1}(k) \,
P_{\rho_1\nu}(k_2)(p+p_2)^\nu \Big ) \, I,
\eeqar{M12C_1}
where in the second step we only keep the dominant parts from triple gluon 
vertices (see~\cite{DH_Inf}).

In Eq.~(\ref{M12C_1}) $I$ corresponds to
\beqar 
I &=& \int \prod_{i=0}^{5}\ dx_i \theta(x_1^+ -x_0^+) \theta(x_2^ + -x_1^+)
\theta(2L-(x_2-x_0)^+)\theta(2L-(x_3-x_5)^+)\theta(x_3^+ -x_4^+)
\theta(x_4^+ -x_5^+)\,  \nonumber \\
&& \; \; \times\ e^{-ip_{1} (x_{1} -x_{0})} e^{-i k_{1}(x_{2}
-x_{1})} e^{-i k(x_{3} -x_{2})}  e^{-i q(x_{3} -x_{2})} e^{-i p
(x_{4} -x_{1})} e^{-ik_2(x_{4} -x_{3})} e^{-i p_2(x_{5} -x_{4})}\,
J(x_0)\, J(x_5) \, \nonumber \\
&=&  |J(p)|^2 \,(2\pi)^3\,
\delta((p_{1}-p-k_1)^+)\delta^2(\bp_{1}-\bp-\bk_1) (2\pi)^3
\delta((k_{1}-k-q)^+)\delta^2(\bk_{1}-\bk-\bq)
\nonumber \\
&& \; \; \times\ (2\pi)^3 \delta((k_2-q-k)^+)\delta^2(\bk_2-\bq-\bk)
(2\pi)^3 \delta((p_2-p-k_2)^+)\delta^2(\bp_2-\bp-\bk_2)\ I_1 \, ,
\eeqar{I_12C_1}
where
\beqar I_1 &=& \int_{0}^{2L} dx_2^{\prime+}
e^{-\frac{i}{2}(k_{1}-k-q)^- x_2^{\prime+}}
\int_{0}^{x_2^{\prime+}} dx_1^{\prime+}\
e^{-\frac{i}{2}(p_{1}-p-k_{1})^- x_1^{\prime+}}
 \nonumber \\
&& \; \; \times\ \int_{0}^{2L} dx_3^{\prime+}\
e^{\frac{i}{2}(k_2-q-k)^- x_3^{\prime+}} \int_{0}^{x_3^{\prime +}}
dx_4^{\prime+}\ e^{\frac{i}{2}(p_2-p-k_2)^-x_4^{\prime+}}\,
\eeqar{I1_12C_1}
Here $x_1^{\prime+}=x_1^+-x_0^+$, $x_2^{\prime+}=x_2^+-x_0^+$, 
$x_3^{\prime+}=x_3^+-x_5^+$ and $x_4^{\prime+}=x_4^+-x_5^+$. 
Furthermore, from $\delta$ functions in Eq.~(\ref{I_12C_1}) it follows 
\beqar
p_2 = p_1,  \, \, \, \,    
k_2 = k_1,  \, \, \, \, 
(p_1 - k_1 -p)^- = (p_2 - k_2 -p)^- = - \zeta \, .
\eeqar{pkq_relations_12C}

After using Eq.~(\ref{pkq_relations_12C}), Eq.(\ref{I1_12C_1}) becomes 
\beqar 
I_1 &=& \frac{16}{\zeta^2}\Big[\frac{4\sin^2((p_1-p-k-q)^{-}L/2)-1}{((p_1-p-k-p)^{-})^2}
\Big\{1-\frac{(p_1-p-k-q)^-}{e^{i(p_1-p-k-q)^{-} L}-1}\cdot\frac{e^{-i(k_1-k-q)^{-} L}-1}{k_1-k-q}
\nonumber\\ &&\;\;-\frac{(p_1-p-k-q)^-}{e^{-i(p_1-p-k-q)^{-} L}-1}\cdot\frac{e^{i(k_1-k-q)^{-} L}-1}{k_1-k-q}\Big\}+\frac{4\sin^2((k_1-k-q)^{-} L/2)-1}{((k_1-k-q)^{-})^2}
\Big]\,\nonumber\\
&\approx& \frac{16}{\zeta^2}\Big[2\pi L\delta((p_1-p-k-q)^{-})
\Big\{1-\frac{e^{i \zeta L}-1}{i\zeta L}-\frac{e^{-i \zeta L}-1}{-i\zeta L}\Big\}+
2\pi L\delta((k_1-k-q)^{-})\Big]\,\nonumber\\
&=& 2\pi L  \frac{16}{\zeta^2}\Big[\Big(1-2 \,\frac{\sin\zeta L}{\zeta
L}\Big) \delta((p_1-p-k-q)^{-})+\delta((k_1-k-q)^{-})\Big]\nonumber \\
&\approx& \frac{64 \pi L}{\zeta^2}\Big(1-\frac{\sin\zeta L}{\zeta L}\Big)
\delta(q_0-q_z) \, .  
\eeqar{I1_12C_2}
As in the previous sections, we here assumed that finite size effects are 
negligible for collisional energy loss. Furthermore, in the last step we 
used Eq.~(\ref{q-xi-zeta}), i.e. 
$(p_1-p-k-q)^{-}\approx (k_1-k-q)^{-} \approx q_0-q_z$.

\medskip
Similarly as in~\cite{DH_Inf}, for highly energetic jets
\beqar
(p+p_{1})^\mu\, P_{\mu\rho_1}(k_{1})
\,P^{\sigma_1\rho_1}(k)\,P_{\sigma\nu}(k_2)(p+p_1)^\nu) &=&
(p+p_{1})^\mu\, P_{\mu\rho_1}(k_{1})
\,P^{\sigma_1\rho_1}(k)\,P_{\sigma_1\nu}(k_1)(p+p_1)^\nu) \approx
-\frac{4(\bk{+}\bq)^2}{x^2}\nonumber \\ 
 (k+k_1)^\rho_2 D_{\rho_2\sigma_2}^{>}(q)(k+k_2)^{\sigma_2} 
&\approx& k^+k_1^+ 
\theta(1-\frac{q_0^2}{\vq^2}) f(q_0) \, \frac{\bq^2}{\vq^2}
\;  2 \, {\rm Im} 
\left( \frac{1}{q^2{-}\Pi_{L}(q)} - \frac{1}{q^2{-}\Pi_{T}(q)} \right)
\eeqar{PQ12C}

Finally, by using Eqs.~(\ref{I_12C_1}), (\ref{I1_12C_2}) and (\ref{PQ12C}), and after performing the same procedure as in the previous appendices,  
Eq.~(\ref{M12C_1}) reduces to
\beqar M_{1,2,C} &=& 8LT \, g^4
\,[t_a,t_c][t_c,t_a] \int 
\frac{ d^{3} p} {(2 \pi)^3 \, 2 E} \, |J(p)|^2 \, \int \,  
\frac{ d^{3} k} {(2 \pi)^3 \, 2 \omega} \,  \frac{d^{2} q} {(2 \pi)^2} \,
\frac{\mu^2}{q^2(q^2+\mu^2)} \nonumber \\
 && \hspace*{3.3cm}
\times\
\frac{(\bk{+}\bq)^2}{((\bk{+}\bq)^2+\chi)^2}
\Big[1-\frac{\sin (\zeta L)}{\zeta L}\Big]\, .
\eeqar{M12C_f}

\section{Computation of diagrams $\bm{M_{1,2,R}}$ and $\bm{M_{1,2,L}}$}
\label{appM12RL}

In this appendix, we will calculate the cut diagrams 
$M_{1,2,R} $ and $M_{1,2,L} $, which are shown in 
Fig.~\ref{DiagM12RL}. We start with $M_{1,2,R} $: 
%
\begin{figure}[ht]
\vspace{4.3cm} 
\includegraphics{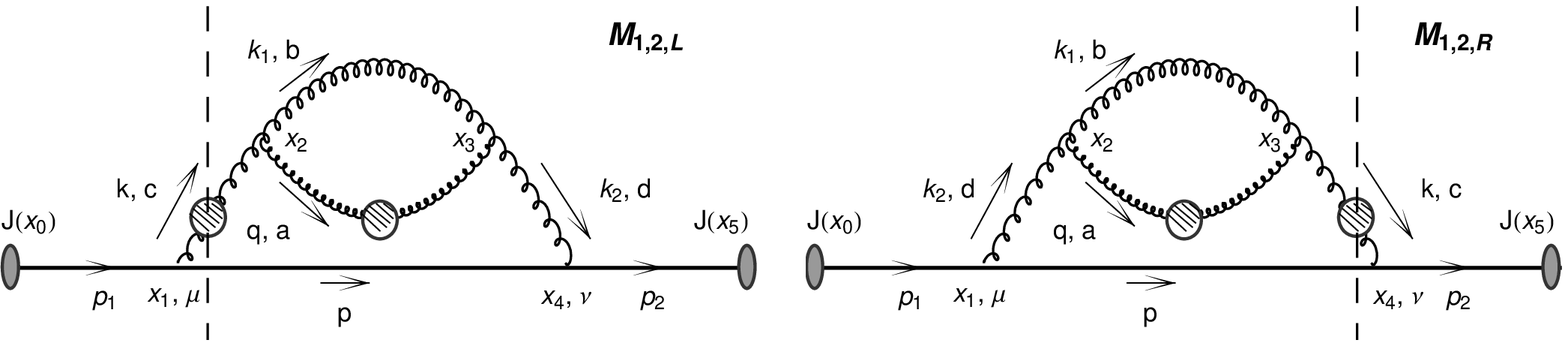}
\caption{Feynman diagram $M_{1,2}$ contributing to the radiative energy loss 
to first order in opacity, labeled in the same way as Fig.~1}
\label{DiagM12RL}
\end{figure}
%

\beqar 
M_{1,2,R}  &=& \int
\prod_{i=0}^{5} \, d x_i \, J(x_0) \, \Delta_{++}^{+} (x_1-x_0) \,
v_\mu^{+} (x_1) \, D_{++}^{+\mu \rho_1} (x_2-x_1) \,
v_{\rho_1\rho_2\rho_3}^{+}(x_2) D_{++}^{+\rho_3\sigma_3} (x_3-x_2)
\,
\nonumber \\
&& \; \; \times\ D_{++}^{+\rho_2 \sigma_2} (x_3-x_2) \,\Delta_{-+}
(x_4-x_1) v_{\sigma_1\sigma_2\sigma_3}^{+} (x_3) \,
D_{-+}^{\sigma_1\nu} (x_4-x_3) \, v_\nu^{-} (x_4) \, \,
\Delta_{--}^{-}(x_5-x_4) \, J(x_5) \,
\nonumber \\
 &&\; \; \times  \theta(x_1^+
-x_0^+) \theta(x_2^+ -x_1^+)\theta(x_3^+ -x_2^+)\theta(x_4^+
-x_5^+)\theta(2L-(x_2-x_0))\nonumber \\
&\approx& - \int_{0}^{\infty}\int_{-\infty}^{\infty}
\prod_{i=1}^{2} \, \frac{d p_i^{+} d^{2} p_i} {(2 \pi)^3 \, 2
p_i^{+}} \frac{ d k_{i}^{+} d^{2} k_{i}} {(2 \pi)^3 \, 2
k_{i}^{+}} \int \frac{ d^{3} k} {(2 \pi)^3 \, 2 \omega} \frac{
d^{3} p} {(2 \pi)^3 \, 2 E} \frac{ d^{4} q} {(2 \pi)^4 } \ 
g^4 \,f^{cba}\ t_c \, f^{abd}t_d
\nonumber \\
&& \; \; \times\ \,
\Big \{(k_{1}+k_{2})^{\rho}\,D_{\rho\sigma}^{>}(q)(k+k_{1})^\sigma \Big \}\, 
\Big \{ (p+p_{1})^\mu\, P_{\mu\rho_1}(k_{2})\,P^{\sigma_1\rho_1}(k_1)\, 
P_{\rho_1\nu}(k)\, (p+p_2)^\nu \Big\} \, I,
\eeqar{M12R_1}
where in the second step we only keep the dominant parts from triple gluon 
vertices (see~\cite{DH_Inf}).

In Eq.~(\ref{M12R_1}) $I$ corresponds to
\beqar I &=& \int \prod_{i=0}^{5}\ dx_i \theta(x_1^+ -x_0^+)
\theta(x_2^ + -x_1^+)\theta(2L-(x_2-x_0)^+)\theta(x_3^+
-x_2^+)\theta(x_4^+
-x_5^+)\,  \nonumber \\
&& \; \; \times \, e^{-ip_{1} (x_{1} -x_{0})} e^{-i k_{2}(x_{2}
-x_{1})} e^{-i k_1(x_{3} -x_{2})}  e^{-i q(x_{3} -x_{2})} e^{-i p
(x_{4} -x_{1})} e^{-ik(x_{4} -x_{3})} e^{-i p_2(x_{5} -x_{4})}\,
J(x_0)\, J(x_5) \, \nonumber \\
&=&  |J(p)|^2 \,(2\pi)^3\,
\delta((p_{1}-p-k_2)^+)\delta^2(\bp_{1}-\bp-\bk_2) (2\pi)^3
\delta((k_{2}-k)^+)\delta^2(\bk_{2}-\bk)
\nonumber \\
&& \; \; \times\ (2\pi)^3 \delta((k_1+q-k)^+)\delta^2(\bk_1+\bq-\bk)
(2\pi)^3 \delta((p_2-k-p)^+)\delta^2(\bp_2-\bp-\bk)\ I_1 \, ,
\eeqar{I_12R_1} and where

\beqar I_1 &=& \int_{0}^{2L} dx_2^{\prime+}
e^{-\frac{i}{2}(k_{2}-k)^- x_2^{\prime+}} \int_{0}^{x_2^{\prime+}}
dx_1^{\prime+}\ e^{-\frac{i}{2}(p_{1}-p-k_{2})^- x_1^{\prime+}}
 \nonumber \\
&& \; \; \times\ \int_{0}^{\infty} dx_3^{\prime+}\
e^{-\frac{i}{2}(k_1+q-k)^- x_3^{\prime+}} \int_{0}^{\infty}
dx_4^{\prime+}\ e^{\frac{i}{2}(p_2-p-k)^-x_4^{\prime+}}\, .
\eeqar{I1_12R_1}
In the last equation, we defined $x_1^{\prime+}=x_1^{+}-x_0^{+}$, 
$x_2^{\prime+}=x_2^{+}-x_0^{+}$,  $x_3^{\prime+}=x_3^{+}-x_2^{+}$ and 
$x_4^{\prime+}=x_4^{+}-x_5^{+}$.

\medskip
After applying  $\delta$ functions
from Eq.~(\ref{I_12R_1}) and by using Eq.~(\ref{pk-}), we obtain $k^-=k_2^-$,
which reduces the 
Eq.~(\ref{I1_12R_1}) to  
\beqar I_1 &=&
\int_{0}^{2L} dx_2^{\prime+}\int_{0}^{x_2^{\prime +}}
e^{\frac{i}{2} \,\xi\, x_1^{\prime+}}dx_1^{\prime +}
\, \int_{0}^{\infty} dx_4^{\prime+}\
e^{-\frac{i}{2}\, \xi\, x_4^{\prime+}} \int_{0}^{\infty}
dx_3^{\prime+}\ e^{-i q^- y}\, \nonumber \\
&& \hspace*{1cm}= \frac{16L}{\xi^2}
\Big[1-\frac{\sin \xi L}{\xi L}+i\frac{1-\cos \xi L}{\xi L}\Big]\int_0^{\infty}dy\,e^{-iq^-y}
\eeqar{I1_12R_2}
In the above equation, we also used $(p_1-k-p)^-=(p_2-p-k)^- = - \xi $, 
$(k_1+q-k)^- \approx q^-$ and $y \equiv \frac{x_3^{\prime+}}{2}$.

Similarly as in~\cite{DH_Inf}, for highly energetic jets
\beqar
&& \hspace{2.5cm}(p+p_{1})^\mu\, P_{\mu\rho_1}(k_{2})
\,P^{\rho_1 \sigma_1}(k)\,P_{\sigma_1\nu}(k)(p+p_2)^\nu) =-\frac{4\bk^2}{\xi^2}
\nonumber \\
&& (k+k_1)^\rho D_{\rho\sigma}^{>}(q)(k+k_1)^{\sigma} \approx 
k^+k_1^+ \theta(1-\frac{q_0^2}{\vq^2}) f(q_0) \, \frac{\bq^2}{\vq^2}
\;  2 \, {\rm Im} 
\left( \frac{1}{q^2{-}\Pi_{L}(q)} - \frac{1}{q^2{-}\Pi_{T}(q)} \right) \, .
\eeqar{PQ12R}

Finally, by using Eqs.~(\ref{I_12R_1}), (\ref{I1_12R_2}) and (\ref{PQ12R}), and after performing the same procedure as in the previous appendices,  
Eq.~(\ref{M12R_1}) reduces to
\beqar M_{1,2,R}  &=& -4 L T \, g^4 \,[t_c,t_a][t_a,t_c]
\int 
\frac{ d^{3} p} {(2 \pi)^3 \, 2 E} \,|J(p)|^2 \, \int 
\frac{ d^{3} k} {(2 \pi)^3 \, 2 \omega} \, 
 \frac{d^{4} q} {(2 \pi)^4} \nonumber \\
 && \;\;
\times\ \frac{\bk^2}{(\bk^2+ \chi)^2} 
\Big[1-\frac{\sin \xi L}{\xi L}+i\frac{1-\cos \xi L}{\xi L}\Big]
\int_0^{\infty}dy\,e^{-iq^-y}
\nonumber\\
&&\;\;\times\ \theta(1-\frac{q_0^2}{\vq^2}) \frac{1}{q_0} \,\frac{\bq^2}{\vq^2}
\;  2 \, {\rm Im} 
\left( \frac{1}{q^2{-}\Pi_{L}(q)} - \frac{1}{q^2{-}\Pi_{T}(q)} \right) \, .
\,\eeqar{M12R_f}

Finally, by using that $M_{1,2,L} $ is a complex conjugate of $M_{1,2,R} $, 
and by performing the same procedure as in Eqs.~(\ref{M102R+L_2})-(\ref{J_q2}), 
one obtains
\beqar 
M_{1,2,R} +M_{1,2,L}  &=& -4LT \, g^4 \,[t_a,t_c][t_c,t_a]
\int \frac{ d^{3} p} {(2 \pi)^3 \, 2 E} \, |J(p)|^2 \, \int
\frac{ d^{3} k} {(2 \pi)^3 \, 2 \omega}  \,
\frac{d^{2} q} {(2 \pi)^2} \,  \frac{\mu^2}{\bq^2(\bq^2+\mu^2)} \nonumber \\
 && \;\;
\times\ \frac{\bk^2}{(\bk^2+\chi)^2}
\Big[1-\frac{\sin \xi L}{ \xi L}\Big] \, .
\eeqar{M12RL_f}

\section{Computation of the tadpole diagrams}
\label{appTadpole}

\begin{figure}[ht]
\vspace{4.3cm} 
\includegraphics{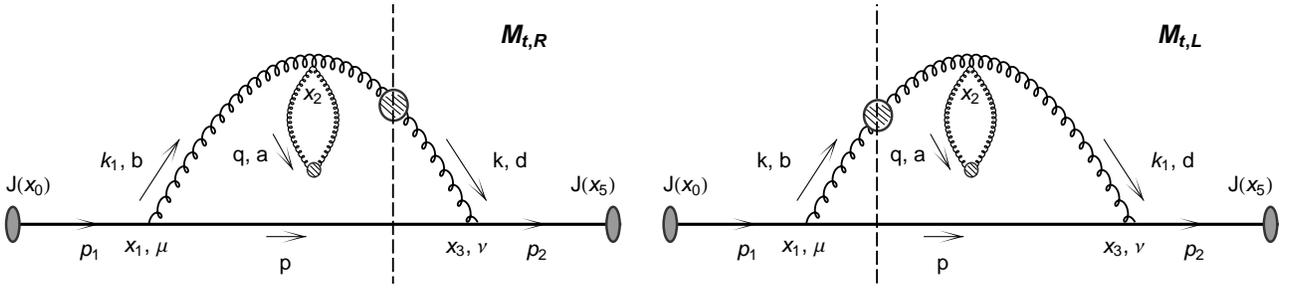}
\caption{Tadpole Feynman diagrams $M_{t,R}$ and $M_{t,L}$, labeled in the 
same way as Fig.~1.}
\label{tadpole}
\end{figure}

{\bf 1.} In this appendix we calculate tadpole diagrams shown in 
Fig.~(\ref{tadpole}), and show that they present negligible contribution 
to the radiative energy loss. 

%
\beqar M_{t,R}  &=& \int
\prod_{i=0}^{4} \, d x_i \, J(x_0) \, \Delta_{++}^{+} (x_1-x_0) \,
v_\mu^{+} (x_1) \, D_{++}^{+\mu \rho} (x_2-x_1) \,
v_{\rho\sigma\lambda\tau}^{+}(x_2) D_{++}^{+\lambda\tau} (x_2) \,
D_{-+}^{\sigma\nu} (x_3-x_2)v_\nu^-(x_3) \nonumber \\
&& \; \; \times\ 
\,\Delta_{-+} (x_3-x_1)  \, \Delta_{--}^{-} (x_4-x_3)  \, J(x_4) \,
\ \theta(x_1^+ -x_0^+)\, \theta(x_2^+ -x_1^+) \, \theta(2L-(x_2-x_0)^+)
\, \theta(x_3^+-x_4^+)\, ,
\eeqar{M_tR_1}
where $v_{\rho\sigma\lambda\tau}^+(x_2) =
-g^2\Big[f_{cea}f_{ceb}(2g_{\rho\sigma}g_{\lambda\tau}-g_{\rho\lambda}g_{\sigma\tau}-g_{\rho\tau}g_{\sigma\lambda})\Big]$, leading to

\beqar
M_{t,R}  &=&  \prod_{i=0}^{4} \, d x_i \, J(x_0) \,
\int_{-\infty}^{\infty}\int_{0}^{\infty}   
\frac{ d p_{1}^{+} d^{2} p_{1}} {(2 \pi)^3 \, 2 p_{1}^{+}}   
e^{-i p_{1} (x_{1} -x_{0})}
\int \frac{ d^{3} p} {(2 \pi)^3 2E } e^{-i p (x_{3} -x_{1})} \, 
(-ig (p+p_1)_\mu t_a) \nonumber \\
&& \; \; \times\ (-1) \int_{-\infty}^{\infty} \int_{0}^{\infty}  \frac{
dk_1^{+} d^2k_1} {(2 \pi)^3 \, 2 k_1^+} e^{-i k_1 (x_{2} -x_{1})}
P^{\mu\rho}(k_1)
\ (-g^2) f^{cea}f_{ceb}(2g_{\rho\sigma}g_{\lambda\tau}-
g_{\rho\lambda}g_{\sigma\tau}-g_{\rho\tau}g_{\sigma\lambda})
 \nonumber \\
 && \; \; \times \ \int \frac{d^4 q} {(2 \pi)^4}\, D^{\lambda\tau >}(q)\,
(-1) \int \frac{ d^{3} k} {(2 \pi)^3 2\omega } e^{-i k (x_{3} -x_{2})}
P^{\sigma\nu}(k) \int \frac{d p_{2}^{+} d^{2} p_{2}} {(2 \pi)^3 \, 2 p_{2}^+} 
e^{-i p_{2} (x_{4} -x_{3})} \, (ig (p+p_{2})_{\nu}t_b) \,\nonumber\\
&&\; \; \times \   \theta(x_1^+ -x_0^+) \, 
\theta(x_2^+ -x_1^+) \, \theta(2L-(x_2-x_0)^+) \, \theta(x_3^+ -x_4^+) \,J(x_4) \
\nonumber \\
&=& \int \frac{ d^{3} k} {(2 \pi)^3 \, 2 \omega} \, 
\frac{ d^{3} p} {(2 \pi)^3 \, 2 E} \, \frac{ d^{4} q} {(2 \pi)^4 } 
\prod_{i=1}^{2} \, \frac{d p_i^{+} d^{2} p_i} {(2 \pi)^3 \, 2 p_i^{+}} 
\frac{ d k_{i}^{+} d^{2} k_{i}} {(2 \pi)^3 \, 2
k_{i}^{+}} \, g^4 (-f^{cea}\, t_a \, f^{ceb}t_b)
\nonumber \\
&&\;\; \times\ (p+p_1)_\mu
P^{\mu\rho}(k_1)(2g_{\rho\sigma}g_{\lambda\tau}-
g_{\rho\lambda}g_{\sigma\tau}-g_{\rho\tau}g_{\sigma\lambda})\, 
D^{\tau\lambda >}(q) P^{\sigma\nu}(k)(p+p_{2})_\nu \, I,
\eeqar{M_tR_2} 
where 
\beqar 
I &=& \int \prod_{i=0}^{4}\ dx_i J(x_0)\,
J(x_4)\,e^{-ip_{1} (x_{1} -x_{0})} e^{-i k_{1}(x_{2} -x_{1})}
e^{-i k(x_{3} -x_{2})}  e^{-i p(x_{3} -x_{1})} e^{-i p_2 (x_{4}
-x_{3})}\nonumber\\
&& \;\;\times\ \theta(x_1^+ -x_0^+) \theta(x_2^ +
-x_1^+)\theta(2L-(x_2-x_1)^+)\theta(x_3^+ -x_4^+)  \nonumber \\
\nonumber \\
&=&  |J(p)|^2 \,(2\pi)^3\,
\delta((p_{1}-p-k_1)^+)\delta^2(\bp_{1}-\bp-\bk_1) (2\pi)^3
\delta((k_{1}-k)^+)\delta^2(\bk_{1}-\bk)
\nonumber \\
&& \; \; \times\ (2\pi)^3 \delta((p_2-p-k)^+)\delta^2(\bp_2-\bp-\bk)
I_1, \,  
\eeqar{I_tR_1} and where

\beqar 
I_1 &=& \int_{0}^{2L} dx_2^{\prime+}
e^{-\frac{i}{2}(k_{1}-k)^- x_2^{\prime+}} \int_{0}^{x_2^{\prime+}}
dx_1^{\prime+}\ e^{-\frac{i}{2}(p_{1}-p-k_{1})^- x_1^{\prime+}}
  \int_{0}^{\infty} dx_3^{\prime+}\
e^{+\frac{i}{2}(p_2-p-k)^- x_3^{\prime+}}
\eeqar{I1_tR_1} 

Here $x_1^{\prime+}=x_1^+-x_0^+$, $x_2^{\prime+}=x_2^+-x_0^+$ and 
$x_3^{\prime+}=x_3^+-x_4^+$. Also, by using $\delta$ functions from
Eq.~(\ref{I_tR_1}) and Eq.~(\ref{pk-}) it follows 
\beqar
p_2 = p_1; \hspace*{0.2cm}  k = k_1; \hspace*{0.2cm} 
(p_1 -p - k)^- = (p_2 -p - k)^-\approx - \xi
\eeqar{pkq_relations_tR}

After using relations from~(\ref{pkq_relations_tR}), Eq.(\ref{I1_tR_1} ) becomes%
\beqar 
I_1 &=&
\int_{0}^{2L} dx_2^{\prime+}\int_{0}^{x_2^{\prime +}}
e^{\frac{i}{2} \xi x_1^{\prime+}}
\int_{0}^{\infty} dx_3^{\prime+}\
e^{-\frac{i}{2} \xi x_3^{\prime+}}=
\frac{8L}{\xi^2}\Big[1-\frac{\sin \xi L}{\xi L}+i\frac{1-\cos
\xi L}{\xi L}\Big] \, .
\eeqar{I1_tR_f}

By using Eqs.~(\ref{I1_tR_f}) and~(\ref{I_tR_1}), Eq.~(\ref{M_tR_2}) becomes
\beqar
M_{tR}&=& 8g^4 L \,[t_a,t_c][t_c,t_a]
\int  \frac{ d^{3} p} {(2 \pi)^3 \, 2 E} \, |J(p)|^2 \,
\int  \frac{ d^{3} k} {(2 \pi)^3 \, 2 \omega} \, 
\frac{ d^{4} q} {(2 \pi)^4}  \frac{1}{(2E^+)^2} \frac{1}{2k_1^+}
\nonumber\\
&&\;\; \times\ \left(8p^\mu P_{\mu\rho}(k)
\,P_\nu^{\rho}(k)\,p^{\nu}D_{\lambda}^{>\lambda}(q)-
8p^\mu P_{\mu\rho}(k) \,D^{\rho\sigma
>}(q)P_{\sigma\nu}(k)\,p^{\nu} \right)\,\nonumber\\
&&\;\;\times\ \frac{x^2E^{+
2}}{(k^2+\chi)^2}\Big[1-\frac{\sin \xi L}{\xi L}+i\frac{1-\cos
\xi L}{\xi L}\Big] \, . 
\eeqar{M_tR_3}

To proceed further, lets first calculate
\beqar 
p^\mu P_{\mu\rho}(k)
\,P_\nu^{\rho}(k)\,p^{\nu}D_{\lambda}^{>\lambda}(q)-
p^\mu P_{\mu\rho}(k) \,D^{\rho\sigma
>}(q)P_{\sigma\nu}(k)\,p^{\nu} = p^\mu P_{\mu\nu}(k)
p^{\nu} \, D_{\lambda}^{>\lambda}(q)-p^\mu P_{\mu\rho}(k)
\,D^{\rho\sigma>}(q)P_{\sigma\nu}(k)\,p^{\nu} \,.
\eeqar{pq_relations_tR} 
We here use
\beqar
p^\mu P_{\mu\rho}(k) \,P_\nu^{\rho}(k)\,p^{\nu} = - \frac{\bk^2}{x^2}\, ; 
\; \; \;  P_\lambda^\lambda(q)=2 \,; 
\; \; \;  Q_\lambda^\lambda(q)=1-\frac{q_0^2}{q^2} \, ,
\eeqar{PQ_lambda}
leading to
\beqar
D_\lambda^{>\lambda}(q)=
\theta\Big(1-\frac{q_0^2}{\vq^2} \Big) \, (1+f(q^0)) \, 2 \, {\rm Im} \, 
\Big(\frac{2}{q^2-\Pi_T(q)}+\frac{1-\frac{q_0^2}{\vq_2}}{q^2-\Pi_L(q)} \Big) 
\,.
\eeqar{D_lambda}
Furthermore, in Coulomb gauge
\beqar
P_{\mu\rho}(k)Q^{\rho\sigma}(q)P_{\sigma\lambda} (k)\equiv 0 \, ,
\eeqar{PQP_tR}
leading to
\beqar
p^\mu P_{\mu\rho}(k) \,D^{\rho\sigma>}(q)P_{\sigma\nu}(k)\,p^{\nu} &=&
p^\mu P_{\mu\rho}(k) \,P^{\rho\sigma}(q)P_{\sigma\lambda}\,p^{\nu} \nonumber \\
&=&-p^2+\frac{(p\, k)^2}{k^2}+\frac{(p\, q)^2}{q^2}-
2\frac{(p\, k)(p\, q)(q\, k)}{k^2\, q^2}
+\frac{(p\, k)(q\, k)^2}{(k^2)^2\, q^2}\nonumber \\
&=& -\frac{\bk^2}{x^2} \, \theta\Big(1-\frac{q_0^2}{\vq^2} \Big)\,
(1+f(q^0)) \, 2 \, {\rm Im} \Big(\frac{1+ \frac{(\bq \dot \bk)^2}
{\bk^2\vq^2}}{q^2-\Pi_T(q)}+\frac{1-\frac{q_0^2}{\vq_2}}{q^2-\Pi_L(q)} \,.
\Big)\eeqar{PDP_tR}

By using Eqs.~(\ref{pq_relations_tR})-(\ref{PDP_tR}), Eq.(\ref{M_tR_3}) becomes
\beqar 
M_{t,R}  &=&  -8  L \, g^4 \,[t_a,t_c][t_c,t_a] 
\int \frac{ d^{3} p} {(2 \pi)^3 \, 2 E}\, |J(p)|^2  \, 
\int \frac{ d^{3} k} {(2 \pi)^3 \, 2 \omega} \,
\frac{ d^{4} q} {(2 \pi)^4}  \,\frac{\bk^2}{(\bk^2+ \chi)^2}\nonumber\\
&&\; \; \times\ \Big[1-\frac{\sin \xi L}{\xi L}+
i \frac{1-\cos \xi L}{\xi L}\Big]\frac{1}{k^+}\int\frac{d^4q}{(2\pi)^4} \,
\theta\Big(1-\frac{q_0^2}{\vq^2} \Big)\,
\frac{T}{q_0}\, 2 \, {\rm Im} \Big(\frac{1- \frac{(\bq \dot \bk)^2}
{\bk^2\vq^2}}{q^2-\Pi_T(q)}\Big) 
\, .
\nonumber\\
\eeqar{M_tR_f}

Since $M_{t,L} $ is a complex conjugate of $M_{t,R} $,
one finally obtains 
\beqar 
M_{t,R} +M_{t,L} =
8LTg^4 \,[t_a,t_c][t_c,t_a]
\int  \frac{d^{3} p} {(2 \pi)^3 \, 2 E} \, |J(p)|^2 \,
\int \frac{ d^{3} k} {(2 \pi)^3 \, 2 \omega}  \frac{\bk^2}{(\bk^2+Mx^2+mg^2)^2}
 \Big[1-\frac{\sin \xi L}{\xi L}\Big] I_t
\eeqar{MtRL}
where
\beqar
I_t=- \frac{2}{\omega}\int\frac{d^4q}{(2\pi)^4} \,
\theta\Big(1-\frac{q_0^2}{\vq^2} \Big)\,
\frac{1}{q_0} \, {\rm Im} \Big(\frac{1- \frac{(\bq \dot \bk)^2}
{\bk^2\vq^2}}{q^2-\Pi_T(q)}\Big) \, . \eeqar{I_t} 
Finally, we want to show that $I_t \ll \int\frac{d \bq^2}{4 \pi} v(\bq^2)$ 
(where $v(\bq^2) =\frac{\mu^2}{\bq^2(\bq^2+\mu^2)}$), leading to the conclusion that 
tadpole contribution is negligible to the first order in opacity radiative 
energy loss. To do this, we first observe that 
$0 \le \frac{(\bq \dot \bk)^2} {\bk^2\vq^2} \le 1$, leading to
\beqar
0 \le I_t \le I_{t,max},
\eeqar{It_min_max}
where
\beqar
I_{t,max}= -\frac{2}{\omega}\int\frac{d^4q}{(2\pi)^4} \,
\theta\Big(1-\frac{q_0^2}{\vq^2} \Big)\,
\frac{1}{q_0} \, 2 \, {\rm Im} \Big(\frac{2}{q^2-\Pi_T(q)}\Big) =
\int \frac{d \vq^2}{4 \pi} \frac{|\vq|}{\omega} \, J_{t,max},(\vq^2) \, ,
\eeqar{I_t_max} 
where we defined $y\equiv \frac{q_0}{|\vq|}$, and 
\beqar
J_{t,max}(\vq^2) = \frac{4}{\pi^2} \, \int_{0}^{1}
 \frac{dx }{x} \, 
 {\rm Im} \Big(\frac{1}{\vq^2 (1-x^2)+\Pi_T(x)}\Big) \, ,
\eeqar{J_t_max}

\begin{figure}[h]
\vspace*{6 cm} 
\includegraphics{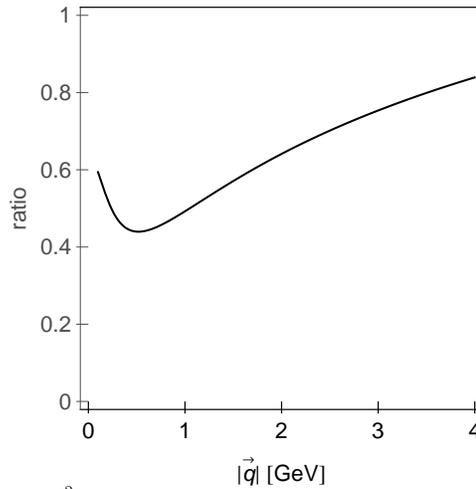}
\caption{Ratio $\frac{J_{t,max}(\vq^2)}{v(\vq^2)}$ is shown as a function of 
momentum $|\vq|$, for Debye mass $\mu=0.5$. }
\label{tadpole_mm}
\end{figure}
%
Furthermore, we want to show that 
$I_t \sim \int\frac{d \vq^2}{4 \pi} \frac{|\vq|}{\omega} v(\vq^2) $. 
To do this, we have to prove that 
$J_{t,max}(\vq^2)$ is comparable with $v(\vq^2)$. We numerically confirmed 
that $\frac{J_{t,max}(\vq^2)}{v(\vq^2)} \lesssim 1 $, as demonstrated in 
Fig.~\ref{tadpole_mm} for typical value of Debye mass $\mu=0.5$. That is, in 
Fig.~\ref{tadpole_mm} we see that the absolute values of the ratios are 
notably smaller than 1. We also checked that the same conclusion is valid 
independently on the value of Debye mass. Having in mind that 
$ \frac{|\vq|}{\omega} \ll 1$, and by using $I_t \sim \int\frac{d \vq^2}{4 \pi} \frac{|\vq|}{\omega} v(\vq^2) $,  it becomes evident that 
$I_t \ll \int\frac{d \vq^2}{4 \pi}  v(\vq^2) $, which leads to the conclusion that tadpoles present a negligible contribution to the $1^{st}$ order in 
opacity radiative energy loss.



\begin{thebibliography}{80}
\bibitem{fn1}
Note that in the widely used static approximation this problem 
is absent: only longitudinally polarized gluons contribute to the static 
potential (see e.g.~\cite{Le_Bellac,Gyulassy_Wang,GLV}), leading to 
individually convergent results for each of the scattering diagrams.
\end{thebibliography}
\end{document}